\documentclass[11pt,a4paper]{article}    
\usepackage{jheppub}
\makeatletter

\usepackage{fontenc}
\usepackage[utf8]{inputenc}
\usepackage{float}
\usepackage{units}
\usepackage{amsmath}
\usepackage{amssymb}
\usepackage{graphicx}
\usepackage{color}
\usepackage{esint}
\usepackage{youngtab}

\usepackage{nicematrix}
\usepackage{bbold}

\hypersetup{%
	pdftitle   = {Strings in the Carroll regime and near black holes},
	pdfkeywords = {Carrollian strings, black holes, non-Lorentzian geometry},
	pdfauthor  = {Arjun Bagchi, Aritra Banerjee, Emil Have, Jelle Hartong, Kedar Kolekar},
}

\usepackage{wrapfig}
\usepackage{slashed}
\usepackage{epsfig}
\usepackage{tensor}

\def\be{\begin{eqnarray}}
\def\ee{\end{eqnarray}}
\newcommand{\nn}{\nonumber}

\def\Dslash{\,\,{\raise.15ex\hbox{/}\mkern-12mu D}}
\def\Dbarslash{\,\,{\raise.15ex\hbox{/}\mkern-12mu {\bar D}}}
\def\delslash{\,\,{\raise.15ex\hbox{/}\mkern-9mu \partial}}
\def\delbarslash{\,\,{\raise.15ex\hbox{/}\mkern-9mu {\bar\partial}}}
\def\pslash{\,\,{\raise.15ex\hbox{/}\mkern-9mu p}}
\def\calDslash{\,\,{\raise.15ex\hbox{/}\mkern-12mu {\cal D}}}

\newcommand{\D}{{\partial}}

\newcommand{\ZZ}{{\mathbb Z}}

\newcommand{\RR}{{\mathbb R}}

\def\lae{\mathrel{\mathop{\smash{\lower .5 ex \hbox{$\stackrel<\sim$}}}}}
\def\lae{\mathrel{\mathop{\smash{\lower .5 ex \hbox{$\stackrel>\sim$}}}}}

\usepackage[useregional]{datetime2}

\usepackage{float}
\usepackage{tikz}
\usepackage{pgfplots}
\pgfplotsset{compat=1.16}
\usetikzlibrary{calc,arrows,cd}
\usetikzlibrary{shapes,decorations}
\usepackage{makecell}
\pgfdeclarelayer{nodelayer}
\pgfdeclarelayer{edgelayer}
\pgfsetlayers{edgelayer,nodelayer,main}
\tikzstyle{ghost}=[fill={rgb,255: red,140; green,76; blue,150}, draw=black, shape=circle,scale=0.6]
\tikzstyle{real_ghost}=[fill=none, draw=none, shape=circle]
\tikzstyle{none}=[fill=none, draw=none, shape=circle]
\tikzstyle{Red_Circle}=[fill=none, draw=red, shape=circle]

\tikzstyle{BlackLine}=[-, draw=black, fill=white]
\tikzstyle{Arrow}=[<-, thick]
\tikzstyle{thin_black_line}=[-, fill=blue]
\tikzstyle{thin_black_line_red}=[-, fill=red]
\tikzstyle{thin_black_line_semi_purple}=[-, fill={rgb,255: red,1; green,1; blue,255}]
\tikzstyle{thin_black_line_purple}=[-, fill={rgb,255: red,200; green,1; blue,255}]
\tikzstyle{thin_black_line_turquoise}=[-, fill={rgb,255: red,1; green,200; blue,200}]
\tikzstyle{thin_black_line_gray}=[-, fill=gray]
\tikzstyle{thin_black_line_null}=[-, fill=orange]
\tikzstyle{Double_Arrow}=[<->]
\tikzstyle{Dashed_arrow}=[->, dashed, draw=red]
\tikzstyle{Dashed_arrow_gray}=[->, dashed, draw=gray]
\tikzstyle{BlackLine_dash}=[-, draw=red, fill=white, dashed]
\tikzstyle{Arrow_order}=[<-, draw=red]

\DeclareMathAlphabet{\mathdutchcal}{U}{dutchcal}{m}{n}
\SetMathAlphabet{\mathdutchcal}{bold}{U}{dutchcal}{b}{n}
\DeclareMathAlphabet{\mathdutchbcal}{U}{dutchcal}{b}{n}
\title{
Strings near black holes are Carrollian - Part II
}

\author[1]{Arjun Bagchi,}
\author[2,3]{Aritra Banerjee,}
\author[4]{Jelle Hartong,}
\author[5]{Emil Have,}
\author[6]{Kedar S.~Kolekar.}
\author{\\}
\affiliation[1]{Indian Institute of Technology Kanpur, Kanpur 208016, India.\\}
\affiliation[2]{Birla Institute of Technology and Science, Pilani Campus, Rajasthan 333031, India.\\}
\affiliation[3]{Asia Pacific Center for Theoretical Physics, Postech, Pohang 37673, Korea.\\}
\affiliation[4]{School of Mathematics and Maxwell Institute for Mathematical Sciences,\\
	University of Edinburgh, Peter Guthrie Tait Road, Edinburgh EH9 3FD, UK.\\}
\affiliation[5]{Niels Bohr International Academy, Niels Bohr Institute,\\ University of Copenhagen, Blegdamsvej 17, DK-2100 Copenhagen Ø, Denmark.\\}
\affiliation[6]{Yau Mathematical Sciences Center, Tsinghua University, Beijing 100084, China.\\}

\emailAdd{abagchi@iitk.ac.in}
\emailAdd{aritra.banerjee@pilani.bits-pilani.ac.in}
\emailAdd{j.hartong@ed.ac.uk}
\emailAdd{emil.have@nbi.ku.dk}
\emailAdd{kedarsk@mail.tsinghua.edu.cn}

\abstract{We study classical closed bosonic strings probing the near-horizon region of a non-extremal black hole and show that this corresponds to understanding string theory in the Carroll regime. This is done by first performing a Carroll expansion and then a near-horizon expansion of a closed relativistic string, subsequently showing that they agree. Concretely, we expand the phase space action in powers of $c^2$, where $c$ is the speed of light, assuming that the target space admits a string Carroll expansion (where two directions are singled out) and show that there exist two different Carroll strings: a magnetic and an electric string. The magnetic string has a Lorentzian worldsheet, whereas the worldsheet of the electric string is Carrollian. The geometry near the horizon of a four-dimensional (4D) Schwarzschild black hole takes the form of a string Carroll expansion (a 2D Rindler space fibred over a 2-sphere). We show that the solution space of relativistic strings near the horizon bifurcates and the two sectors precisely match with the magnetic/electric Carroll strings with an appropriate target space.
Magnetic Carroll strings near a black hole shrink to a point on the two-sphere and either follow null geodesics or turn into folded strings on the 2D Rindler spacetime. Electric Carroll strings 
wrap the two-sphere and follow a massive geodesic in the Rindler space. Finally, we show that 4D non-extremal Kerr and Reissner-Nordstr\"om black holes also admit string Carroll expansions near their outer horizons, indicating that our formulation extends to generic non-extremal black holes.}

\usepackage{todonotes}
\usepackage{mathrsfs}
\usepackage{amscd}

\usepackage{amsthm}

\theoremstyle{remark}

\usepackage{tcolorbox}
\usepackage{cancel}
\usepackage{xcolor}

\usepackage{tikz}

\DeclareFontFamily{U}{skulls}{}
\DeclareFontShape{U}{skulls}{m}{n}{ <-> skull }{}

\renewcommand{\i}{\text{i}}

\begin{document}
\pagestyle{plain} \setcounter{page}{1}
\newcounter{bean}
\baselineskip16pt \setcounter{section}{0}
\maketitle
\flushbottom
\section{Introduction}
\label{sec:intro}

We are in the era of black hole detection, both directly and indirectly. With the first photographs of black holes and the advent of gravitational wave spectroscopy, the notion of a black hole is no longer just a theoretical construct. This makes it imperative to construct a theory of Quantum Gravity that allows us to describe all aspects of black holes. One of the very important questions to answer in this context is how black holes are perceived by different probes and observers.

\subsection*{Strings near black holes}
String theory is perhaps the most viable of the currently available frameworks for Quantum Gravity, and the question of how a string behaves in the region of spacetime near a black hole horizon is an important aspect of understanding the interplay between string theory and black holes. A theory of Quantum Gravity should lead to the resolution of spacetime singularities, and so formulating string theory near black hole singularities, or near the initial Big Bang singularity, is an important direction of research which deserves further study.\footnote{See \cite{deVega:1987veo, Dijkgraaf:1991ba, Susskind:1993if,Susskind:1994sm,Susskind:1994uu} and references therein for a non-exhaustive list of earlier works on this topic.} Before reaching the singularity of a black hole, the description of which requires a formulation of string theory \textit{inside} the horizon of the black hole, one should first understand what happens when a string approaches the horizon from the outside. This is the question we concern ourselves with in the present manuscript. 

\subsection*{Near-horizon regions and Carrollian geometry}
It is well-known that the near-horizon (NH) region of a generic non-extremal black hole contains a 2D Rindler spacetime. However, the near-horizon region for a non-extremal black hole is not a solution of the same equations of motion as are solved by the black hole spacetime. Rather, it is an approximate solution that can be used as the starting point for a Taylor expansion that approximates the original black hole spacetime. This is in strong contrast with the case of an extremal black hole where the near-horizon region contains an $\mathsf{AdS}_2$ factor~\cite{Kunduri:2007vf} 
which does solve the same equations as the extremal black hole -- a throat develops that becomes an asymptotic region with associated decoupling features. 

For non-extremal black holes, the situation is thus more difficult. The metric evaluated on the horizon of a non-extremal black hole is degenerate and is known to form a Carrollian geometry.\footnote{Carrollian geometry describes null hypersurfaces \cite{hartong:2015xda} and therefore, in particular, the horizon of a black hole is a Carrollian manifold \cite{Donnay:2019jiz}.} However, if we include the (radial) direction away from the horizon the metric remains non-degenerate and sufficiently close to the horizon the metric can be viewed as admitting a string Carroll expansion in which a 2-dimensional Rindler spacetime is fibred over a 2-sphere (for the case of a spherically symmetric black hole\footnote{In appendix \ref{Kerr} we show that a non-extremal Kerr black hole also admits a string Carroll expansion near the horizon but in the bulk of this paper we restrict our attention to the spherically symmetric case.}) \cite{Bagchi:2023cfp}. Even though the full 4-dimensional near-horizon metric remains Lorentzian it is quite advantageous to write the geometry and the probes on this geometry in the language of non-Lorentzian geometry which in this case is string Carroll geometry~(see \cite{Bergshoeff:2022eog,Oling:2022fft,Hartong:2022lsy,Baiguera:2023fus} for a set of review articles on non-Lorentzian geometry, strings and field theory).\footnote{The most familiar example of a non-Lorentzian geometry is Newton--Cartan geometry, which arises from the degeneration of the metric that occurs when the speed of light is sent to infinity. If the speed of light is instead taken to zero, the resulting degeneration of the metric leads to the emergence of Carrollian geometry.} In this paper, we build on our initial observations of \cite{Bagchi:2023cfp} and paint a more complete picture of strings near black holes and the importance of Carroll structures in this framework. 

\subsection*{Carroll regime}
As we alluded to above, Carrollian physics generically appears on null hypersurfaces or when taking the speed of light $c$ to zero in a relativistic theory. Recent years have borne witness to a surge of interest in Carrollian geometry, due in large part to its relevance in the context of flat space holography~\cite{Bagchi:2010zz, Barnich:2012aw, Bagchi:2012xr,Barnich:2012xq,Bagchi:2012cy,Bagchi:2014iea,Hartong:2015usd, Bagchi:2016bcd,Donnay:2022aba,Bagchi:2022emh,Figueroa-Ofarrill:2021sxz,Donnay:2022wvx, Bagchi:2023fbj, Saha:2023hsl, Bagchi:2023cen, Mason:2023mti,Have:2024dff}, since the conformal version of Carrollian structures have been shown to be isomorphic to the Bondi--Metzner--Sachs (BMS) algebra \cite{Bondi:1962px,Sachs:1962wk, duval:2014uoa,Duval:2014lpa}, which in turn is the asymptotic symmetry algebra for flat Minkowski spacetime. However, in recent times, Carrollian physics has also appeared in condensed matter systems such as fractons~\cite{Bidussi:2021nmp,Marsot:2022imf, Figueroa-OFarrill:2023vbj,Figueroa-OFarrill:2023qty} and systems with flat bands~\cite{Bagchi:2022eui} as well as cosmology \cite{deBoer:2021jej}. It also plays a role in fluid models for the quark-gluon plasma~\cite{Bagchi:2023ysc, Bagchi:2023rwd}, where Carrollian fluids~\cite{Ciambelli:2018wre,Redondo-Yuste:2022czg, Ciambelli:2018xat, Campoleoni:2018ltl, Petkou:2022bmz, Freidel:2022bai, Freidel:2022vjq, deBoer:2023fnj,Armas:2023dcz} were used to describe Bjorken flow and its conformal generalisation. Finally, one can define interesting Carroll approximations of GR \cite{Bergshoeff:2017btm,Grumiller:2020elf,Hansen:2021fxi,Campoleoni:2022ebj,Perez:2021abf} which has a rich solutions space including Kasner spacetimes \cite{Hansen:2021fxi,deBoer:2023fnj} and Carroll black holes \cite{Ecker:2023uwm,Aggarwal:2024gfb}.

In many cases, instead of the strict $c\to 0$ limit, it is advantageous to consider the \textit{Carroll regime}, where the speed of light is small but finite. This naturally leads to the notion of an expansion in powers of\footnote{For simplicity we expand in even powers of $c$ only. In the Galilean regime, where an expansion in inverse powers of the speed of light is considered (see, e.g.,~\cite{VandenBleeken:2017rij,Hansen:2018ofj,Hansen:2019vqf,Bergshoeff:2019ctr,Hansen:2020pqs,Hartong:2023yxo}), the inclusion of odd powers was considered in~\cite{Ergen:2020yop,Hartong:2023ckn}.} $c^2$ around the $c\to 0$ Carrollian theory.

The string Carroll expansion, in contrast to the \textit{particle} Carroll expansion~\cite{Hansen:2021fxi,deBoer:2023fnj}, singles out two special directions that get rescaled by factors of $c^2$. In this way, the Minkowski metric $\eta(c)$ of $(d+2)$-dimensional flat Lorentzian spacetime in Cartesian coordinates can be written as
\begin{equation}
\label{eq:Minkowski-metric-intro}
    \eta(c) = dx^idx^i -c^2dx^0dx^0 + c^2 dx^1 dx^1\,,
\end{equation}
where $\mu=0,\dots,d+2$ and $i = 2,\dots,d+1$. The two-dimensional subspace $(x^0,x^1)$ is the longitudinal space, while the remaning $x^i$-directions parameterise the transverse space. Generalising this to a general Lorentzian metric $g(c)$, which is analytic in $c^2$ by assumption, we may decompose $g(c)$ into vielbeine as
\begin{equation}
\label{eq:metric-intro}
    g(c) = E^i(c)E^i(c) - c^2 E^0(c) E^0(c) + c^2 E^1 E^1 \,,
\end{equation}
which generalises the flat space decomposition~\eqref{eq:Minkowski-metric-intro}. The vielbeine $E(c)$ themselves admit expansions in $c^2$ of the form $E(c) = e+c^2e_{(1)} + \mathcal{O}(c^4)$, which means that the string Carroll expansion of the metric $g(c)$ in~\eqref{eq:metric-intro} can be written as:
\begin{equation}
    g(c) = e^i e^i + c^2(-  e^0 e^0 + e^1 e^1) + 2c^2 e^ie_{(1)}^i + \mathcal{O}(c^4) = h + c^2 \tau + c^2 \Phi + \mathcal{O}(c^4)\,, 
\end{equation}
where we have defined $h = e^i e^i$, $\tau = -  e^0 e^0 + e^1 e^1$ and $\Phi = 2e^i e^i_{(1)}$. The corresponding gauge symmetries are obtained by expanding the local Lorentz transformations and diffeomorphisms in powers of $c^2$.

\subsection*{String theory in the Carroll regime}
Before discussing strings probing black hole NH regions, we will formulate string theory in the Carroll regime. This general formulation has potential wider applications, although we will focus exclusively on strings near black holes later in the paper. 

In the first half of the paper, where we develop the formalism, we begin by considering closed bosonic strings in the Carroll regime by performing a string Carroll expansion in powers of $c^2$ of relativistic string theory in the phase space formulation using the techniques developed in~\cite{Hartong:2021ekg,Hartong:2022dsx,Hartong:2024ydv}. Building on previous work~\cite{Bagchi:2023cfp}, where we expanded the Polyakov action to obtain a theory of magnetic Carroll strings, we here demonstrate that the expansion of the phase space action reproduces this magnetic Carroll string, with a Lorentzian worldsheet, as well as an electric Carroll string which has a Carrollian worldsheet.\footnote{See also~\cite{Cardona:2016ytk,Blair:2023noj,Gomis:2023eav,Harksen:2024bnh,Casalbuoni:2024jmj} for other works on Carrollian string theory.} This electric Carroll string was previously encountered in the duality web of~\cite{Blair:2023noj,Gomis:2023eav}, where it was obtained via a series of T-dualities. 

In order to expand the relativistic string action $S(c) = \int d\sigma^0\,L(c)$, where $\sigma^\alpha$ for $\alpha = 0,1$ are worldsheet coordinates, we must both expand the background as in~\eqref{eq:metric-intro} and the fields on which $L(c)$ depends, e.g., the embedding fields $X^\mu(c)$ would expand as
\begin{equation}
\label{eq:X-exp-intro}
    X^\mu(c) = x^\mu + c^2 y^\mu + \mathcal{O}(c^4)\,.
\end{equation}
The magnetic Carroll string may be obtained by expanding the relativistic Polyakov Lagrangian~\cite{Bagchi:2023cfp} or equivalently from a particular expansion of the phase space action, which is obtained by demanding that the Lagrange multipliers and the tension scale in a particular way with $c$.
The LO theory in the expansion of the Polyakov Lagrangian in flat space~\eqref{eq:Minkowski-metric-intro} is simply
\begin{equation}
    \tilde L_{\text{LO}} = -\frac{\tilde T}{2} \oint d\sigma^1\, \sqrt{-\gamma_{(0)}}\gamma^{\alpha\beta}_{(0)} \D_\alpha x^i\D_\beta x^i\,,
\end{equation}
where $\tilde T$ is the magnetic tension, and where $\gamma_{(0)\alpha\beta}$ is a Lorentzian worldsheet metric. Since this theory seeks to embed a Lorentzian worldsheet in $d$-dimensional Euclidean space, the solutions $x^i$ are necessarily trivial. The NLO Polyakov theory gives equations for $y^i$ and $x^A$ for $A = 0,1$, which can have nontrivial solutions. 

The electric Carroll string, on the other hand, is obtained by assuming a different scaling for the Lagrange multipliers that appear in the phase space action and the tension. In that case, the phase space Lagrangian expands as $L = \hat L_{\text{LO}} + \mathcal{O}(c^2)$, where, unlike for the magnetic Carroll string, the LO theory already gives rise to interesting dynamics. We will refer to this LO theory as the electric Carroll string, and in flat space~\eqref{eq:Minkowski-metric-intro} the electric Carroll string Lagrangian turns out to be:
\begin{equation}
\label{eq:electric-theory-intro}
    \hat L_{\text{LO}} = \oint d\sigma^1\, \mathbb e\left[ \hat T^2 \D_\alpha x^i \tilde P_i + \frac{1}{2} \eta_{AB}\mathbb v^\alpha\mathbb v^\beta \D_\alpha x^A \D_\beta x^B - \frac{\hat T^2}{2}\mathbb e^\alpha\mathbb e^\beta\D_\alpha x^i\D_\beta x^i \right]\,,
\end{equation}
where $\hat T$ is the electric string tension, $\tilde P_i$ is a Lagrange multiplier, $\eta_{AB}$ is the two-dimensional Minkowski metric in the longitudinal directions, and $(\mathbb{v}^\alpha,\mathbb{e}^\alpha)$ is a Carroll structure on the worldsheet, while $\mathbb e = (\det(\mathbb v^\alpha,\mathbb e^\alpha))^{-1}$ is the measure. This latter property is a distinguishing feature of the electric Carroll string: as mentioned earlier, while both the magnetic and electric string have target spaces that are obtained by the string Carroll expansion, the electric string has a Carrollian worldsheet, while the magnetic string has a Lorentzian worldsheet.

\subsection*{Near-horizon strings are Carrollian}
After developing a general theory of strings in the Carroll regime by considering the string $c^2$ expansion of the relativistic phase space action, we will use the fact that the near-horizon expansion of a non-extremal black hole, and concretely a four-dimensional Schwarzschild black hole, takes the form of a string Carroll expansion, where the role of $c^2$ is now played by the (dimensionless) distance to the horizon $\epsilon$. For a stationary observer at infinity, the expansion for the near-horizon geometry of a Schwarzschild black hole takes the form~\cite{Bagchi:2023cfp}:
\begin{equation}
    ds^2 = r^2_hd\Omega^2 + \epsilon\left[ ds^2_{\text{Rindler}} + 2\mathdutchcal r^2 d\Omega^2 \right] + \mathcal{O}(\epsilon^2)\,,
\end{equation}
where $r_h$ is the Schwarzschild radius, $d\Omega^2$ is the metric on a $2$-sphere, and $(t,\mathdutchcal r)$ are coordinates on the 2D Rindler space with metric $ds^2_{\text{Rindler}}$. Comparing this with~\eqref{eq:metric-intro} with $\epsilon = c^2$, we may read off the corresponding string Carroll structures. The near-horizon expansion of strings again leads to the magnetic and electric Carroll strings as discussed above\footnote{However, one cannot simply replace $c^2 \to \epsilon$ in the expansions of the action, since certain factors of $c$ appear for dimensional reasons unrelated to the string Carroll form of the near-horizon expansion.}, and we will explicitly construct solutions to both the magnetic and electric Carroll strings on this background.

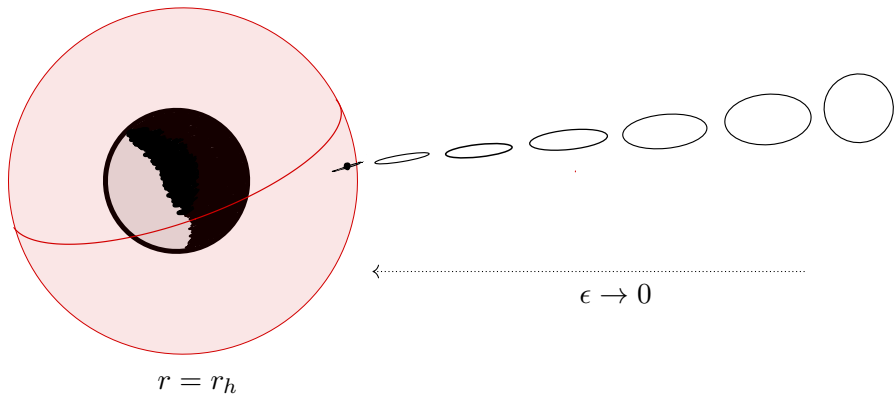
\begin{figure}[ht!]
	\centering
	\includegraphics[width=0.8\textwidth]{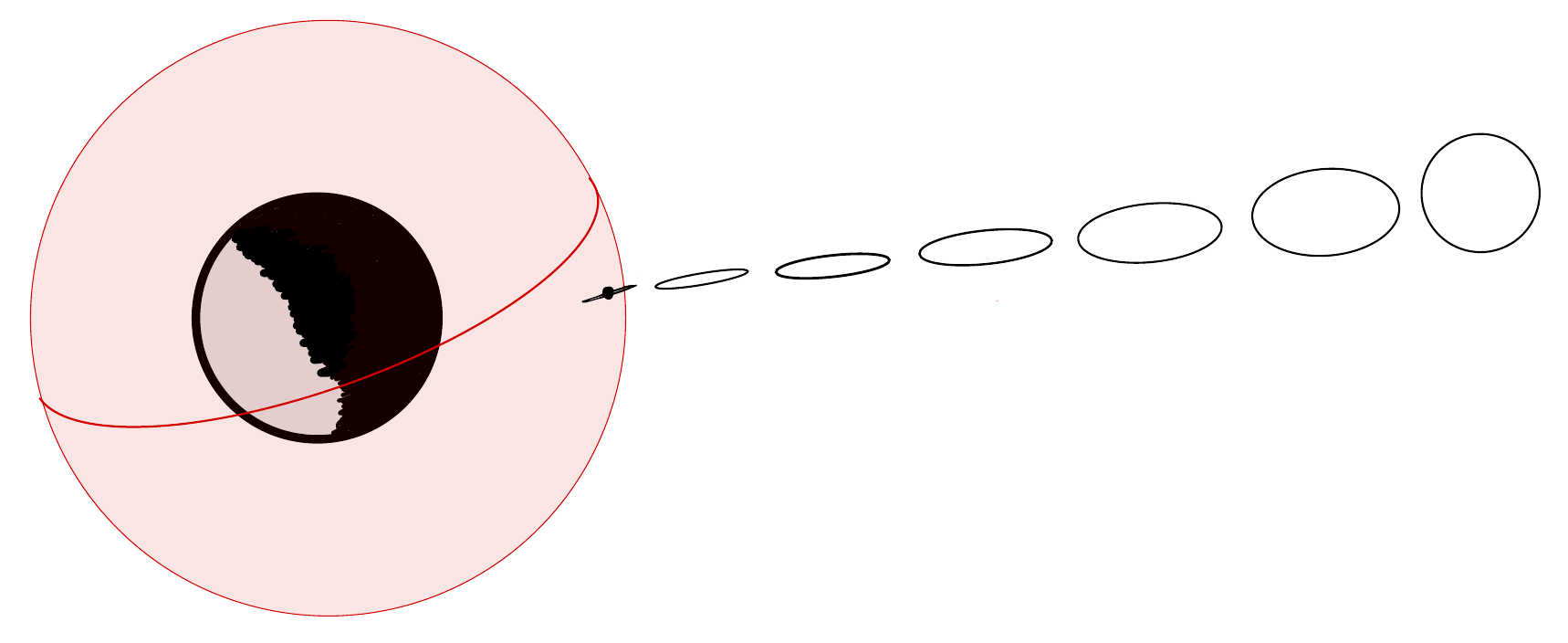}
	\begin{tikzpicture}[overlay]
	\begin{pgfonlayer}{nodelayer}
			\node [style=real_ghost] (0) at (-4, 1) {{$\epsilon\rightarrow 0$}};
			\node [style=real_ghost] (2) at (-9.5, -0.2) {$r=r_h$};
			\draw [densely dotted,<-] (-7.2,1.3) to (-1.5,1.3);
	\end{pgfonlayer}
	\end{tikzpicture}
	\caption{A scenario where an asymptotic observer sees a closed string collapse to a line as it approaches the horizon of the black hole, leading to a folded string. }
	\label{fig:NH-Carroll-string-fold}
\end{figure}

In the case of the magnetic Carroll string, we find two classes of solutions. For both of these classes, as discussed above, the LO embedding into the transverse $2$-sphere is trivial, but the string in the longitudinal 2D Rindler space either shrinks to a point and follows a null geodesic, or it becomes a folded string in order to preserve closed string boundary conditions. The string retains transverse oscillations in the subleading embedding fields $y^i$ (cf.~\eqref{eq:X-exp-intro}). We described the first of these solutions in~\cite{Bagchi:2023cfp}.  

\begin{figure}[ht!]
	\centering
	\includegraphics[width=0.9\textwidth]{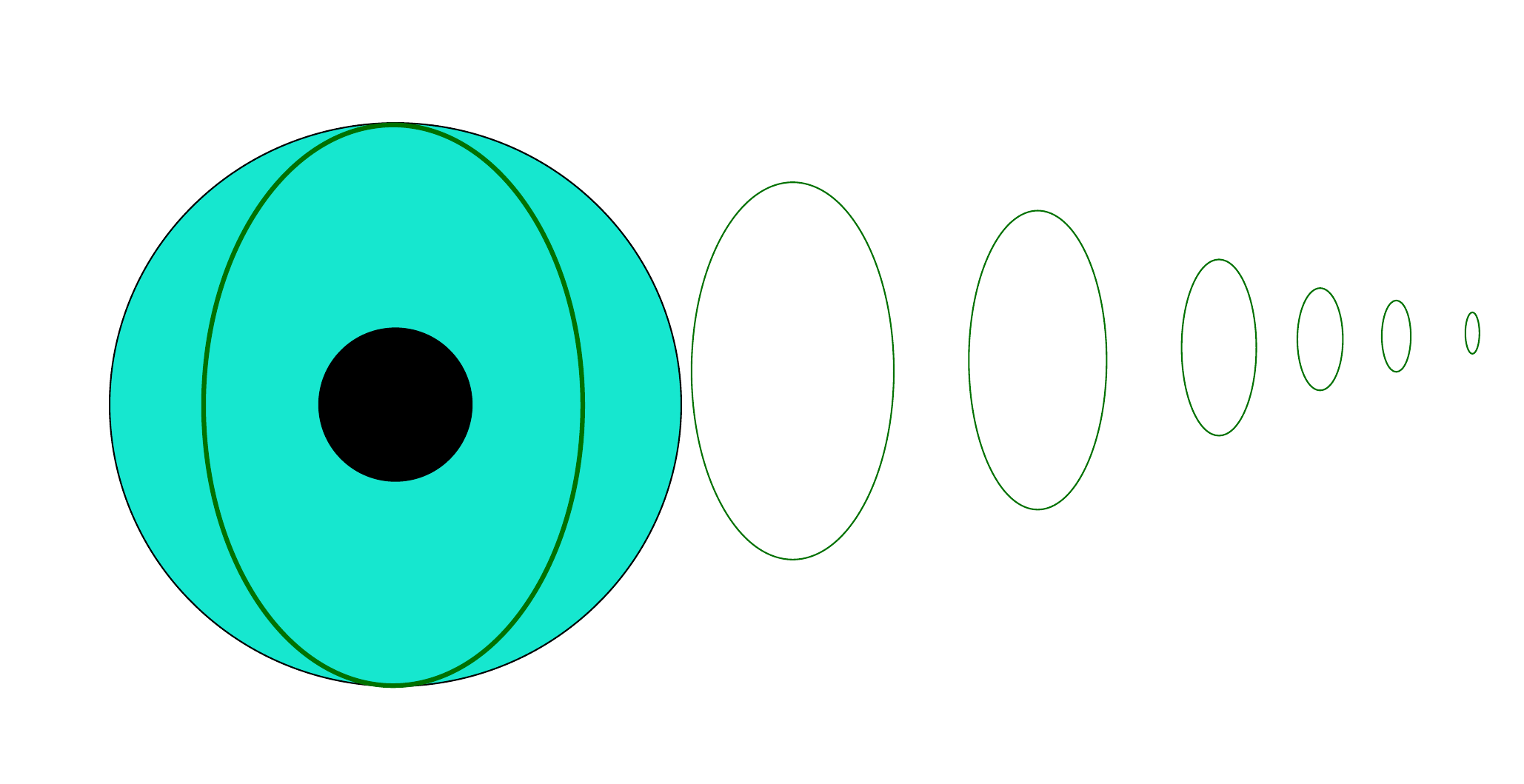}
 	\begin{tikzpicture}[overlay]
	\begin{pgfonlayer}{nodelayer}
			\node [style=real_ghost] (0) at (-4, 1) {{$\epsilon\rightarrow 0$}};
			\node [style=real_ghost] (2) at (-10., 0.3) {$r=r_h$};
			\draw [densely dotted,<-] (-7.4,1.3) to (-1.5,1.3);
	\end{pgfonlayer}
	\end{tikzpicture}
	\caption{The electric string has a solution where the string wraps the sphere.}
	\label{fig:NH-Carroll-electric-wrap}
\end{figure}
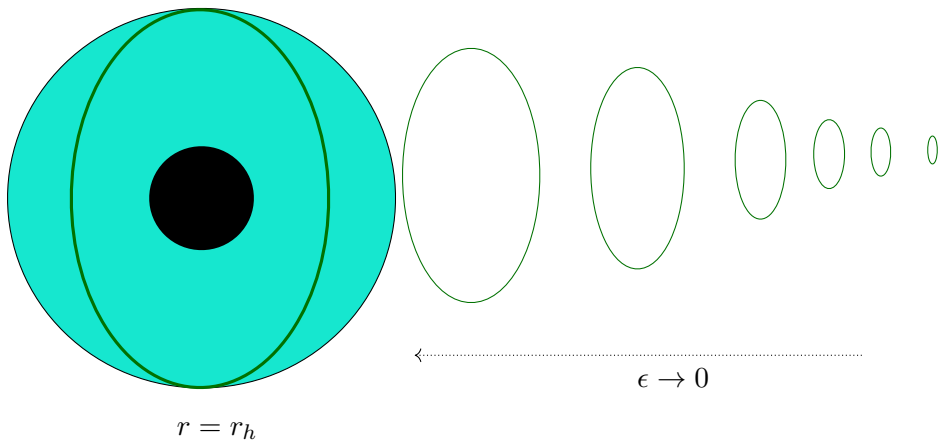

For the electric string, %we also find two classes of solutions: the first of these is closely related to the first solution of the magnetic string and takes the form of a one-parameter family of null geodesics in the 2D Rindler space, while the transverse embedding is trivial. The other class of
the solutions describe timelike geodesics in the 2D Rindler space, while the string wraps the transverse $2$-sphere, cf.~Figure~\ref{fig:NH-Carroll-electric-wrap}. Note that unlike the magnetic string solutions we considered above, the electric string solutions do not involve the subleading transverse embedding field (cf.~Eq.~\eqref{eq:X-exp-intro}).

\subsection*{Outline of the paper}
This paper is structured as follows. In Section~\ref{sec:string-carroll-exp}, we discuss the string Carroll expansion. We begin in Section~\ref{sec:string-carroll-geom} with a discussion of the string Carroll expansion of Lorentzian geometry, followed by an analysis of the string Carroll expansion of the phase space action for the relativistic string, which we show leads to two distinct theories: the magnetic and electric Carroll strings, described in detail in Section~\ref{sec:magnetic-and-electric-strings}. In Section~\ref{sec:magnetic-string}, we begin with a discussion of the magnetic Carroll string and we show how the phase space analysis recovers previous results obtained by expanding the relativistic Polyakov formulation. Then, in Section~\ref{sec:electric-strings}, we develop the theory of electric Carroll strings and demonstrate that they come equipped with a Carrollian worldsheet. We then turn our attention to strings near non-extremal black holes in Section~\ref{sec:near-horizon-strings}, where we first discuss the equivalence between string Carroll expansions and near-horizon expansions in Section~\ref{sec:near-horizon-is-string-Carroll}. We then show in Section~\ref{sec:near-horizon-exp-of-strings} how the near-horizon expansion of strings leads to the same electric and magnetic theories we obtained previously, and we study their solutions on the background obtained by string Carroll expanding the Schwarzschild geometry in Sections~\ref{sec:magnetic-string-on-BH-bg} and~\ref{sec:electric-strings-on-BH-bg} for the magnetic and electric string, respectively. In addition, we have included a number of appendices. The first of these, Appendix~\ref{app:trafos}, contains a detailed exposition of the string Carroll expansion and its symmetries. In order to contrast the string Carroll expansion with the more familiar particle Carroll expansion we discuss in Appendix~\ref{app:particle-exps} the particle Carroll expansion and derive the coupling of point particles to these geometries.  Appendix~\ref{app:phase-space-formulation-magnetic-string} provides the details of the matching between the phase space and Polyakov forms of the magnetic Carroll string. Finally, in Appendix~\ref{Kerr}, we present the near-horizon expansion of a non-extremal 4-dimensional Kerr black hole.

\section{String Carroll expansion}
\label{sec:string-carroll-exp}
In this section, we elaborate on the string Carroll expansion we introduced in~\cite{Bagchi:2023cfp}, which in turn builds on its nonrelativistic counterpart that was developed in~\cite{Hartong:2021ekg,Hartong:2022dsx,Hartong:2024ydv}. While relegating a more detailed analysis to Appendix~\ref{app:trafos}, we derive the geometric fields that arise in the string Carroll expansion up to $\mathcal{O}(c^2)$, and we examine the causal structure of the leading order string Carroll geometry. We then discuss the string Carroll expansion of string theory and elucidate how magnetic and electric Carroll strings emerge from a $c^2$ expansion of the relativistic string phase space action.
\subsection{String Carroll geometry and the \texorpdfstring{$c^2$}{c} expansion}
\label{sec:string-carroll-geom}
As we outlined in the introduction, the string Carroll expansion of a $(d+2)$-dimensional Lorentzian geometry with metric $g$ as defined in~\cite{Bagchi:2023cfp} is an expansion in powers of $c^2$ of the form
\begin{equation}
\label{eq:metric-exp}
    \begin{split}
        g_{\mu\nu} &= h_{\mu\nu} + c^2  \tau_{\mu\nu} + c^2\Phi_{\mu\nu} + \mathcal{O}(c^4)\,,\\
        g^{\mu\nu} &= \frac{1}{c^2}v^{\mu\nu} + \mathcal{O}(1)\,,
    \end{split}
\end{equation}
where $h_{\mu\nu}$ has signature $(0,0,+,\dots,+)$, while both $\tau_{\mu\nu}$ and $v^{\mu\nu}$ have signature $(-,+,0,\dots,0)$. In particular, we have the relation 
\begin{equation}
    h_{\mu\nu}v^{\nu\rho} = 0\,.
\end{equation}
We will refer to the geometry defined by the fields $(v^{\mu\nu},h_{\mu\nu})$ in~\eqref{eq:metric-exp} as string Carroll geometry, while geometries that include subleading fields generically will be referred as string Carroll expanded geometries.
Together, $(v^{\mu\nu},h_{\mu\nu})$ form a ``string Carroll structure'' and play the role of the metric in a string Carroll geometry. Since the combination $h_{\mu\nu} + \tau_{\mu\nu}$ has full rank it is invertible, and introducing\footnote{This field appears, among other objects, at $\mathcal{O}(1)$ in the expansion of $g^{\mu\nu}$ in~\eqref{eq:metric-exp}. See Appendix~\ref{app:trafos} for details.} $h^{\mu\nu}$ satisfying $\tau_{\mu\nu} h^{\nu\rho}= 0$, we can write this inverse as $v^{\mu\nu} + h^{\mu\nu}$, i.e.,
\begin{equation}
\label{eq:OG-completeness}
    \delta^\mu_\nu = v^\mu_\nu + h^\mu_\nu\,,\quad \text{where}\quad v^\mu_\nu :=  v^{\mu\rho}\tau_{\rho\nu}\quad \text{and}\quad h^\mu_\nu := h^{\mu\rho}h_{\rho \nu }\,.
\end{equation}
We will refer to $v^\mu_\nu$ as the \textit{longitudinal} projector, and to $h^\mu_\nu$ as the \textit{transverse} projector. We emphasise that $\tau_{\mu\nu} + h_{\mu\nu}$ is \textit{not} a metric since $\tau_{\mu\nu}$ is not invariant under the $c^2$-expanded local Lorentz transformations that turn into the string Carroll transformations; see Appendix~\ref{app:trafos} for more details.

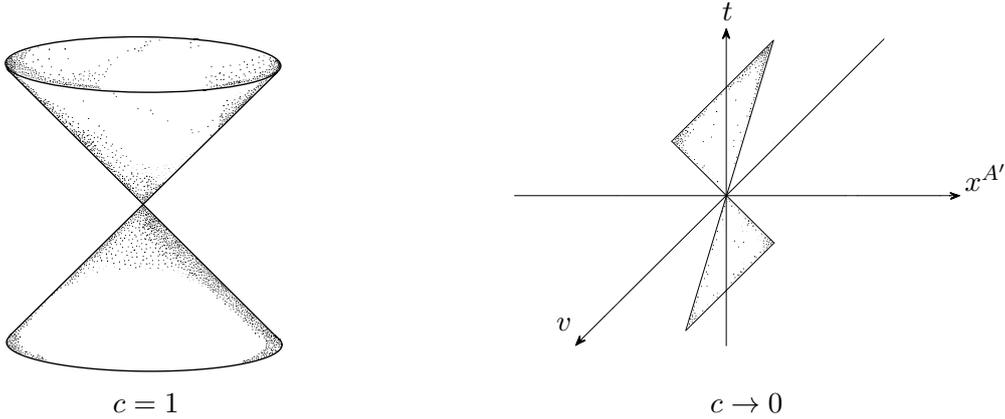
\begin{figure}[t!]
    \centering
    \includegraphics[width=0.9\textwidth]{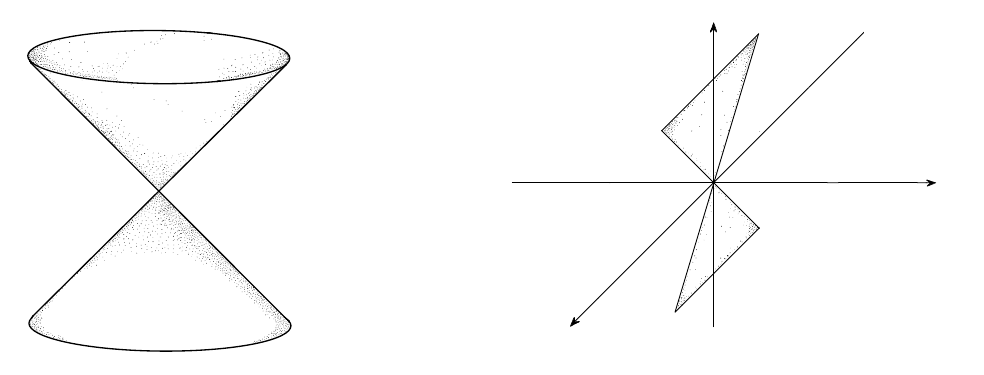}
    \begin{tikzpicture}[overlay]
        \begin{pgfonlayer}{nodelayer}
	        \node [style=real_ghost] (0) at (-6, 0.95) {$v$};
            \node [style=real_ghost] (1) at (-0.45, 2.9) {$x^{A'}$};
            \node [style=real_ghost] (2) at (-3.85, 5.1) {$t$};
            \node [style=real_ghost] (1) at (-11.5, -0.1) {$c=1$};
            \node [style=real_ghost] (1) at (-3.6, -0.1) {$c\to 0$};
\end{pgfonlayer}
\end{tikzpicture}
    \caption{The causal structure of string Carroll geometry. On the left is the standard Minkowski lightcone centred at the origin, which in the string Carroll limit $c\to 0$ becomes a two-dimensional lightcone in the $(v,t)$ plane as depicted on the right.}
    \label{fig:2D-lightcone}
\end{figure}

It is often useful to decompose the longitudinal objects $\tau_{\mu\nu}$ and $v^{\mu\nu}$ into ``longitudinal vielbeine'' by writing
\begin{equation}
    \tau_{\mu\nu} = \eta_{AB}\tau_\mu^A \tau_\nu^B\,,\qquad v^{\mu\nu} = \eta^{AB} v^\mu_A v^\nu_B\,, 
\end{equation}
where $A,B=0,1$ are longitudinal tangent space indices, and where $\eta_{AB} = \text{diag}(-1,1)$ is the Minkowski metric in the two-dimensional longitudinal tangent space. The longitudinal vielbeine satisfy the relations
\begin{equation}
\label{eq:vielbein-rels}
    v^\mu_A h_{\mu\nu} = \tau_\mu^A h^{\mu\nu} = 0\,,\qquad \tau_\mu^A v^\mu_B = \delta^A_B\,.
\end{equation}
Similarly, we may split $h_{\mu\nu}$ and $h^{\mu\nu}$ into transverse vielbeine via
\begin{equation}
\label{eq:transverse-vielbeine-def}
    h_{\mu\nu} = \delta_{A'B'}e^{A'}_\mu e^{B'}_\nu\,,\qquad h^{\mu\nu} = \delta^{A'B'} e^\mu_{A'}e^\nu_{B'}\,,
\end{equation}
where $A',B'=2,\dots,d+1$ are transverse tangent space indices. The transverse vielbeine are such that
\begin{equation}
    e^\mu_{A'}\tau_\mu{^A} = e_\mu^{A'}v^\mu_A = 0\,,\qquad e^\mu_{A'}e_\mu^{B'} = \delta^{B'}_{A'}\,.
\end{equation}
We now turn our attention to the causal structure of string Carroll geometry. In \textit{particle} Carroll geometry, which we briefly discuss in Appendix~\ref{app:particle-exps}, the lightcone collapses to a line when taking the limit $c\to 0$. In a string Carroll geometry, two longitudinal directions are rescaled with factors of $c$, cf.~the metric expansion~\eqref{eq:metric-exp}, which in the Lorentzian tangent space corresponds to the lightcone centred at the origin given by the quadric
\begin{equation}
\label{eq:quadric}
    0 = -c^2 t^2 + c^2 v^2 + x^{A'}x^{A'}\,,
\end{equation}
where $(t,v,x^{A'})$ are the $(d+2)$ coordinates of the tangent space, with $(t,v)$ forming the longitudinal directions. In the limit $c\to 0$, the projections of the lightcone in both the $(x^{A'},t)$ and the $(x^{A'},v)$ planes collapse to the line $x^{A'} = 0$, while the projection of the lightcone in the $(v,t)$ plane remains unchanged. The result of this is that the lightcone collapses to a two-dimensional lightcone in the $(v,t)$ plane with $x^{A'} = 0$ as depicted in Figure~\ref{fig:2D-lightcone}. 

\subsection{String Carroll expansion of string theory}
\label{sec:string-carroll-exp-of-ST}
String theory provides dynamics for a set of embedding fields $X^\mu(\sigma)$ that describe the embedding of a two-dimensional Lorentzian manifold, i.e., the worldsheet with coordinates $\sigma^\alpha$ for $\alpha=0,1$, into the target space, which is the ambient Lorentzian spacetime. On the worldsheet, the ambient metric $g_{\mu\nu}$ becomes a function of the embedding fields: $g_{\mu\nu} = g_{\mu\nu}(X)$. In this section, we set up the string Carroll expansion of closed bosonic strings, while we review the Carroll expansion of particle actions in Appendix~\ref{app:particle-exps}. The embedding fields $X^\mu$ have a $c^2$ expansion of the form
\begin{equation}
\label{eq:emb-field-exp}
    X^\mu(\sigma) = x^\mu(\sigma) + c^2 y^\mu(\sigma) + c^4 z^\mu(\sigma) + \mathcal{O}(c^6)\,,
\end{equation}
which means that the string Carroll expansion of the metric $g_{\mu\nu}(X)$ %and its inverse 
takes the form (cf.~\eqref{eq:metric-exp})
\begin{equation}
\label{eq:metric-with-embedding-fields}
    \begin{split}
        g_{\mu\nu}(X) &= h_{\mu\nu }(x) + c^2  \tau_{\mu\nu}(x) + c^2 \Phi_{\mu\nu}(x,y) + \mathcal{O}(c^4)\,,
    \end{split}
\end{equation}
where the $y$-dependence in $\Phi_{\mu\nu}(x,y)$ comes from Taylor expanding $g_{\mu\nu}(x + c^2 y)$; explicitly, we have that
\begin{equation}
\label{eq:stringy-objects}
\begin{split}
    \Phi_{\mu\nu}(x,y) &= \Phi_{\mu\nu}(x) + y^\rho\D_\rho h_{\mu\nu}(x)\,.
\end{split}
\end{equation}
In what follows, we will write, e.g., $\Phi_{\mu\nu} = \Phi_{\mu\nu}(x)$; in other words, we will refrain from explicitly indicating the dependence on $x$ when the quantity in question only depends on $x$. 

The dynamics of a string moving in a Lorentzian geometry with metric $g_{\mu\nu}$ is described by the phase space action $S = \int d\sigma^0\, L$, where
\begin{equation}
\label{eq:rel-phase-space-action}
    L = \oint d\sigma^1\,\left[ \dot X^\mu P_\mu - \frac{1}{2}e\left( c^2g^{\mu\nu}(X)P_\mu P_\nu + (c^2T)^2g_{\mu\nu}(X)X'^\mu X'^\nu \right) - u X'^\mu P_\mu\right] \,,
\end{equation}
is the phase space Lagrangian. Here, the Lagrange multiplier field $e$ has dimensions of time$^3/($mass$\times$length$^2)$, while the Lagrange multiplier field $u$ is dimensionless. We take the worldsheet coordinates to be dimensionless. In addition to the $c^2$-expansion of the embedding field $X^\mu$ in~\eqref{eq:emb-field-exp}, the other variables appearing in~\eqref{eq:rel-phase-space-action} expand in powers of $c^2$ according to
\begin{equation}
\label{eq:field-expansions}
    \begin{split}
        P_\mu &= P_{(0)\mu} + c^2 P_{(2)\mu} + c^4 P_{(4)\mu} + \mathcal{O}(c^6)\,,\qquad u = u_{(0)} + c^2u_{(2)} + c^4 u_{(4)} + \mathcal{O}(c^6)\,,
    \end{split}
\end{equation}
while the expansion of $e$ changes depending on whether we want to arrive at a theory of magnetic or electric Carroll strings.  In particular, if we vary $u, e, P_\mu$ in the phase space Lagrangian~\eqref{eq:rel-phase-space-action}, we may solve for these variables to get
\begin{subequations}
\label{eq:some-eq}
    \begin{align}
    u & = \frac{\dot X\cdot X'}{X'\cdot X'}\,,\label{eq:u}\\
    e^2 & = -\frac{1}{c^6 T^2}\frac{\dot X\cdot \dot X-u \dot X\cdot X'}{X'\cdot X'}\,,\\
    P_\mu & = \frac{1}{e c^2}g_{\mu\nu}\left(\dot X^\nu-u X'^\nu\right)\,,\label{eq:P}
\end{align}
\end{subequations}
where the dot indicates an inner product with respect to $g_{\mu\nu}$, e.g., $\dot X\cdot X' := g_{\mu\nu}\dot X^\mu X'^\nu$. The vectors $\dot X^\mu$ and $X'^\mu$ are tangent vectors to the worldsheet of the string which is a $(1+1)$-dimensional Lorentzian manifold. We will require (choice of orientation) $\dot X\cdot \dot X<0$ and $X'\cdot X'>0$. Let us choose a gauge such that $\dot X\cdot X'=0$, or equivalently $u=0$. This is always possible, at least locally. Since 
$\dot X\cdot \dot X<0$ and $X'\cdot X'>0$ it is guaranteed that $e^2>0$. Using the $c^2$ expansion of  the metric~\eqref{eq:metric-with-embedding-fields}, we conclude that in order for $\dot X\cdot \dot X<0$ we need $\dot X\cdot \dot X=\mathcal{O}(c^2)$ due to the signature of $h_{\mu\nu}$. 

For the electric Carroll string, which we discuss in detail in Section~\ref{sec:electric-strings}, we will take $c^2T$ to be constant, while for the magnetic Carroll string, which we discuss in Section~\ref{sec:magnetic-string} (see also~\cite{Bagchi:2023cfp}), we will take $cT$ to be constant. We do this is in order that the particle limits correspond to the electric and magnetic particle actions of~\cite{deBoer:2021jej}, respectively. We discuss the derivation of the electric and magnetic particle theories from the perspective of $c^2$ expansions in Appendix~\ref{app:particle-exps}. 

The arguments above lead us to conclude that the power of $c^2$ in the expansion of $e^2$ is directly related to the power of $c^2$ in the expansion of $X'\cdot X'$. For the electric theory we will take $e=\mathcal{O}(1)$ and for the magnetic theory $e=\mathcal{O}(c^{-2})$. At this stage, one might wonder why the magnetic theory cannot have $X'\cdot X'=\mathcal{O}(1)$: for this to be possible we would need $e=\mathcal{O}(c^{-1})$ which leads to odd powers in the expansion of the Lagrangian (alongside even powers). Since we only consider expansions in even powers in this work, this option is excluded.

\section{Magnetic and Electric Carroll Strings}
\label{sec:magnetic-and-electric-strings}
In this section, we derive the magnetic and electric Carroll string theories by applying the string Carroll expansion to the relativistic phase space action for the string. We begin by re-deriving the results of~\cite{Bagchi:2023cfp} from a phase space perspective, followed by the formulation of the electric Carroll string. 
\subsection{Magnetic Carroll strings}
\label{sec:magnetic-string}
We start by analysing the structure of the magnetic Carroll string as defined in~\cite{Bagchi:2023cfp} to NNLO, though we depart from the phase space formulation~\eqref{eq:rel-phase-space-action} rather than the Polyakov formulation. We then derive the Polyakov form of the corresponding theories by integrating out the momenta, thereby recovering the results for the LO and NLO Polyakov actions obtained in~\cite{Bagchi:2023cfp}. We relegated the details of this calculation to Appendix~\ref{app:phase-space-formulation-magnetic-string}. 

To set up the string Carroll expansion of the phase space Lagrangian~\eqref{eq:rel-phase-space-action} that leads to the magnetic Carroll string, we define the ``magnetic quantities''
\begin{equation}
\label{eq:magnetic-string-combos}
    \tilde e := c^{2} e\,,\qquad \tilde T := cT\,,
\end{equation}
which are finite in the limit $c\to 0$. This is the string counterpart of the magnetic particle limit \eqref{eq:magneticparticle} and ensures that the magnetic string in the particle limit becomes the magnetic particle. We thus see that the starting point for performing the string Carroll expansion that leads to the magnetic Carroll string is
\begin{equation}
\label{eq:magnetic-start-Lagrangian}
    \tilde L = \oint d\sigma^1\,\left[ \dot X^\mu P_\mu - \frac{1}{2}\tilde e\left( g^{\mu\nu}(X)P_\mu P_\nu + \tilde T^2g_{\mu\nu}(X)X'^\mu X'^\nu \right) - u X'^\mu P_\mu\right] \,,
\end{equation}
where $\tilde e$ expands in powers of $c^2$ according to
\begin{equation}
\label{eq:alternative-expansion}
    \tilde e = \tilde e_{(0)} + c^2\tilde e_{(2)} + c^4 \tilde e_{(4)} + \mathcal{O}(c^6)\,,
\end{equation}
while the remaining fields are expanded as in~\eqref{eq:emb-field-exp} and~\eqref{eq:field-expansions}. This leads to an expansion of the form 
\begin{equation}
\label{eq:magnetic-L-exp}
    \tilde L = c^{-2}\tilde L_{\text{LO}} + \tilde L_{\text{NLO}} + c^2\tilde L_{\text{NNLO}} + \mathcal{O}(c^4)\,.
\end{equation}
A detailed study of the Lagrangians that appear at the various orders in $c^2$ may be found in Appendix~\ref{app:phase-space-formulation-magnetic-string}. The LO Lagrangian $\tilde L_{\text{LO}}$ involves only $v^\nu_\mu P_{(0)\nu}$ (cf.~\eqref{eq:field-expansions}) and is ``trivial'' in the sense that the associated equation of motion just sets $v^\nu_\mu P_{(0)\nu}=0$. It also has no Polyakov counterpart, since the Polyakov Lagrangian is obtained by integrating out $P_{(0)_\mu}$. However, the NLO phase space Lagrangian $\tilde L_{\text{NLO}}$ does have a Polyakov form, which is obtained by integrating out the momenta $P_{(0)\mu}$ and $P_{(2)\mu}$. The resulting Lagrangian appears at leading order in the expansion of the relativistic Polyakov Lagrangian~\cite{Bagchi:2023cfp} and reads
\begin{equation}
\label{eq:LO-Polyakov-Lagrangian}
    \tilde L_{\text{P-LO}} = - \frac{\tilde T}{2} \oint d\sigma^1 \sqrt{-\gamma_{(0)}} \gamma^{\alpha\beta}_{(0)} \D_\alpha x^\mu \D_\beta x^\nu h_{\mu\nu}\,,
\end{equation}
where $\gamma_{(0)\alpha\beta}$ is a Lorentzian metric on the worldsheet with $\alpha,\beta = 0,1$ worldsheet indices, and where we defined $\gamma_{(0)} = \det(\gamma_{(0)\alpha\beta})$. This Lorentzian metric is built from the Lagrange multipliers $\tilde e_{(0)}$ and $u_{(0)}$ (cf.~\eqref{eq:tilde-gamma}). The Virasoro constraints obtained by integrating out the worldsheet metric 
are
\begin{equation}
    h_{\alpha\beta}=\frac{1}{2}\gamma_{(0)}^{\gamma\delta}h_{\gamma\delta}\gamma_{(0)\alpha\beta}\,.
\end{equation}
The pullback $\D_\alpha x^\mu \D_\beta x^\nu h_{\mu\nu} = h_{\alpha\beta}$ does not have Lorentzian signature while $\gamma_{(0)\alpha\beta}$ does. Hence, the embedding is trivial in the sense that $h_{\alpha\beta} = 0$. 

The NNLO phase space Lagrangian gives rise to the NLO Polyakov Lagrangian of~\cite{Bagchi:2023cfp} when integrating out the momenta. This Polyakov Lagrangian reads
\begin{equation}
\label{eq:P-NLO-lagrangian}
    \begin{split}
        \tilde{L}_{\text{P-NLO}} &= - \frac{\tilde T}{2} \oint d\sigma^1\, \sqrt{-{\gamma_{(0)}}} \left( \gamma_{(0)}^{\alpha\beta}\hat\Phi_{\alpha\beta}(x,y) - \frac{1}{2}G_{(0)}^{\alpha\beta\gamma\delta} h_{\alpha\beta}(x)\gamma_{(2)\gamma\delta} \right)\,, 
    \end{split}
\end{equation}
where we defined the combination
\begin{equation*}
    \hat\Phi_{\alpha\beta}(x,y) := \D_\alpha x^\mu \D_\beta x^\nu \tau_{\mu\nu} + \D_\alpha x^\mu \D_\beta x^\nu \Phi_{\mu\nu} +  2h_{\mu\nu}(x)\D_{(\alpha} x^\mu \D_{\beta)}y^\nu + y^\rho\D_\alpha x^\mu \D_\beta x^\nu  \D_\rho h_{\mu\nu}(x)\,,
\end{equation*}
where worldsheet indices are raised and lowered using the Lorentzian metric $\gamma_{(0)}$, and where $G_{(0)}^{\alpha\beta\gamma\delta}$ is the Wheeler-DeWitt metric 
\begin{equation}
    G_{(0)}^{\alpha\beta\gamma\delta} = \gamma^{\alpha\gamma}_{(0)} \gamma^{\delta\beta}_{(0)} + \gamma^{\alpha\delta}_{(0)} \gamma^{\gamma\beta}_{(0)} - \gamma^{\alpha\beta}_{(0)} \gamma^{\gamma\delta}_{(0)}\,.
\end{equation}
The expression for $\gamma_{(2)\alpha\beta}$ in terms of the Lagrange multipliers $\tilde e_{(0)},\tilde e_{(2)},u_{(0)}$ and $u_{(2)}$ appears in~\eqref{eq:gamma-2}. 

In the Polyakov formalism, where the LO and NLO Lagrangians are given by~\eqref{eq:LO-Polyakov-Lagrangian} and~\eqref{eq:P-NLO-lagrangian}, respectively, the worldsheet symmetries are obtained by expanding relativistic worldsheet symmetries consisting of Weyl rescalings and two-dimensional diffeomorphisms in powers of $c^2$~\cite{Bagchi:2023cfp}. These can be gauge-fixed order-by-order, which leaves as the residual gauge freedom the familiar Virasoro symmetries. This procedure is formally equivalent to the nonrelativistic expansion in powers of $1/c^2$~\cite{Hartong:2021ekg,Hartong:2022dsx,Hartong:2024ydv}, where the residual symmetries also reduce to Virasoro. 

\subsection{Electric Carroll strings}
\label{sec:electric-strings}
In this section, we study ``electric'' Carroll strings, which arise from a different expansion of~\eqref{eq:rel-phase-space-action}. We will see that the electric Carroll string shares certain similarities with the null string (see, e.g.,~\cite{Bagchi:2015nca} for a discussion of null strings). 

\subsubsection{LO electric Carroll string action}
Just as for the magnetic Carroll string expansion that we discussed in Section~\ref{sec:magnetic-string}, the starting point for the expansion that leads to the electric Carroll string involves scaling the tension and the Lagrange multiplier $e$ that appears in~\eqref{eq:rel-phase-space-action}:
\begin{equation}
\label{eq:electric-string-combos}
    \hat T := c^2T\,,\qquad \hat e := e\,,
\end{equation}
i.e., we do not scale $e$ at all, though we still decorate it with a hat to emphasise that we are considering the Lagrange multiplier in an electric context. This is the string counterpart of the electric particle limit \eqref{eq:electricparticle} and ensures that the electric string in the particle limit becomes the electric particle. The Lagrangian that forms the starting point for the electric Carroll expansion takes the form
\begin{equation}
\label{eq:elec-starting-point}
    \hat L = \oint d\sigma^1\,\left[ \dot X^\mu P_\mu - \frac{1}{2}\hat e\left( c^2g^{\mu\nu}(X)P_\mu P_\nu + \hat T^2g_{\mu\nu}(X)X'^\mu X'^\nu \right) - u X'^\mu P_\mu\right] \,,
\end{equation}
which has an expansion in powers of $c^2$ of the form
\begin{equation}
    \hat L = \hat L_{\text{LO}} + \mathcal{O}(c^2)\,.
\end{equation}
Unlike for the magnetic string, we will not consider the subleading theories, since the LO theory itself exhibits interesting solutions. In particular, the LO theory has a ``true''\footnote{In the sense that it may be obtained as a $c \to 0$ limit.} string Carroll target space, while the NLO Polyakov theory~\eqref{eq:P-NLO-lagrangian} that we considered in the magnetic case has a target space that involves additional fields (such as $\Phi_{\mu\nu}$) that appear in the string-Carroll expansion of the Lorentzian metric. 

The worldsheet fields are expanded according to~\eqref{eq:emb-field-exp} and~\eqref{eq:field-expansions}, while the electric Lagrange multiplier $\hat e$ expands according
\begin{equation}
\label{eq:hat-e-exp}
    \hat e = \hat e_{(0)} + \mathcal{O}(c^2)\,.
\end{equation}
This leads to
\begin{equation}
\label{eq:electriclimit}
\hat L_{\text{LO}} = \oint d\sigma^1\,\bigg[ \dot x^\mu P_{(0)\mu} - \frac{1}{2}\hat e_{(0)}(v^{\mu\nu} P_{(0)\mu} P_{(0)\nu} + \hat T^2 h_{\mu\nu} x'^\mu x'^\nu ) - u_{(0)} x'^\mu P_{(0)\mu} \bigg]\,.
\end{equation}
Since we will only consider the LO electric theory, we will henceforth drop the ``$(0)$'' subscript for notational simplicity. Just as for the magnetic string, it will be useful to split the momemtum $P_{\mu}$ into its string Carroll tangent space components $P_A$ and $P_{A'}$ defined as
\begin{equation}
P_A:=\tau^\mu_AP_\mu\,,\qquad P_{A'}:=e^\mu_{A'}P_\mu\,.
\end{equation}
When expressed in terms of $P_A$ and $P_{A'}$, we see that the electric Lagrangian~\eqref{eq:electriclimit} is linear in $P_{A'}$ and quadratic in $P_A$. We can thus only integrate out $P_A$ and we must keep $P_{A'}$ as a Lagrange multiplier field. This leads to
\begin{equation}
\label{eq:electriclimit2}
\hat L_{\text{LO}} = \oint d\sigma^1\,\bigg[ \left(\dot x^\mu-u x'^\mu\right)e^{A'}_\mu P_{A'}+  \frac{1}{2 \hat e}\tau_{\mu\nu}\left(\dot x^\mu-u x'^\mu\right)\left(\dot x^\nu-u x'^\nu\right)-\frac{\hat T^2}{2}\hat e h_{\mu\nu} x'^\mu x'^\nu \bigg]\,.
\end{equation}
 
\subsubsection{Carrollian worldsheet}
\label{sec:Carroll-worldsheet}
To make two-dimensional worldsheet diffeomorphism invariance manifest, we introduce zweibeine $\mathbb v^\alpha$ and $\mathbb e^\alpha$ defined via 
\begin{equation}
u=-\frac{\mathbb v^1}{\mathbb v^0}\,,\qquad \hat e=\frac{1}{\mathbb e( \mathbb v^0)^2}\,,
\end{equation}
where $\mathbb e=\text{det}\,\left(\mathbb t_\alpha\,, \mathbb e_\alpha\right)$ with $\left(\mathbb t_\alpha\,, \mathbb e_\alpha\right)$ the inverse of $\left(\mathbb v^\alpha\,, \mathbb e^\alpha\right)$. In two dimensions we have
\begin{equation}
\mathbb v^0=-\frac{\mathbb e_1}{\mathbb e}\,,\qquad \mathbb v^1=\frac{\mathbb e_0}{\mathbb e}\,,\qquad \mathbb e^0=-\frac{\mathbb t_1}{\mathbb e}\,,\qquad \mathbb e^1=\frac{\mathbb t_0}{\mathbb e}\,,\qquad \mathbb e=\mathbb t_0 \mathbb e_1- \mathbb t_1 \mathbb e_0\,.
\end{equation}
We can then express the electric Carroll string Lagrangian~\eqref{eq:electriclimit2} as
\begin{equation}\label{eq:electriclimit3}
\hat L_{\text{LO}} = \oint d\sigma^1\mathbb e\bigg[\hat T^2 \mathbb v^\alpha\partial_\alpha x^\mu e^{A'}_\mu \tilde P_{A'}+  \frac{1}{2}\tau_{\mu\nu}\mathbb v^\alpha\partial_\alpha x^\mu \mathbb v^\beta\partial_\beta x^\nu-\frac{\hat T^2}{2}h_{\mu\nu}\mathbb e^\alpha\partial_\alpha x^\mu \mathbb e^\beta\partial_\beta x^\nu \bigg]\,,
\end{equation}
where we defined $\tilde P_{A'}$ as
\begin{equation}
\tilde P_{A'} := -\frac{1}{\hat T^2 \mathbb e_1}P_{A'}\,.
\end{equation}
Let us decode the above Lagrangian a bit more. The first thing to notice here is that if we set $\hat T=0$, this becomes the action of a null string \cite{Isberg:1993av,Bagchi:2013bga,Bagchi:2015nca} in the longitudinal 2D Rindler spacetime. This can be thought of a tensile deformation of a tensionless string. As we will show below, the symmetries on the worldsheet are Carrollian. The deformation thus keeps Carroll invariance and this can be already read off from the Lagrangian as the null string part is an electric Carroll scalar theory in the 2D Rindler space, while the tensile deformation takes the form of a magnetic scalar theory in the transverse directions.\footnote{The Lagrangian~\eqref{eq:electriclimit3} also appears in Eq.~(3.65) in~\cite{Gomis:2023eav} for $q=8$ (or $p=-9$) and $B = 0$, where it arises in a decoupling limit after a series of T-dualities (see also~\cite{Blair:2023noj}).} 

In order to recover the Lagrangian in~\eqref{eq:electriclimit2} we must be able to set $\mathbb e^0=0$ without loss of generality. This is possible due to the emergence of a new gauge symmetry on the worldsheet, namely a local Carroll boost that acts on the zweibeine according to\begin{equation}
\delta \mathbb t_\alpha=\lambda \mathbb e_\alpha\,,\qquad\delta \mathbb e_\alpha=0\,,\qquad\delta \mathbb v^\alpha=0\,,\qquad\delta \mathbb e^\alpha=\lambda\mathbb v^\alpha\,,
\end{equation}
where $\lambda(\sigma)$ is an arbitrary function on the worldsheet. In particular, this allows us to interpret the worldsheet geometry as a Carrollian geometry. In order for this to be a symmetry we must also transform $\tilde P_{A'}$ 
\begin{equation}
\delta\tilde P_{A'} = \lambda \mathbb e^\beta\partial_\beta x^\mu e_{\mu\, A'}\,.
\end{equation}
As stated above, we may understand the form of the electric string action~\eqref{eq:electriclimit3} as a combination of a two-dimensional magnetic Carroll scalar field theory in the transverse directions and an electric Carroll scalar field theory in the longitudinal directions. A second symmetry that emerges in this zweibeine (or Polyakov-type) formulation is a Weyl symmetry acting as
\begin{equation}
\delta \mathbb t_\alpha=\omega \mathbb t_\alpha\,,\qquad\delta \mathbb e_\alpha=\omega \mathbb e_\alpha\,,\qquad\delta \mathbb v^\alpha=-\omega \mathbb v^\alpha\,,\qquad\delta \mathbb e^\alpha=-\omega \mathbb e^\alpha\,.
\end{equation}
This is a symmetry provided we transform $\tilde P_{A'}$ as
\begin{equation}
\delta\tilde P_{A'}=-\omega\tilde P_{A'}\,.
\end{equation}
Finally, the couplings in the Lagrangian \eqref{eq:electriclimit3} have an ambiguity that originates from a target space string Carroll boost transformation with parameter $\lambda^A_{(0)B'}$ (cf.~\eqref{eq:vielbein-trafos}), which acts on $\tilde P_{A'}$ and $\tau^A_{\mu}$ as
\begin{equation}
\delta\tilde P_{A'}=-\eta_{AB} \mathbb v^\beta\partial_\beta x^\mu\tau_\mu^B\lambda^A_{(0)A'}\,,\qquad \delta\tau_\mu^A=\lambda^A_{(0)B'} e^{B'}_\mu\,.
\end{equation}
In addition to the symmetries we described above, the Carroll geometry on the worldsheet is reparameterisation invariant. In~\eqref{eq:electriclimit3} the action of such a worldsheet diffeomorphism generated by $\zeta^\alpha$ is given by
\begin{equation}
\delta \mathbb t_\alpha=\pounds_\zeta \mathbb t_\alpha\,,\qquad\delta\mathbb  e_\alpha=\pounds_\zeta \mathbb e_\alpha\,,\qquad \delta X^\mu = \zeta^\alpha\partial_\alpha X^\mu\,,\qquad\delta\tilde P_a=\zeta^\alpha\partial_\alpha\tilde P_a\,,
\end{equation}
and similarly for $\mathbb v^\alpha$ and $\mathbb  e^\alpha$. The Lagrangian density for the electric Carroll string $\hat{\mathcal{L}}$, which is defined such that  
\begin{equation}
\hat L_{\text{LO}} = \oint d\sigma^1\hat{\mathcal{L}}\,,    
\end{equation}
then transforms as a density, i.e.,
\begin{equation}
\delta\hat{\mathcal{L}}=\partial_\alpha\left(\hat{\mathcal{L}}\zeta^\alpha\right)\,.
\end{equation}
In the gauge-fixed version of the electric Carroll string action~\eqref{eq:electriclimit2} these transformations act as follows. First we choose the gauge $\mathbb e^0=0$ and consider the residual gauge transformations. Using that the total gauge transformation acting on $\mathbb e^\alpha$ is given by
\begin{equation}
\delta\mathbb e^\alpha=\pounds_\zeta\mathbb e^\alpha+\lambda\mathbb v^\alpha-\omega\mathbb e^\alpha\,,
\end{equation}
we set $\mathbb e^0=0$ and $\delta \mathbb e^0=0$ to obtain an expression for the worldsheet Carroll boost parameter
\begin{equation}
\label{eq:lambda}
\lambda = \frac{\mathbb e^1}{\mathbb v^0}\partial_1\zeta^0\,.
\end{equation}
The total worldsheet gauge transformation consists of reparameterisations $\zeta^\alpha$, Carroll boosts $\lambda$ and Weyl rescalings $\omega$, and it acts on the fields $\mathbb t_\alpha$, $\mathbb e_\alpha$ and $\tilde P_{A'}$ as 
\begin{subequations}
\begin{align}
\delta\mathbb t_\alpha & =  \pounds_\zeta \mathbb t_\alpha+\lambda \mathbb e_\alpha+\omega \mathbb t_\alpha\,,\label{eq:totvar1}\\
\delta \mathbb e_\alpha & =  \pounds_\zeta \mathbb e_\alpha+\omega \mathbb e_\alpha\,,\label{eq:totvar2}\\
\delta \tilde P_{A'}  &= \zeta^\alpha\partial_\alpha\tilde P_{A'}+ \lambda \mathbb e^\beta\partial_\beta X^\mu e_\mu^{A'}-\omega\tilde P_{A'}\,.\label{eq:totvar3}
\end{align}
\end{subequations}
In order to find the gauge transformations acting on $u$, $\hat e$ and $P_{A'}$ we use \eqref{eq:totvar1}--\eqref{eq:totvar3} as well as \eqref{eq:lambda} to find
\begin{eqnarray}
\delta u & = & \zeta^\alpha\partial_\alpha u -u^2\partial_1\zeta^0+u\left(\partial_0\zeta^0-\partial_1\zeta^1\right)+\partial_0\zeta^1\,,\\
\delta\hat e & = & \zeta^\alpha\partial_\alpha \hat e -2\hat e u\partial_1\zeta^0+\hat e\left(\partial_0\zeta^0-\partial_1\zeta^1\right)\,,\\
\delta P_{A'} & = & \zeta^\alpha\partial_\alpha P_{A'}+u P_{A'}\partial_1\zeta^0+P_{A'}\partial_1\zeta^1+\hat T^2\hat e e_{\mu\,A'}\partial_1 X^\mu\,.
\end{eqnarray}
It can be checked by explicit calculation that these transformations act on the action in~\eqref{eq:electriclimit2} such that it transforms into a total derivative.

Finally, we remark that we may fix the gauge redundancy by setting the Lagrange multipliers equal to $\hat e = 1$ and $u=0$, though this is only possible locally.\footnote{The worldsheet is a cylinder parametrised by coordinates $(\sigma^0, \sigma^1)$ where $-\infty<\sigma^0<\infty$ and $0<\sigma^1<2\pi$ in any coordinate system we use to parametrise the cylinder. Infinitesimally, a worldsheet diffeomorphism is of the form $\delta\sigma^\alpha=-\zeta^\alpha$. In order to respect that $0<\sigma^1<2\pi$ we must require that $\zeta^1$ is periodic with period $2\pi$ and that its zero mode vanishes so as not to displace the origin. This latter restriction does not mean that we do not have a gauge symmetry of the form $\sigma'^1=\sigma^1+f(\sigma^0)$ but that we cannot fix this gauge transformation because of our insisting on keeping $0<\sigma^1<2\pi$. We impose this so that $\sigma^1$ integrals runs from $0$ to $2\pi$ as opposed to from, say, $f(\sigma^0)$ to $f(\sigma^0)+2\pi$. This means globally we cannot do better than the following gauge fixing
\begin{equation}\label{eq:gaugechoice}
\hat e=1\,,\qquad u=u(\sigma^0)\,.
\end{equation}
This is a well-known fact from string theory. In terms of the worldsheet metric this means that we cannot choose a metric that is everywhere flat.}

\section{Strings near a black hole horizon}
\label{sec:near-horizon-strings}
Having laid out the detailed formalism of string theory in the Carroll regime over the previous sections, we now turn to the main focus of the paper, namely understanding string theory in the near-horizon region of non-extremal black holes. To be specific, and for computational simplicity, we will focus on the Schwarzschild black hole in four-dimensional asymptotically flat spacetimes.

It was demonstrated in~\cite{Bagchi:2023cfp} that the near-horizon expansion of the Schwarzschild black hole takes the form of a string Carroll expansion as discussed in Section~\ref{sec:string-carroll-exp}, with the only difference being that the role of the speed of light is now played by the distance away from the horizon (see also~\cite{Fontanella:2022gyt} for a discussion of non-extremal near-horizon geometries). In that same paper, we studied the behaviour of strings near the black hole horizon using the magnetic string of Section~\ref{sec:magnetic-string}. We begin this section with a review of the results of~\cite{Bagchi:2023cfp}, which we generalise slightly, followed by a discussion of the properties of near-horizon strings using the electric string theory that we derived in Section~\ref{sec:electric-strings}. 

In the discussion in Section~\ref{sec:discussion} and in Appendix \ref{Kerr} we show that the near-horizon geometries of both Reissner--Nordstr\"om and Kerr black holes also admit string Carroll expansions.

\subsection{Near-horizon expansion is string Carroll}
\label{sec:near-horizon-is-string-Carroll}
In the remainder of this article, we shall set $c=1$. In Schwarzschild coordinates, the four-dimensional Schwarzschild metric is
\begin{equation}
ds^2 = -\left(1 - \frac{2GM}{r}\right)d  t^2  + \left(1 - \frac{2GM}{r}\right)^{-1} dr^2 + r^2d\Omega^2\,,
\end{equation}
where $d\Omega^2 = d\theta^2 + \sin^2\theta\,d\phi^2$ is the metric on the two-sphere. We are interested in the near horizon region $r\sim 2GM$. Following~\cite{Bagchi:2021ban,Bagchi:2023cfp}, we change coordinates according to
\begin{equation}
    r = r_h + \frac{1}{r_h}\epsilon\mathdutchcal{r}^2\,, \qquad r_h := 2GM\,,
\end{equation}
where $\epsilon$ is a positive dimensionless parameter, which goes to zero as we approach the horizon.\footnote{For nonzero values $\epsilon$ can be removed by a coordinate transformation and is thus not physical. It is however a useful formal expansion parameter.} The near-horizon region is thus characterised by $\epsilon \ll 1$. This brings the Schwarzschild metric to the form
\begin{equation}
\label{eq:schwarzschild-exp}
    ds^2 = r_h^2d\Omega^2 + \epsilon\left[ -\frac{\mathdutchcal{r}^2}{r_h^2}(dt)^2 + 4d\mathdutchcal{r}^2 + 2\mathdutchcal{r}^2d\Omega^2 \right] + \epsilon^2\left[ \frac{\mathdutchcal{r}^4}{r_h^2}(dt)^2 + \frac{4\mathdutchcal{r}^2}{r_h^2}d\mathdutchcal{r}^2 + \frac{\mathdutchcal{r}^4}{r_h^2}d\Omega^2 \right] + \mathcal{O}(\epsilon^3)\,.
\end{equation}
Hence, the near-horizon region to first order in $\epsilon$ is a 2D Rindler spacetime times a 2-sphere. At order $\epsilon^2$, additional corrections appear.

The point that we wish to stress is that this near-horizon expansion of the metric is essentially the same as the string Carroll expansion that we discussed in Section~\ref{sec:string-carroll-exp}. Comparing the above with the string Carroll expansion of the metric~\eqref{eq:metric-exp}
with the identification\footnote{While this identification is true at the level of the backgrounds, it ceases to be valid at the level of the Lagrangians, where one cannot just replace factors of $c^2$ with $\epsilon$. This is due to factors of $c$ appearing in the string Lagrangians for dimensional reasons, which therefore no longer appear after we set $c=1$.  }
\begin{equation}
\label{eq:epsilon-replaces-c^2}
\epsilon \equiv c^2\,,    
\end{equation}
the expansion in~\eqref{eq:schwarzschild-exp} corresponds to (cf.~\eqref{eq:combinations})
\begin{equation}
\label{eq:schwarzschild-string-carroll-bg}
    \begin{split}
        h_{\mu\nu}dx^\mu dx^\nu = r_h^2d\Omega^2\,,\qquad
        \tau_{\mu\nu} dx^\mu dx^\nu = -\frac{\mathdutchcal{r}^2}{r_h^2}dt^2 + 4d\mathdutchcal{r}^2\,,\qquad
        \Phi_{\mu\nu} dx^\mu dx^\nu = 2\mathdutchcal{r}^2d\Omega^2 \,.
    \end{split}
\end{equation}
The two-dimensional longitudinal space is $(1+1)$-dimensional Rindler spacetime with metric $ds^2 = -\frac{\mathdutchcal{r}^2}{r_h^2}dt^2 + 4d\mathdutchcal{r}^2$, while the LO transverse space is a two-sphere of radius $r_h$ with metric $ds^2 = r_h^2d\Omega^2$.   

All our previous result regarding string Carroll expansions in terms of $c^2$ that we derived in Section~\ref{sec:string-carroll-exp} now carry over to the near-horizon expansion with the replacement~\eqref{eq:epsilon-replaces-c^2}; in particular, the expansion of the embedding field~\eqref{eq:emb-field-exp} now takes the form
\begin{equation}
\label{eq:emb-field-exp-epsilon}
    X^\mu(\sigma) = x^\mu(\sigma) + \epsilon y^\mu(\sigma) + \epsilon^2 z^\mu(\sigma) + \mathcal{O}(\epsilon^3)\,,
\end{equation}
which for the metric $g_{\mu\nu}(X)$ leads to~\eqref{eq:metric-with-embedding-fields} with the replacement~\eqref{eq:epsilon-replaces-c^2}. 

\subsection{Near-horizon expansion of strings}
\label{sec:near-horizon-exp-of-strings}

The purpose of this subsection is to show that relativistic strings in the near-horizon geometry (as viewed by a stationary observer at infinity) fall into two categories depending on whether the string has an extent around the near-horizon 2-sphere or not. We will argue this from the equations of motion and constraints for the string propagating in this region. We will then show in subsequent subsections that these two sectors are reproduced by the magnetic and electric Carroll strings discussed in Section~\ref{sec:magnetic-and-electric-strings}. We have summarised these results in the flowchart depicted in Figure~\ref{fig:square} below.  

\begin{figure}[ht!]
	\centering
	\includegraphics[width=0.8\textwidth]{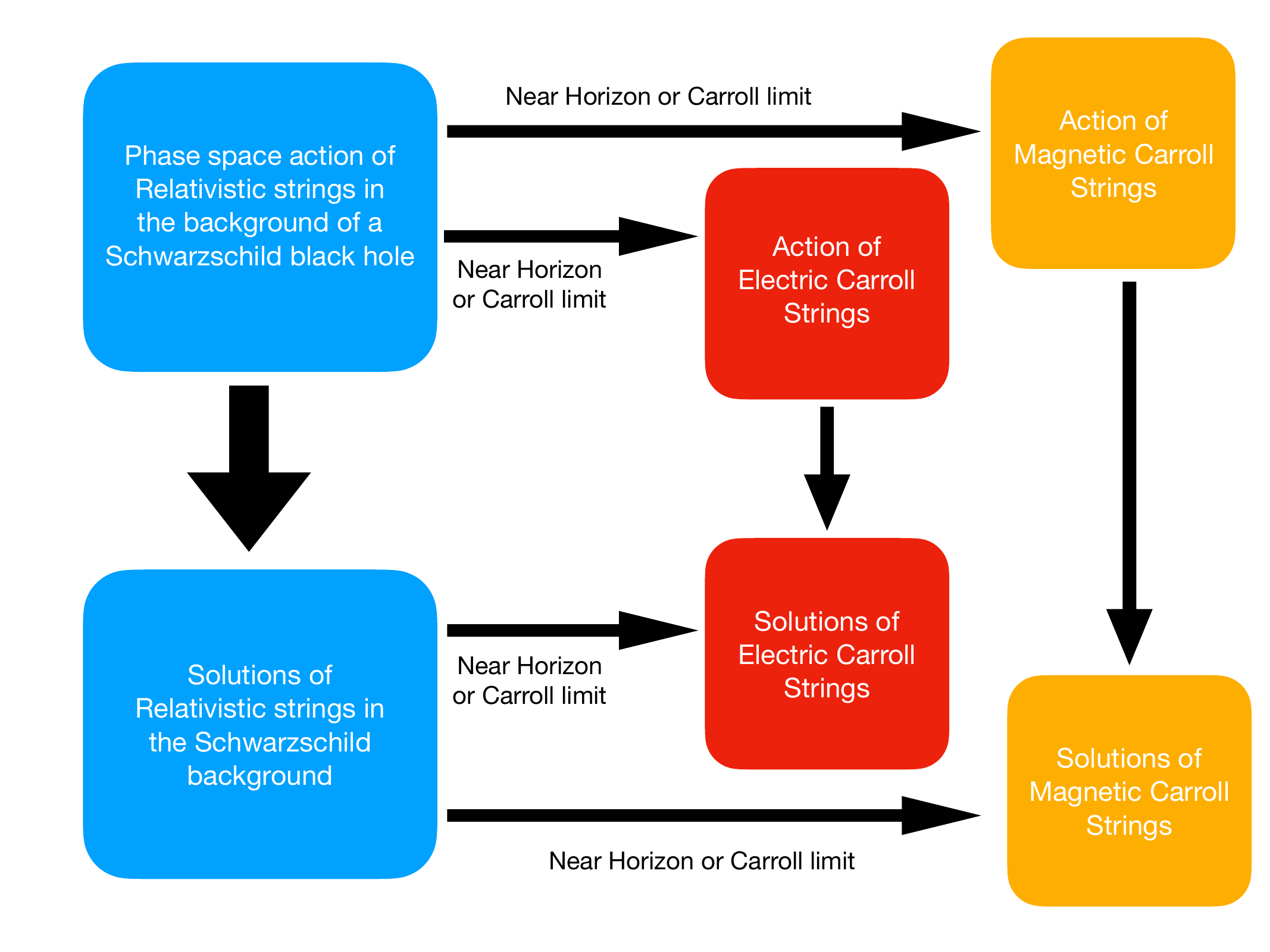}
	\caption{Closing squares: different ways to obtain solutions to strings near black holes.}
	\label{fig:square}
\end{figure}

We begin with the relativistic phase space Lagrangian
\begin{equation}
L = \oint d\sigma^1\,\left[ \dot X^\mu P_\mu - \frac{1}{2}e\left( g^{\mu\nu}(X)P_\mu P_\nu + T^2g_{\mu\nu}(X)X'^\mu X'^\nu \right) - u X'^\mu P_\mu\right] \, ,
\end{equation}
where now we have set $c=1$ (cf.~\eqref{eq:rel-phase-space-action}). Varying this with respect to $X^\mu$ leads to
\begin{equation}
    -\dot P_\mu+\partial_\sigma\left(u P_\mu\right)-\frac{e}{2}T^2\partial_\mu g_{\nu\rho}X'^\rho X'^\nu+\partial_\sigma\left(T^2 e g_{\mu\nu}X'^\nu\right)-\frac{e }{2}\partial_\mu g^{\nu\rho}P_\nu P_\rho=0\,.
\end{equation}
Using
\begin{equation}
P_\mu = \frac{1}{e}g_{\mu\nu}\left(\dot X^\nu-u X'^\nu\right)\,,
\end{equation}
and choosing a gauge in which $u=0$ (locally) and $e$ is a nonzero constant, the equation above can be recast as 
\begin{equation}
\label{eq:XEOM}
    \ddot X^\rho+\Gamma^\rho_{\mu\nu}\dot X^\mu \dot X^\nu-e^2 T^2\left(X''^\rho+\Gamma^\rho_{\mu\nu}X'^\mu X'^\nu\right)=0\,.
\end{equation}
In this gauge, the constraints imposed by the Lagrange mulitpliers $e$ and $u$ take the form
\begin{equation}
    \dot X\cdot X'=0\,,\qquad \dot X\cdot\dot X+ e^2 T^2 X'\cdot X'=0\,.
\end{equation}
We choose an orientation such that $\dot X\cdot\dot X<0$ and $X'\cdot X'>0$.

Now, as we have seen before, the near-horizon metric arranges itself in an $\epsilon$-expansion as in~\eqref{eq:schwarzschild-exp}. For the purposes of this subsection, it useful to split the coordinates used in that expression into two sets
\begin{equation}
    \label{eq:coordinate-split}
    x^{a}=(\tau, \mathdutchcal r)\,,\qquad x^{i}=(\theta,\phi)\,,
\end{equation}
i.e., we split the spacetime index $\mu = (a,i)$ into longitudinal components labelled by $a,b,c,\dots$ from the beginning of the Latin alphabet and transverse components labelled by $i,j,k,\dots$. We then define 
\begin{equation}
\begin{split}
    g_{ab}dx^{a} dx^{b} & = \epsilon\tau_{ab}dx^{a} dx^{b}+\mathcal{O}(\epsilon^2)\,,\\
g_{ij}dx^i dx^{j} & =  h_{ij}dx^{i} dx^{j}+\epsilon\Phi_{ij}dx^{i} dx^{j}+\mathcal{O}(\epsilon^2)\,,
\end{split}
\end{equation}
where $\tau_{ab}$, $h_{ij}$ and $\Phi_{ij}$ are part of the string Carroll tensors $\tau_{\mu\nu}$, $h_{\mu\nu}$ and $\Phi_{\mu\nu}$ as shown in the expansion of \eqref{eq:metric-exp}. In particular, this is useful since, as we may read off from~\eqref{eq:schwarzschild-string-carroll-bg}, $h_{\mu\nu}$ and $\Phi_{\mu\nu}$ only have nonzero legs in the transverse directions, while $\tau_{\mu\nu}$ only has nonzero longitudinal components. Including the choice of orientation discussed below~\eqref{eq:some-eq}, we have for the vectors $\dot X^\mu$ and $X'^\mu$ the following consistency conditions
\begin{subequations}
\label{eq:string-eom}
\begin{align}
        h_{ij}\dot X^{i } \dot X^{j}+\epsilon\tau_{a b}\dot X^{a } \dot X^{b}+\epsilon\Phi_{i j}\dot X^{i }\dot X^{j}+\mathcal{O}(\epsilon^2) & <  0\,,\label{eq:first-eq}\\
    h_{i j} X'^{i } X'^{j}+\epsilon\tau_{a b}X'^{a } X'^{b}+\epsilon\Phi_{i j}X'^{i } X'^{j} +\mathcal{O}(\epsilon^2)& >  0\,,\label{eq:second-eq}\\
    h_{i j}\dot X^{i } X'^{j}+\epsilon\tau_{a b}\dot X^{a } X'^{b}+\epsilon \Phi_{i j}\dot X^{i } X'^{j}+\mathcal{O}(\epsilon^2) & =  0\,, \label{eq:third-eq} \\
    \begin{split}
    \label{eq:fourth-eq}
    h_{i j}\dot X^{i } \dot X^{j}+\epsilon\tau_{a b}\dot X^{a }\dot X^{b}+\epsilon \Phi_{i j}\dot X^{i } \dot X^{j}\\
    +e^2 T^2\left(h_{i j}X'^{i } X'^{j}+ \epsilon\tau_{a b}(X)X'^{a } X'^{b}+\epsilon \Phi_{i j}X'^{i } X'^{j}\right) +\mathcal{O}(\epsilon^2) & =  0\,,    
    \end{split}
    \end{align}
\end{subequations}
The embedding scalars $X^\mu$ themselves admit a Taylor expansion in $\epsilon$ as in~Eq.~\eqref{eq:emb-field-exp-epsilon}, and we note that all geometric fields, e.g., $\Phi_{ij}$, in princple are functions of $X$ in the above. In what follows, we will denote the LO component of the embedding fields $X^\mu$ by $X^\mu\big|_{\text{LO}}$ rather than $x^\mu$ as we did above (cf.~Eq.~\eqref{eq:emb-field-exp-epsilon}) to avoid confusion with the coordinates that appear in~\eqref{eq:coordinate-split}.

Let us start with the first equation~\eqref{eq:first-eq}: here we see that since the first term (after taking care of the expansion of $X^i$) is of order $\epsilon^0$ and non-negative it must be that $h_{i j}\dot X^{i } \dot X^{j}$ vanishes at leading order in $\epsilon$, which is only possible if 
\begin{equation}
\label{eq:dot-x-i-zero}
\dot X^{i }\big|_{\text{LO}}=0\,.    
\end{equation}
At NLO the same equation then reveals that $\dot X^{a }\big|_{\text{LO}}$ is non-spacelike. Given that this holds, the effective constraints imposed on the string motion by the set of above equations then must be dependent on the behaviour of the spatial derivatives. The second equation~\eqref{eq:second-eq}, for example, is already obeyed at leading order if $X'^{i }\big|_{\text{LO}}\neq 0$. If, however, it is the case that $X'^{i }\big|_{\text{LO}}=0$, then we must have that $X'^{a }\big|_{\text{LO}}$ is non-timelike. 

Since $\dot X^{i }\big|_{\text{LO}}=0$ for the full set of equations, the third equation~\eqref{eq:third-eq} vanishes identically at LO. At NLO it depends on whether $X'^{i }\big|_{\text{LO}}$ is zero or not. If 
$X'^{i }\big|_{\text{LO}}=0$, Eq.~\eqref{eq:third-eq} reduces to
\begin{equation}
\label{eq:magconstraint1}
\tau_{a b}\dot X^{a }\big|_{\text{LO}} X'^{b}\big|_{\text{LO}}=0\,.    
\end{equation}
If $X'^{i }\big|_{\text{LO}}\neq 0$ then the equation involves NLO embedding scalars. Finally, the fourth equation~\eqref{eq:fourth-eq} is slightly more involved, and the possible conditions again split into two cases: either $X'^{i }\big|_{\text{LO}}=0$ or $X'^{i }\big|_{\text{LO}}\neq 0$. In the former case, due to~\eqref{eq:dot-x-i-zero}, we find the condition
\begin{equation}
\label{eq:magconstraint2}
    \tau_{a b}\dot X^{a }\big|_{\text{LO}}\dot X^{b}\big|_{\text{LO}}+e^2 T^2 \tau_{a  b}X'^{a }\big|_{\text{LO}} X'^{b}\big|_{\text{LO}} = 0\,.
\end{equation}
Since $\dot X^{a }\big|_{\text{LO}}$ and $X'^{a }\big|_{\text{LO}}$ are independent and orthogonal it follows that $\dot X^{a }\big|_{\text{LO}}$ is timelike with respect to $\tau_{a  b}$, while $X'^{a }\big|_{\text{LO}}$ is spacelike with respect to $\tau_{a  b}$. In this case we can choose a gauge such that $eT=1$.

If, on the other hand, we take $X'^{i }\big|_{\text{LO}}\neq 0$, the fourth equation~\eqref{eq:fourth-eq} reduces to 
\begin{equation}
\begin{split}
        &\epsilon\tau_{a b}\dot X^{a }\big|_{\text{LO}}\dot X^{b}\big|_{\text{LO}}\\
    &\quad+e^2 T^2\left(h_{i  j}X'^{i }\big|_{\text{LO}} X'^{j}\big|_{\text{LO}}+ \epsilon\tau_{a  b}X'^{a }\big|_{\text{LO}}X'^{b}\big|_{\text{LO}}+\epsilon \Phi_{i  j}X'^{i }\big|_{\text{LO}} X'^{j}\big|_{\text{LO}}\right)+\mathcal{O}(\epsilon^2) = 0\,.
\end{split}
\end{equation}
This implies that we can only have a cancellation when $e^2 T^2$ is order $\epsilon$ and so we will choose a gauge such that $e^2 T^2=\epsilon$. This immediately means that the last two terms in the above expression are of order $\epsilon^2$ and we may therefore discard them. We thus have at order $\epsilon$ the reduced equation
\begin{equation}
    \tau_{a b}\dot X^{a }\big|_{\text{LO}}\dot X^{b}\big|_{\text{LO}}+ h_{i  j}X'^{i }\big|_{\text{LO}} X'^{j}\big|_{\text{LO}} = 0\,,
\end{equation}
so that $\dot X^{a}\big|_{\text{LO}}$ is timelike.

Hence, there are two cases to consider: in the first we have $X'^{i }\big|_{\text{LO}}=0$ and so at LO the string just becomes a point on the 2-sphere (since we always have $\dot X^{i }\big|_{\text{LO}}=0$). Furthermore, with $e^2 T^2=1$, the $X^\mu$ equation of motion \eqref{eq:XEOM} at LO simplifies to
\begin{equation}
    \left(\ddot X^{a }+\Gamma^{a }_{b c}\dot X^{b}\dot X^{c}-X''^{a }-\Gamma^{a }_{b c}X'^{b} X'^{c}\right)\Big|_{\text{LO}} = 0\,,
\end{equation}
for the longitudinal coordinates.
In this expression the Levi-Civita connection is that of the 2D Rindler spacetime.  This should be supplemented with the constraints \eqref{eq:magconstraint1} and \eqref{eq:magconstraint2}. These are the equations of a string theory with a 2-dimensional (Rindler) target spacetime.

In the second case, we have that $X'^{i }\big|_{\text{LO}}\neq 0$ and $e^2 T^2=\epsilon$. Now the $X^\mu$ equation of motion \eqref{eq:XEOM} at LO simplifies to terms containing only temporal derivatives
\begin{equation}
    \left(\ddot X^{a }+\Gamma^{a }_{bc}\dot X^{b}\dot X^{c}\right)\Big|_{\text{LO}} = 0\,,
\end{equation}
which is the geodesic equation for a 2D Rindler spacetime.

Thus, by considering the full set of equations of motion~\eqref{eq:string-eom}, we have shown that the leading order solutions for the relativistic string near the horizon split into two distinct sectors:
\begin{subequations}
\label{eq:electric-magnetic-split}
\begin{align}
    &\text{Magnetic:} &X'^{i }\big|_{\text{LO}}&= 0\,,& e^2 T^2&=1\,,\label{eq:magnetic-case}\\
    &\text{Electric:} &X'^{i }\big|_{\text{LO}}&\neq 0\,,& e^2  T^2&=\epsilon\,,\label{eq:electric-case}
\end{align}    
\end{subequations}
and these are precisely the two cases we recognise as the magnetic and electric Carroll strings described in Section~\ref{sec:magnetic-and-electric-strings}. In particular, this shows that the behaviour of strings in the near-horizon region is governed by the same dynamical equations as those we obtain from the string Carroll expansion of relativistic string theory. This equivalence is one of the main results of this work.
To explicitly compare these defining conditions in \eqref{eq:electric-magnetic-split} to the earlier notion of $c$ scaling for magnetic and electric Carroll strings as shown in \eqref{eq:magnetic-string-combos} and \eqref{eq:electric-string-combos}, one can simply take the corresponding non-trivial string actions (\eqref{eq:P-NLO-lagrangian} and \eqref{eq:electriclimit}), and verify that the equations of motion for longitudinal coordinates reduce to the magnetic and electric equations of motion above.

We now demonstrate how to match the $c^2$ expansion and the near-horizon expansion discussed in this section off-shell. The off-shell $\epsilon$-expansion, i.e., the expansion of the Lagrangian, goes as follows. The starting point is the relativistic phase space Lagrangian~\eqref{eq:rel-phase-space-action} with $c=1$
\begin{equation}
\label{eq:rel-phase-space-action-c-is-one}
    L = \oint d\sigma^1\,\left[ \dot X^\mu P_\mu - \frac{1}{2}e\left( g^{\mu\nu}(X)P_\mu P_\nu + T^2g_{\mu\nu}(X)X'^\mu X'^\nu \right) - u X'^\mu P_\mu\right] \,.
\end{equation}
The embedding fields expand as in~\eqref{eq:emb-field-exp-epsilon}. The remaining $\epsilon$-expansions may be deduced from the on-shell expressions in~\eqref{eq:u}--\eqref{eq:P}. In particular, Eq.~\eqref{eq:u} tells us that $u = u_{(0)} + \mathcal{O}(\epsilon)$, while the scaling of $P$, which is determined by~\eqref{eq:P}, is dictated by the scaling we impose for $e$. In the magnetic case, we have that $e = \mathcal{O}(1)$, in which case $P = \mathcal{O}(1)$. This exactly reproduces the magnetic Carroll expansion of string theory as discussed in Section~\ref{sec:magnetic-string}, where the $\epsilon$-expansion of~\eqref{eq:rel-phase-space-action-c-is-one} takes the form 
\begin{equation}
    L = \epsilon^{-1} \tilde L_{\text{LO}} + \tilde L_{\text{NLO}} + \epsilon \tilde L_{\text{NNLO}} + \mathcal{O}(\epsilon^2)\,,
\end{equation}
in exact agreement with~\eqref{eq:magnetic-L-exp}. Note, however, that the tension that appears in the Lagrangians above is not rescaled; in other words, $\tilde T = T$. We can play the same game for the electric theory: there, the requirement that $e = \mathcal{O}(\sqrt{\epsilon})$ in~\eqref{eq:P} leads us to conclude that $P_\mu = \mathcal{O}(\sqrt{\epsilon})$. Writing
\begin{equation}
    P_{\mu} = \sqrt{\epsilon}\,P_{(0)\mu} + \mathcal{O}(\epsilon^{3/2})\,,\qquad e = \sqrt{\epsilon}\,\hat e_{(0)} +  \mathcal{O}(\epsilon^{3/2})\,,
\end{equation}
the $\epsilon$-expansion of~\eqref{eq:rel-phase-space-action-c-is-one} takes the form
\begin{equation}
    L = \sqrt{\epsilon} \hat L_{\text{LO}}  + \mathcal{O}(\epsilon^{3/2})\,,
\end{equation}
where $\hat L_{\text{LO}}$ is the electric Carroll string Lagrangian of~\eqref{eq:electriclimit} with $\hat T = T$. 

In this way, we have shown that the on-shell bifurcation into the magnetic and electric string solutions observed in~\eqref{eq:electric-magnetic-split} extends off-shell, where the associated $\epsilon$-expansions exactly reproduces the magnetic Lagrangians of section~\ref{sec:magnetic-string} and Appendix~\ref{app:phase-space-formulation-magnetic-string} when $\epsilon = \mathcal{O}(1)$, and the electric Lagrangian of Section~\ref{sec:electric-strings} when $e = \mathcal{O}(\sqrt{\epsilon})$. 

\subsection{Magnetic Carroll strings near a black hole horizon and folded strings}
\label{sec:magnetic-string-on-BH-bg}
The string solutions of the magnetic theory on the background given by~\eqref{eq:schwarzschild-string-carroll-bg} was studied from the perspective of the Polyakov formalism in~\cite{Bagchi:2023cfp}. In this section, we apply the results we derived in Section~\ref{sec:near-horizon-exp-of-strings} for the equations of motion for the near-horizon string. For the magnetic string, the LO transverse embedding fields $X^i\big|_{\text{LO}} = x^i$ are constants since $\dot x^i = x'^i = 0$, i.e.,
\begin{equation}
    x^\theta = x^\theta_0\,,\qquad x^\phi = x^\phi_0\,,
\end{equation}
where $x^{\theta,\phi}_0$ are constants. To obtain the equations of motion for the longitudinal LO embedding fields $X^a\big|_{\text{LO}} = x^a$, the analysis of Section~\ref{sec:near-horizon-exp-of-strings} implies that we need to solve the equations
\begin{equation}
\label{eq:Magnetic-LO-eqs}
    \begin{split}
    \tau_{ab}\dot x^a x'^b&=0\,,\\
    \tau_{ab}\dot x^a \dot x^b + \tau_{ab}x'^a x'^b &= 0\,,\\
    \ddot x^a + \Gamma^a_{bc}\dot x^b\dot x^c-x''^a-\Gamma^a_{bc}x'^b x'^c &= 0\,.
    \end{split}
\end{equation}
The solutions to these equations depend on whether $x'^{\mathdutchcal r}$ is zero or nonzero. When $x'^{\mathdutchcal r}=0$, the equations collapse to
\begin{equation}
    \tau_{ab}\dot x^a \dot x^b = 0\,,\qquad \ddot x^a + \Gamma^a_{bc}\dot x^b\dot x^c = 0\,,
\end{equation}
which describe a null geodesic in two-dimensional Rindler spacetime. Explicitly solving these equations leads to
\begin{equation}
    x^t = x_0^t \pm r_h \log (2\sigma^0)\,,\qquad x^{\mathdutchcal r} = r_h\sqrt{2\sigma^0}\,,
\end{equation}
which we can combine to find that
\begin{equation}
\label{eq:null-geodesic-rindler}
    x^t = x_0^t \pm 2r_h \log(x^{\mathdutchcal r}/r_h)\,.
\end{equation}
For the behaviour of the NLO embedding fields $y^i$, which appear in the NLO action, we solve~\eqref{eq:XEOM} at $\mathcal{O}(\epsilon)$. Using that $\Gamma^i_{ab} = 0$ and $x^i = x^i_0$, we get wave equations 
\begin{equation}
    \ddot y^i = y''^i\,,
\end{equation}
implying that
\begin{equation}
    y^\theta = y_-^\theta(\sigma^-) + y_+^\theta(\sigma^+)\,,\qquad y^\phi = y_-^\phi(\sigma^-) + y_+(\sigma^+)\,,
\end{equation}
where $\sigma^\pm = \sigma^0 \pm \sigma^1$. In other words, this branch of solutions corresponds to a string shrinking to a point as it approaches the horizon, after which it behaves as a massless particle in a two-dimensional Rindler spacetime. In particular, since the solution at LO does not depend on $\sigma^1$, we did not have to worry about imposing periodic boundary conditions for the LO embedding fields. This branch of solutions was described in~\cite{Bagchi:2023cfp} from the perspective of the Polyakov formalism.

There is, however, another branch where $x'^{\mathdutchcal r}$ is nonzero. There we must solve the full system of equations in~\eqref{eq:Magnetic-LO-eqs}, which was considered for Rindler space in general dimension in~\cite{deVega:1987um}. In that case, the string orients itself radially and has no angular extent to leading order: this is the folded string we briefly mentioned in~\cite{Bagchi:2023cfp}. For a discussion of folds in two-dimensional string theories, we refer to~\cite{Ganor:1994rm,Bars:1994sv}. 

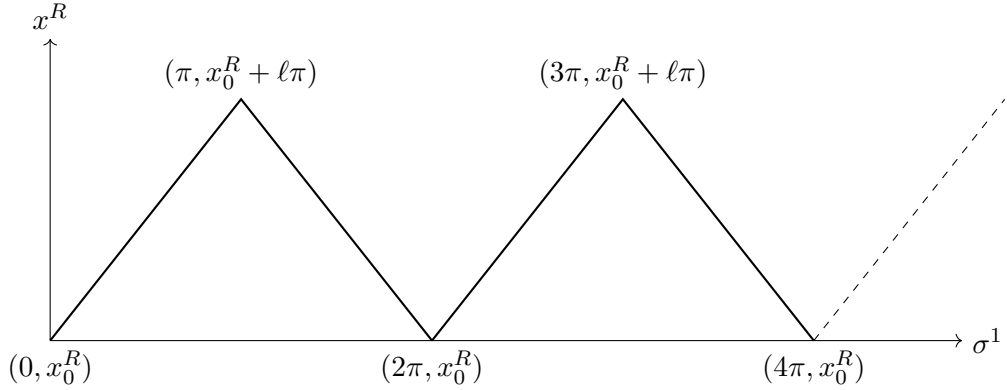
\begin{figure}
    \centering 
    	\begin{tikzpicture}[scale=0.8]
		% Axes
		\draw[->] (0,0) -- (15,0) node[right] {$\sigma^1$};
		\draw[->] (0,0) -- (0,5) node[above] {$x^R$};
		
		% Saw-toothed graph
		\foreach \i in {0,1} {
			\draw[thick] (\i*2*3.14,0) -- ({\i*2*3.14+3.14},4) -- ({\i*2*3.14+2*3.14},0);
		}
        \draw[dashed] (4*3.14,0) -- (5*3.14,4);	
		% Labels
		\node[below] at (0,0) {$(0,x^R_0)$};
		\node[above] at (3.14,4) {$(\pi,x_0^R + \ell\pi)$};
		\node[below] at (2*3.14,0) {$(2\pi,x_0^R)$};
        \node[above] at (3*3.14,4) {$(3\pi,x_0^R + \ell\pi)$};
        \node[below] at (4*3.14,0) {$(4\pi,x_0^R)$};
	\end{tikzpicture}
    \caption{The graph of the yo-yo solution~\eqref{eq:yo-yo-sol} that shows $x^R$ as a function of $\sigma^1$.}
    \label{fig:yoyo-sol-graph}
\end{figure}

To explicitly solve the equations~\eqref{eq:Magnetic-LO-eqs}, it is useful to go to target space coordinates $(T,R)$ defined via
\begin{equation}
    \begin{split}
        T = 2\mathdutchcal r \sinh(t/2r_h)\,,\qquad R = 2\mathdutchcal r \cosh(t/2r_h)\,.
    \end{split}
\end{equation}
Since $t\in (-\infty,\infty)$ and $\mathdutchcal r \in (0,\infty)$, we find that $T \in (-\infty,\infty)$ and $R\in (0,\infty)$. In terms of $(T,R)$, the longitudinal line element $ds^2_{\parallel}$ takes the form of two-dimensional Minkowski space
\begin{equation}
    ds^2_{\parallel} = -dT^2 + dR^2\,.
\end{equation}
The inverse transformations are
\begin{equation}
\label{eq:inverse-trafo}
    t = 2r_h\tanh^{-1}(T/R)\,,\qquad \mathdutchcal r = \frac{1}{2}\sqrt{R^2 - T^2}\,.
\end{equation}
Defining $x^\pm = x^T \pm x^R$, the equations for the embedding fields in~\eqref{eq:Magnetic-LO-eqs} simplify to
\begin{equation}
\label{eq:folded-string-eqs}
    \begin{split}
        \D_+\D_- x^a &= 0\,,\\
        \D_+ x^+\D_+x^- &= 0\,,\\
        \D_-x^+ \D_-x^- &= 0\,.
    \end{split}
\end{equation}
The first equation tells us that
\begin{equation}
\label{eq:x+andx-}
    x^+ = f(\sigma^+) + g(\sigma^-)\,,\qquad x^- = h(\sigma^+) + j(\sigma^-)\,,
\end{equation}
while the two last equations tell us that either $g$ and $h$ are constants or $f$ and $j$ are constants. By absorbing these constants into $f$ and $j$ or $g$ and $h$, respectively, we may without loss of generality set those constants to zero. Another branch of solutions that solves the equations~\eqref{eq:folded-string-eqs} has the form
\begin{equation}
\begin{split}
    x^+ &= x^+_0\qquad\text{and}\qquad x^- = a(\sigma^+) + b(\sigma^-)\,,\\
    x^- &= x^-_0\qquad\text{and}\qquad x^+ = c(\sigma^+) + d(\sigma^-)\,,
\end{split}
\end{equation}
where $x^\pm_0$ are constants.  In summary, we find four classes of solutions
\begin{align}
\label{eq:folded-string-cases}
\begin{aligned}
\textsc{I}&:x^+ = f(\sigma^+)\,,&&& x^- = j(\sigma^-)\,,\\
\textsc{I\hspace{-1pt}I}&:x^+=g(\sigma^-)\,,&&& x^- = h(\sigma^+)\,,\\
\textsc{I\hspace{-1pt}I\hspace{-1pt}I}&: x^+ =x^+_0 \,,&&& x^- = a(\sigma^+) + b(\sigma^-)\,,\\
\textsc{{I\hspace{-1.5pt}V}}&: x^- = x^-_0 \,,&&& x^+ = c(\sigma^+) + d(\sigma^-)\,,
\end{aligned}
\end{align}
where the functions $f,g,h,j,a,b,c,d$ are all $2\pi$-periodic, though not continuously differentiable. In particular, let us consider the ``yo-yo solution''~\cite{Bars:1994sv} (see also~\cite{Bardeen:1975gx})
\begin{equation}
\label{eq:yo-yo-sol}
\begin{split}
x^T &= \ell \sigma^0\,,\\ 
    x^R &= x^R_0 + \ell \sum_{n\in \ZZ }\Big[ \mathbb{1}_{[2\pi n,\pi(1+2n)]}(\sigma^1)(\sigma^1-2\pi n) + \mathbb{1}_{(\pi(1+2n),2\pi(1+n) )}(\sigma^1)(2\pi(1+n)-\sigma^1)\Big]\,,    
\end{split}
\end{equation}
where $\ell$ and $x_0^R$ are constants, and where, by construction, $x^R \geq 0$. The graph of the yo-yo solution~\eqref{eq:yo-yo-sol} in the $R$-direction features in Figure~\ref{fig:yoyo-sol-graph}. We also used indicator functions $\mathbb{1}_I(z)$, where $I \subset \RR$ is an interval. These are defined by
\begin{equation}
    \mathbb{1}_I(z) = \begin{cases}
        1\quad\text{if}\quad z\in I \\
        0\quad\text{if}\quad z\notin I
    \end{cases}\,.
\end{equation}
In all intervals $[2\pi n,\pi(1+2n)]$ for $n\in \ZZ$, the solution~\eqref{eq:yo-yo-sol} corresponds to case $\textsc{I}$ in~\eqref{eq:folded-string-cases} with $f(\sigma^+) = \ell\sigma^+ + x_0^R$ and $j(\sigma^-) = \ell\sigma^- - x_0^R$, where $\sigma^\pm=\sigma^0\pm\sigma^1$ with $\sigma^1$ evaluated mod $2\pi$. In all intervals $(\pi(1+2n),2\pi(1+n) )$ for $n\in \ZZ$ the solution corresponds to case $\textsc{I\hspace{-1pt}I}$ with $g(\sigma^-) = \ell \sigma^- + x_0^R$ and $h(\sigma^+) = \ell\sigma^+ - x_0^R$. The folded string solution in Rindler coordinates $(t,\mathdutchcal r)$ is obtained by substituting~\eqref{eq:yo-yo-sol} into the inverse relations~\eqref{eq:inverse-trafo}.

\subsection{Electric Carroll strings near a black hole horizon and wrapped strings}
\label{sec:electric-strings-on-BH-bg}
In this section we will study the electric string~\eqref{eq:electriclimit2} on a string Carroll target space with $\tau_{\mu\nu}$ and $h_{\mu\nu}$ given in~\eqref{eq:schwarzschild-string-carroll-bg}. On this background, the Lagrangian density of~\eqref{eq:electriclimit2} becomes
\begin{eqnarray}
\hat{\mathcal{L}}_{\text{LO}} & = & r_h\left(\dot x^\theta-u x'^\theta\right)P_1+r_h\sin(x^\theta)\left(\dot x^\phi-ux'^\phi\right)P_2\\&&+\frac{1}{2\hat e}\left[-\frac{(x^{\mathdutchcal r})^2}{r_h^2}\left(\dot x^t-u x'^t\right)^2+4\left(\dot x^{\mathdutchcal r}-u x'^{\mathdutchcal r}\right)^2\right]
-\frac{ T^2}{2}r_h^2\hat e\left[(x'^\theta)^2+(x'^\phi)^2\sin^2(x^\theta)\right]\,.\nn
\end{eqnarray}
The equations of motion of $P_1$, $P_2$, $u$, $\hat e$, $x^\theta$, $x^\phi$, $x^t$, $x^{\mathdutchcal r}$ are, respectively,
\begin{subequations}
\label{eq:electric-eoms}
\begin{align}
0 & =  \dot x^\theta-u x'^\theta\,,\label{eq:EOM1}\\
0 & =  \sin(x^\theta)\left(\dot x^\phi-ux'^\phi\right)\,,\label{eq:EOM2}\\
0 & =  -r_hx'^\theta P_1-r_h\sin(x^\theta)x'^\phi P_2+\frac{1}{\hat e}\frac{(x^{\mathdutchcal r})^2}{r_h^2}\left(\dot x^t-ux'^t\right)x'^t-\frac{4}{\hat e}\left(\dot x^{\mathdutchcal r}-u x'^{\mathdutchcal r}\right)x'^{\mathdutchcal r}\,,\label{eq:EOM3}\\
0 & =  \frac{1}{2\hat e^2}\left[-\frac{(x^{\mathdutchcal r})^2}{r_h^2}\left(\dot x^t-u x'^t\right)^2+4\left(\dot x^{\mathdutchcal r} - ux'^{\mathdutchcal r} \right)^2\right]+\frac{ T^2}{2}r_h^2\left[(x'^\theta)^2+\sin^2(x^\theta)(x'^\phi)^2\right]\,,\label{eq:EOM4}\\
0 & =  -r_h\dot P_1+r_h\left(u P_1\right)'+r_h\cos(x^\theta)\left(\dot x^\phi-u x'^\phi\right)P_2+  T^2 r_h^2\left(\hat e x'^\theta\right)'\nn\\
&\qquad- T^2r_h^2\hat e\sin (x^\theta)\cos (x^\theta) (x'^\phi)^2\,,\label{eq:EOM5}\\
0 & =  -r_h\partial_0\left(\sin(x^\theta) P_2\right)+r_h\left(u\sin(x^\theta) P_2\right)'+ T^2 r_h^2\left(\hat e\sin^2(x^\theta) x'^\phi\right)'\,,\label{eq:EOM6}\\
0 & =  \partial_0\left(\frac{(x^{\mathdutchcal r})^2}{\hat e}\left(\dot x^t-u x'^t\right)\right)-\left(\frac{u}{\hat e}(x^{\mathdutchcal r})^2\left(\dot x^t-u x'^t\right)\right)'\,,\label{eq:EOM7}\\
0 & =  -\frac{x^{\mathdutchcal r}}{r_h^2\hat e}\left(\dot x^t-u x'^t\right)^2-\partial_0\left(\frac{4}{\hat e}\left(\dot x^{\mathdutchcal r} -u x'^{\mathdutchcal r}\right)\right)+\left(\frac{4u}{\hat e}\left(\dot x^{\mathdutchcal r} - u x'^{\mathdutchcal r}\right)\right)'\,.\label{eq:EOM8}
\end{align}
\end{subequations}
In order to solve these equations we will choose the gauge \eqref{eq:gaugechoice}. For the sake of solving the equations of motion it is helpful to perform the following coordinate transformation on the worldsheet
\begin{equation}\label{eq:2Ddiffeo}
\tilde\sigma^0=\sigma^0\,,\qquad\tilde\sigma^1=\sigma^1+U(\sigma^0)\,,
\end{equation}
where $\tilde\partial_0 U(\tilde\sigma^0)=u(\tilde\sigma^0)$. In the tilded coordinates we have
\begin{equation}
\tilde\partial_0=\partial_0-u\partial_1\,,\qquad\tilde\partial_1=\partial_1\,,
\end{equation}
where we notice that the equations of motion~\eqref{eq:electric-eoms} simplify substantially when using these tilded coordinates.

Equations \eqref{eq:EOM1} and \eqref{eq:EOM2} are solved by
\begin{equation}
x^\theta=x^\theta(\tilde\sigma^1)\,,\qquad x^\phi=x^\phi(\tilde\sigma^1)\,.
\end{equation}
Equation \eqref{eq:EOM7} can be integrated once to give
\begin{equation}\label{eq:tau}
\tilde\partial_0 x^t=\frac{f}{(x^{\mathdutchcal r})^2}\,,
\end{equation}
where $f$ is a nonzero function of $\tilde\sigma^1$. We want the string embedding to be such that the relation between worldsheet time and target space time is invertible. Using this result equation \eqref{eq:EOM8} becomes
\begin{equation}
\tilde\partial_0^2x^{\mathdutchcal r}=-\frac{f^2}{4 r_h^2}(x^{\mathdutchcal r})^{-3}\,.
\end{equation}
This equation is solved by
\begin{equation}
x^{\mathdutchcal r}=\left(a(\tilde\sigma^0)^2+b\tilde\sigma^0+c\right)^{1/2}\,,
\end{equation}
where $a,b,c$ are functions of $\tilde\sigma^1$ such that 
\begin{equation}
\label{eq:f-rel}
b^2-4ac=\frac{f^2}{r_h^2}\,.
\end{equation}
If $a$ is nonzero we may, since $f$ is also nonzero, write $x^{\mathdutchcal r}$ as
\begin{equation}
(x^{\mathdutchcal r})^2=a\left(\tilde\sigma^0-\lambda_+\right)\left(\tilde\sigma^0-\lambda_-\right)\,,
\end{equation}
where $\lambda_+$ and $\lambda_-$ are real and never equal to each other. In this case, we may recast~\eqref{eq:f-rel} as
\begin{equation}
\label{eq:f-lambda-rel}
    f = \pm r_h a (\lambda_+ - \lambda_-)\,.
\end{equation}
Equation \eqref{eq:EOM4} tells us that
\begin{equation}\label{eq:a}
\frac{T^2r_h^2}{4}\left((x'^\theta)^2+\sin^2(x^\theta) (x'^\phi)^2\right)=-a\,,
\end{equation}
so that $a < 0$. The case $a=0$ is not allowed due to the first equation of \eqref{eq:electric-case}. Hence the solution to \eqref{eq:tau} is
\begin{equation}
x^t=\pm r_h\log\frac{\tilde\sigma^0-\lambda_+}{\tilde\sigma^0-\lambda_-}+H(\tilde\sigma^1)\,,
\end{equation}
where we used~\eqref{eq:f-lambda-rel}. We will choose the plus sign in both cases. 

Consider next equations \eqref{eq:EOM3}, \eqref{eq:EOM5} and \eqref{eq:EOM6}. First of all it can be shown that if we differentiate \eqref{eq:EOM3} with respect to $\tilde\sigma^0$ then the resulting equation is automatically satisfied on account of equations \eqref{eq:EOM5} and \eqref{eq:EOM6} as well as the above solutions for $x^t$ and $x^{\mathdutchcal r}$. This means that as soon as \eqref{eq:EOM3} is satisfied at a certain instant in $\tilde\sigma^0$ it will be satisfied for all times. 
Differentiating equations \eqref{eq:EOM5} and \eqref{eq:EOM6} with respect to $\tilde\sigma^0$ tells us that $P_1$ and $P_2$ are both linear functions of $\tilde\sigma^0$, i.e $\dot P_1$ and $\dot P_2$ are functions of $\sigma^1$ only. In our gauge equations \eqref{eq:EOM5} and \eqref{eq:EOM6} can be obtained from the following effective 2-dimensional Lagrangian
\begin{equation}
    \mathcal{L}[P_1,P_2,x^\theta, x^\phi]=r_h P_1 \tilde\partial_0 x^\theta+r_h P_2(\sin x^\theta)\tilde\partial_0 x^\phi-\frac{ T^2}{2}r_h^2\left[(x'^\theta)^2+(x'^\phi)^2\sin^2(x^\theta)\right]\,.
\end{equation}
This is a magnetic Carroll theory for $x^\theta$ and $x^\phi$ with Lagrange multipliers $P_1$ and $P_2$. When $\dot P_1$ and $\dot P_2$ are both zero then equations  \eqref{eq:EOM5} and \eqref{eq:EOM6} are the geodesic equations on the 2-sphere whose solutions are great circles. In this case $a$ is constant. 

The equation that remains to be solved is \eqref{eq:a}. We will assume that $\dot P_1=\dot P_2=0$ so that the solution for $x^\theta$ and $x^\phi$ is a geodesic. We can use spherical symmetry to set without loss of generality $x^\theta=\pi/2$, so that the strings wrap the equator. If we define $a=-A^2$ where $A$ is a nonzero constant then along the equator equation \eqref{eq:a} becomes
\begin{equation}
 x'^\phi=\pm \frac{2A}{ T r_h}\,.
\end{equation}
The right-hand side must be an integer (winding number) in order for the solution to be globally well-defined.
An example of a string solution that wraps the sphere is to take $x^\phi=\tilde\sigma^1=\sigma^1$, so that $A= T r_h/2$ and $U=0$ (for the definition of $U$ see equation \eqref{eq:2Ddiffeo}). The $a<0$ solutions (for $x^t$ and $x^{\mathdutchcal r}$) are timelike geodesics in 2D Rindler spacetime. From the point of the Rindler space the string moves along a geodesic and the winding of the string along the 2-sphere gives the Rindler particle a mass.

As mentioned earlier the fields $x^t$ and $x^{\mathdutchcal r}$ are electric Carroll scalars and so their spatial, i.e., $\sigma^1$-dependence, is not fixed at LO. This will be fixed by going to subleading orders (see, e.g.,~\cite{deBoer:2021jej}). 

\section{Discussion}
\label{sec:discussion}
\subsection*{Summary of the paper}
In this work, we developed the theory of Carrollian strings by expanding the relativistic phase space action in powers of $c^2$. We demonstrated that this string Carroll expansion leads to two different Carrollian string theories: the magnetic and the electric Carroll string, which are distinguished by the $c^2$-scaling of the relativistic Lagrange multiplier $e$ appearing in the phase space action. While certain aspects of the magnetic Carroll string appeared in~\cite{Bagchi:2023cfp}, the electric Carroll string has not previously been obtained from a $c^2$ expansion (although it featured in~\cite{Blair:2023noj,Gomis:2023eav} as explained in the main text). In particular, the electric Carroll string comes with a Carrollian worldsheet. 

We then showed that the near-horizon expansion of the four-dimensional Schwarzschild black hole takes the form of a string Carroll expansion, where the longitudinal space is a 2D Rindler spacetime and the transverse space a $2$-sphere, and so strings near a Schwarzschild black hole behave as either magnetic or electric Carroll strings. For both of these cases, we constructed the solutions on the background that arises from the near-horizon expansion. For the magnetic case, we find two classes of solutions: in the first, the string shrinks to a point on the sphere. This solution describes null geodesics in the 2D Rindler spacetime. The other class of solutions describe folded strings. In the electric case the string wraps the transverse $2$-sphere, where the wrapping number essentially renders the string massive, and hence it will follow timelike geodesics in the 2D Rindler spacetime.

\subsection*{Questions and Future directions}
The formalism developed in this work is expected to have a wide range of applications, and we have only scratched the surface in the present instalment. To set the stage for future explorations, let us conclude with a brief description of several interesting avenues of related future research. 

\begin{description}

\item[Is the near-horizon string electric or magnetic?]\hfill \\ 
Our paper set out to explore the question of strings probing black holes and we have found intriguing answers. The classical string either grows enormously and wraps the sphere in the near-horizon region (electric string), or it either folds or shrinks to a point on the sphere as it enters the near-horizon region (electric/magnetic string). The split of the solution space into the electric/magnetic branches is dictated by \eqref{eq:electric-magnetic-split}. However, physically it seems unclear what exactly dictates the behaviour of the string or, more precisely, which branch of the solution space the string chooses as it comes close to the black hole and why. It is likely that this is tied to the initial conditions of how the string approaches the near-horizon region. It would be of interest to clarify this rather important question.

\item[Observer-dependence and the picture in Eddington--Finkelstein coordinates. ] \hfill 
One might wonder if the near-horizon expansion of the Schwarzschild black hole only takes the form of a string Carroll expansion for a stationary asymptotic observer. 
To allay any such fears, we briefly consider the near-horizon expansion in infalling Eddington--Finkelstein coordinates, where the Schwarzschild metric reads
\begin{equation}
    ds^2 = -f(r)dv^2 + 2dvdr + r^2d\Omega^2\,,
\end{equation}
where $f(r) = 1-\frac{r_h}{r}$. As above, we change coordinates in the near-horizon region according to
\begin{equation}
    r = r_h + \frac{1}{r_h}\epsilon\mathdutchcal{r}^2\,,
\end{equation}
where $\epsilon$ is a positive dimensionless parameter. Expanding the metric in terms of $\epsilon$ again leads to a string Carroll expansion
\begin{equation}
\label{eq:schwarzschild-exp-EF}
    ds^2 = r_h^2d\Omega^2 + \epsilon\left[ -\frac{\mathdutchcal{r}^2}{r_h^2}dv^2 + \frac{4\mathdutchcal{r}}{r_h}dv d\mathdutchcal{r} + 2\mathdutchcal{r}^2d\Omega^2 \right] + \mathcal{O}(\epsilon^2)\,.
\end{equation}
Comparing this with the string Carroll expansion of the metric~\eqref{eq:metric-exp}
with the expansion in~\eqref{eq:schwarzschild-exp-EF} reveals that the corresponding string Carroll structure is given by
\begin{equation}
\label{eq:schwarzschild-string-carroll-bg-EF}
    \begin{split}
        h_{\mu\nu}dx^\mu dx^\nu = r_h^2d\Omega^2\,,\qquad
        \tau_{\mu\nu} dx^\mu dx^\nu = -\frac{\mathdutchcal{r}^2}{r_h^2}dv^2 + \frac{4\mathdutchcal{r}}{r_h}dvd\mathdutchcal{r}\,,\qquad
        \Phi_{\mu\nu} dx^\mu dx^\nu = 2\mathdutchcal{r}^2d\Omega^2 \,.
    \end{split}
\end{equation}
Thus, the near-horizon expansion again takes the form of a string Carroll expansion for an infalling observer, and is not an artefact of the Schwarzschild coordinates we employed in Section~\ref{sec:near-horizon-is-string-Carroll}. 

\item[Near-horizon regions of more general black holes and extremality.] \hfill \\ We expect that our analysis naturally extends to all non-extremal black holes which all contain a 2D Rindler space in their near-horizon region. The case of the Reissner--Nordstr\"om (RN) black hole is a simple generalisation, as we discuss next. Consider a metric of the form
\begin{equation}\label{eq:sphericalBHs}
    ds^2=-f(r)dt^2+\frac{dr^2}{f(r)}+r^2d\Omega^2\,,
\end{equation}
where $f$ is some function of $r$ with a first order real zero at $r_h$ and $f'(r_h)>0$. Defining $r=r_h+f'(r_h)\varepsilon\mathdutchcal{r}^2$ and using that $f=(f'(r_h))^2\varepsilon\mathdutchcal{r}^2+\mathcal{O}(\varepsilon^2)$ we obtain the approximate near-horizon metric
\begin{equation}
    ds^2=\varepsilon\left(-(f'(r_h))^2\mathdutchcal{r}^2dt^2+4d\mathdutchcal{r}^2 + 2r_hf'(r_h)\mathdutchcal{r}^2 d\Omega^2\right)+r_h^2d\Omega^2+\mathcal{O}(\varepsilon^2)\,,
\end{equation}
which is again of the form of a string Carroll expansion. This includes the case of a RN black hole. In that case we also need to include the Maxwell potential and relate it to either an electric or a magnetic Carrollian gauge field. For the family of BH metrics \eqref{eq:sphericalBHs} the string Carroll geometry is a trivial fibration, namely a product manifold of 2-dimensional Rindler space and a 2-sphere. 

For the Kerr metric, the 2D Rindler space comes with a non-trivial fibre-bundle structure. We provide details of the near-horizon expansion of the non-extremal Kerr solution and how it too can be viewed in terms of string Carroll geometry in Appendix \ref{Kerr}. It would be of interest to explore the solution space for these more general black holes to see how our analysis is generalised, while maintaining the basic features. 

For these more generic black holes, it is particularly intriguing to explore the approach to extremality. It is of course well known that the near-horizon region of an extremal black hole has a metric that contains an $\mathsf{AdS}_2$ factor \cite{Kunduri:2007vf}. One might wonder if and how we can study the extremal limit using a string Carroll expansion of the non-extremal black hole. We hope to come back to this topic in the near future.

\item[Quantisation and the spreading of strings.] \hfill \\ An obvious next step involves the quantisation of the near-horizon string theories developed in this work. 

The electric Carroll string arises at LO and can be considered a tensionful deformation of the null string, which nevertheless retains the Carrollian conformal algebra as its residual symmetry algebra, as described in the main text. It would be interesting to investigate its spectrum and quantum properties along the lines of~\cite{Bagchi:2020fpr}, where it was found that there are three consistent quantum theories arising from a single classical null string.

It has long been known in the literature \cite{Susskind:1993ki} that strings tend to spread over black hole horizons as their effective tension goes to zero. This is an inherently quantum process, since the diffusion of the fundamental string over the horizon is controlled by the evolution of the string wave function. As argued in \cite{Susskind:1993ki} this effect is particular to an asymptotic observer, and this observer will see the string become infinitely diffused and spread across the horizon in finite time. This phenomenon has deep implications for, e.g., information loss during the process of black hole evaporation. It would be interesting to explore if the string solutions discussed in this work, when quantised, help us understand some of these effects better.

More generally, it would be interesting to consider Carrollian open strings and D-branes in both the magnetic and electric Carroll string theories using the methods of~\cite{Hartong:2024ydv}, and to investigate the worldvolume theories that arise on the Carrollian D-branes. 

\item[$p$-brane Carroll expansions, $\beta$-functions and Carrollian gravity] \hfill \\ The magnetic and electric Carroll string theories as described in Sections~\ref{sec:magnetic-string} and~\ref{sec:electric-strings}, respectively, readily generalise to ``$p$-brane Carroll expansions'', where $(p+1)$ directions are singled out rather than just two as in the string Carroll expansion that is relevant for the near-horizon expansion. In particular, the particle ($p=0$) expansion discussed in Appendix~\ref{app:particle-exps} leads to an electric Carroll string Lagrangian of the form
    \begin{equation*}
        L_{\text{e-particle}} = \oint d\sigma^1\mathbb e\bigg[\hat T^2 \mathbb v^\alpha\partial_\alpha x^\mu e^{A'}_\mu \tilde P_{A'} -  \frac{1}{2}\tau_{\mu}\tau_\nu\mathbb v^\alpha\partial_\alpha x^\mu \mathbb v^\beta\partial_\beta x^\nu-\frac{\hat T^2}{2}h_{\mu\nu}\mathbb e^\alpha\partial_\alpha x^\mu \mathbb e^\beta\partial_\beta x^\nu \bigg]\,,
    \end{equation*}
    while the forms of the magnetic Lagrangians in~\eqref{eq:LO-Polyakov-Lagrangian} and~\eqref{eq:P-NLO-lagrangian} remain unchanged, though the target space geometry is now particle Carroll geometry (see Appendix~\ref{app:particle-exps} for more details). Computing the $\beta$-functions for these particle Carroll string theories would be a worthwhile endeavour, since they could be compared with the equations of motion that arise in Carrollian gravity~\cite{Hansen:2021fxi,figueroa-ofarrill:2022mcy}. More generally, one could also do this for $p$-brane Carrollian target spaces and match with the results of~\cite{Bergshoeff:2023rkk}.

\end{description}

\section*{Acknowledgements} 
We are grateful to Mangesh Mandlik for initial collaboration on the project and to Johannes Lahnsteiner and Niels Obers for useful discussions. AB would also like to acknowledge previous discussions with Joan Sim\'on which initiated this line of inquiry. 

\smallskip

AB is partially supported by a Swarnajayanti Fellowship from the Science and Engineering Research Board
(SERB) under grant SB/SJF/2019-20/08 and also by SERB grant CRG/2022/006165. ArB is supported in part by an OPERA grant and a seed grant NFSG/PIL/2023/P3816 from BITS-Pilani. He also acknowledges financial support from the Asia Pacific Center for Theoretical Physics via an Associate Fellowship. JH was supported by the Royal Society University Research Fellowship
Renewal “Non-Lorentzian String Theory” (grant number URF\textbackslash
R\textbackslash 221038). AB and JH were also supported by the Royal Society International Exchange Grant `Carrollian Symmetry and String Theory' (IES\textbackslash R3\textbackslash 213165). The work of EH is supported by Villum Foundation Experiment project 00050317, ``Exploring the wonderland of Carrollian physics''. KK was partially supported by the SERB grant SB/SJF/2019-20/08. KK thanks the hospitality of the University of Edinburgh while this work was in progress, and this visit was supported by the Royal Society International Exchange grant (IES\textbackslash R3\textbackslash 213165). Finally, AB, JH and EH thank the participants and organisers of the workshop ``Carrollian Physics and Holography'' organised at the Erwin Schr{\"o}dinger Institute (ESI), University of Vienna, for interesting discussions, and the ESI for hospitality during the visit.

\bigskip \bigskip

\appendix
\section*{APPENDICES}
\section{String Carroll expansion and its symmetries}
\label{app:trafos}
In this appendix, we provide additional details about the string Carroll expansion of a $(d+2)$-dimensional Lorentzian geometry up to NNLO, complementing the discussion of Section~\ref{sec:string-carroll-geom}. 
\subsection{String Carroll geometry from the string \texorpdfstring{$c^2$}{c} expansion}
The Lorentzian geometry is described by a metric $g_{\mu\nu}$, where $\mu,\nu = 0,\dots,d+1$ are spacetime indices. Like its Galilean counterpart~\cite{Hartong:2021ekg,Hartong:2022dsx,Hartong:2024ydv}, the string Carroll expansion singles out two ``longitudinal'' directions.\footnote{One could single out more directions, leading to a ``$p$-brane'' Carroll expansion''. See~\cite{Barducci:2018wuj,Bergshoeff:2020xhv,Bergshoeff:2023rkk} for a description of $p$-brane Carrollian geometry.} It is useful to express the metric $g_{\mu\nu}$ and its inverse $g^{\mu\nu}$ in terms of vielbeine $\hat E^{\bar A}_\mu$ and $\hat E^\mu_{\bar A}$, where $\bar A = 0,\dots, d+1$ is a Lorentzian tangent space index, leading to
\begin{equation}
\label{eq:lorentzian-vielbeins}
    g_{\mu\nu} = \eta_{\bar A\bar B} \hat E^{\bar A}_\mu \hat E^{\bar B}_\nu\,,\qquad g^{\mu\nu} = \eta^{\bar A \bar B} \hat E^\mu_{\bar A} \hat E^\nu_{\bar B}\,,
\end{equation}
where $\eta_{\bar A\bar B} = \text{diag}(-1,1,\dots,1)$ is the $(d+2)$-dimensional Minkowski metric in tangent space, and $\eta^{\bar A \bar A}$ its inverse. The Lorentzian vielbeine satisfy the relations
\begin{equation}
\label{eq:lorentzian-vielbein-rels}
    \hat E^\mu_{\bar A} \hat E_\mu^{\bar B} = \delta^{\bar A}_{\bar B}\,,\qquad \hat E^\mu_{\bar A} \hat E_\nu^{\bar A} = \delta^{\mu}_{\nu}\,.
\end{equation}
To set up the string Carroll expansion, we split off a two-dimensional longitudinal subspace that scales with $c$ according to
\begin{equation}
    \hat E^{\bar A}_\mu = c T^A_\mu \delta^{\bar A}_A + \mathcal{E}^{A'}_\mu \delta^{\bar A}_{A'}\,,\qquad \hat E^\mu_{\bar A} = c^{-1} V^\mu_A \delta^A_{\bar A} + \mathcal{E}_{A'}^\mu \delta^{A'}_{\bar A}\,,
\end{equation}
where $A = 0,1$ is a longitudinal tangent space index, while $A'=2,\dots,d+1$ is a transverse tangent space index. We will refer to the objects $(T^A_\mu, V^\mu_A,\mathcal{E}^{A'}_\mu,\mathcal{E}^\mu_{A'})$ as ``pre-Carrollian variables''. The relations~\eqref{eq:lorentzian-vielbein-rels} for the Lorentzian vielbeine imply that the pre-Carrollian variables satisfy the relations 
\begin{equation}
\label{eq:pre-carroll-rels}
      \delta^\mu_\nu = T_\nu^A V^\mu_A + \mathcal{E}^{A'}_\nu\mathcal{E}^\mu_{A'}\,,\qquad T_\mu^A V^\mu_B = \delta_B^A\,,\qquad \mathcal{E}^{A'}_\mu\mathcal{E}^\mu_{B'} = \delta^{A'}_{B'}\,,\qquad \mathcal{E}^{A'}_\mu V^\mu_A = \mathcal{E}^\mu_{A'}T_\mu^A = 0\,.
\end{equation}
Hence, the metric and its inverse may be expressed in terms of pre-Carrollian variables as
\begin{equation}
\label{eq:metric-and-inverse-pre-Carroll}
    g_{\mu\nu} = c^2 T_\mu^AT_\nu^B\eta_{AB} + \Pi^\perp_{\mu\nu}\,,\qquad g^{\mu\nu} = \frac{1}{c^2}V^\mu_A V^\nu_B\eta^{AB} + \Pi^{\perp\,\mu\nu}\,,
\end{equation}
where we defined
\begin{equation}
    \Pi^\perp_{\mu\nu} := \delta_{A'B'}\mathcal{E}^{A'}_\mu \mathcal{E}^{B'}_\nu\,,\qquad \Pi^{\perp\,\mu\nu} := \delta^{A'B'}\mathcal{E}^\mu_{A'}\mathcal{E}^\nu_{B'}\,.
\end{equation}
The pre-Carrollian variables themselves expand in powers of $c^2$ according to
\begin{equation}
\begin{split}
\label{eq:string-carroll-exp}
    V^\mu_A &= v^\mu_A + c^2 M^\mu_A + c^4 N^\mu_A + \mathcal{O}(c^6)\,,\\
    T^A_\mu &= \tau^A_\mu + c^2 m^A_\mu + \mathcal{O}(c^4)\,,\\
    \mathcal{E}^\mu_{A'} &= e^\mu_{A'} + c^2 \pi^\mu_{A'} +\mathcal{O}(c^4)\,,\\
    \mathcal{E}_\mu^{A'} &= e_\mu^{A'} + c^2\varpi_\mu^{A'} +\mathcal{O}(c^4)\,.
\end{split}
\end{equation}
After expanding the pre-Carrollian variables in powers of $c^2$ as above, the metric and its inverse~\eqref{eq:metric-and-inverse-pre-Carroll} expand according to
\begin{equation}
\label{eq:metric-exp-app}
    \begin{split}
        g_{\mu\nu} &= h_{\mu\nu} + c^2  \tau_{\mu\nu} + c^2 \Phi_{\mu\nu} + \mathcal{O}(c^4)\,,\\
        g^{\mu\nu} &= \frac{1}{c^2}v^{\mu\nu} + \bar h^{\mu\nu} + c^2\Psi^{\mu\nu} + \mathcal{O}(c^4)\,,
    \end{split}
\end{equation}
where we defined the combinations
\begin{equation}
\label{eq:combinations}
\begin{split}
    h_{\mu\nu} &= \delta_{A'B'}e_\mu^{A'} e_\nu^{B'}\,,\qquad
    \tau_{\mu\nu} = \eta_{AB}\tau_\mu^A\tau_\nu^B\,,\qquad
    \Phi_{\mu\nu} = 2\delta_{A'B'} e^{A'}_{(\mu} \varpi^{B'}_{\nu)}\,,\\
    v^{\mu\nu} &= \eta^{AB} v^\mu_A v^\nu_B\,,\qquad
    \bar h^{\mu\nu} = h^{\mu\nu} + 2\eta^{AB} v^{(\mu}_{A} M^{\nu)}_{B}\,,\\
    \Psi^{\mu\nu} &= \eta^{AB}M^\mu_A M_B^\nu + 2\eta^{AB}v^{(\mu}_A N^{\nu)}_B + 2\delta^{A'
    B'}e^{(\mu}_{A'}\pi^{\nu)}_{B'}\,.
\end{split}
\end{equation}
The LO fields in~\eqref{eq:string-carroll-exp} define a string Carroll geometry. The properties enjoyed by the pre-Carrollian variables~\eqref{eq:pre-carroll-rels} can be used to derive relations for the geometric fields that appear at each order in the string Carroll expansion~\eqref{eq:string-carroll-exp}, which reproduce the relations in Eqs.~\eqref{eq:OG-completeness} and~\eqref{eq:vielbein-rels}. To determine the transformation properties of the fields that appear in the string Carroll expansion, we consider the expansion of the local Lorentz transformations that act on the pre-Carrollian variables following the procedure of~\cite{Hartong:2022dsx}. As we show in the next subsection, the LO combinations $v^{\mu\nu}$ and $h_{\mu\nu}$ (cf.~\eqref{eq:combinations}), which form the string Carroll structure, are invariant. The other LO geometric fields $\tau_{\mu\nu}$ and $h^{\mu\nu}$ transform under string Carroll boosts.

\subsection{Symmetries of the string Carroll expansion}
In this subsection, we derive the transformations of the fields that arise in the string Carroll expansion described in Sec.~\ref{sec:string-carroll-exp} using the methods developed in~\cite{Hartong:2022dsx}. The Lorentzian vielbeine $\hat E^{\bar A}_\mu$ and $\hat E^\mu_{\bar A}$ transform under diffeomorphisms $\Xi$ and local Lorentz transformations $\hat L^{\bar A}{_{\bar B}}$ according to
\begin{equation}
    \delta \hat E_\mu^{\bar A} = \pounds_\Xi \hat E_\mu^{\bar A} + \hat L^{\bar A}{_{\bar B}} \hat E_\mu^{\bar B}\,,\qquad \delta \hat E^\mu_{\bar A} = \pounds_\Xi \hat E^\mu_{\bar A} + \hat L_{\bar A}{^{\bar B}} \hat E^\mu_{\bar B}\,.
\end{equation}
The local Lorentz transformations satisfy $\hat L_{\bar A}{^{\bar C}}\eta_{\bar C\bar D} \hat L^{\bar D}{_{\bar B}} = \eta_{\bar A\bar B}$ as well as $\eta^{\bar A\bar B}\hat L^{\bar C}{_{\bar B}} = -\eta^{\bar B\bar C}\hat L^{\bar A}{_{\bar B}}$, and we may decompose it into longitudinal and transverse components as~\cite{Hartong:2022dsx}
\begin{equation}
\label{eq:LLT-exp}
\begin{split}
    \hat{L}^{\bar A}{_{\bar B}} &= \delta^{\bar{A}}_A \delta^B_{\bar B} \hat L^A{_B} + \delta^{\bar A}_{{A}'} \delta^B_{\bar{B}} \hat L^{{A'}}{_B} + \delta^{{B'}}_{\bar B}\delta^{\bar{A}}_A \hat L^A{_{B'}} + \delta^{\bar A}_{A'}\delta^{B'}_B \hat L^{A'}{_{B'}}\\
    &= \delta^{\bar{A}}_A \delta^B_{\bar B} L \varepsilon^A{_B} + c\delta^{\bar A}_{{A}'} \delta^B_{\bar{B}} \lambda^{{A'}}{_B} + c\delta^{{B'}}_{\bar B}\delta^{\bar{A}}_A \lambda^A{_{B'}} + \delta^{\bar A}_{A'}\delta^{B'}_B \lambda^{A'}{_{B'}} \,,
\end{split}
\end{equation}
where we introduced factors of $c$ on the second line consistent with the requirement that string Carroll symmetries should emerge in the limit $c\rightarrow 0$. Note that we wrote $\hat L^A{_B} = L\varepsilon^A{_B}$ for the parameters of the local Lorentz transformation in the longitudinal sector, where $\varepsilon^A{_B} = \eta^{AC}\varepsilon_{CB}$ is the two-dimensional Levi-Civita symbol. We also note the relation
\begin{equation}
    \lambda^A{_{B'}} = \eta^{AB} \delta_{A'B'} \lambda_B{^{A'}}\,.
\end{equation}
This implies the following transformation properties for the pre-Carrollian variables 
\begin{equation}
\label{eq:pre-carroll-trafos}
    \begin{split}
        \delta T_\mu^A &= \pounds_\Xi T^A_\mu + L \varepsilon^{A}{_B} T^B_\mu + \lambda^A{_{B'}}\mathcal{E}^{B'}_\mu\,,\\
        \delta \mathcal{E}^{A'}_\mu &= \pounds_\Xi \mathcal{E}^{A'}_\mu + \lambda^{A'}{_{B'}}\mathcal{E}^{B'}_\mu + c^2 \lambda^{A'}{_B} T^B_\mu\,,\\
        \delta V^\mu_A &=\pounds_\Xi V^\mu_A + L \varepsilon_A{^B} V_B^\mu + c^2 \lambda_A{^{B'}}\mathcal{E}^\mu_{B'}\,,\\
        \delta \mathcal{E}_{A'}^\mu &=\pounds_\Xi \mathcal{E}_{A'}^\mu +  \lambda_{A'}{^B} V^\mu_B + \lambda_{A'}{^{B'}}\mathcal{E}^\mu_{B'}\,.
    \end{split}
\end{equation}
The string Carroll expansion of the diffeomorphisms and the local Lorentz transformations takes the form\footnote{Rather than rescaling the components of the local Lorentz transformation as in~\eqref{eq:LLT-exp}, we could instead have started the expansion of $\lambda^{A'}{_B}$ at $\mathcal{O}(c)$.}
\begin{equation}
\label{eq:exp-trafo-params}
    \begin{split}
        \Xi^\mu &= \xi^\mu + c^2\zeta^\mu + c^4\vartheta + \mathcal{O}(c^6)\,,\\
        L &= L_{(0)} + c^2 L_{(2)} + c^4 L_{(4)}\mathcal{O}(c^6)\,,\\
        \lambda^{A'}{_{B'}} &= \lambda^{A'}_{{(0)B'}} + c^2\lambda^{A'}_{{(2)B'}} + \mathcal{O}(c^4)\,,\\
        \lambda^{A'}{_{B}} &= \lambda^{A'}_{{(0)B}} + c^2\lambda^{A'}_{{(2)B}} + \mathcal{O}(c^4)\,.
    \end{split}
\end{equation}
Here, we have decomposed the diffeomorphism generated by $\Xi^\mu$ into LO diffeomorphisms $\xi^\mu$, NLO diffeomorphisms $\zeta^\mu$ and NNLO diffeomorphisms $\vartheta^\mu$. We will refer to $L_{(0)}$ as a longitudinal local Lorentz transformation, while $L_{(2)}$ and $L_{(4)}$ are subleading local Lorentz transformtions. Similarly, $\lambda^{A'}_{(0)B}$ represents a string Carroll boost, while $\lambda^{A'}_{(2)B}$ is a subleading string Carroll boost, while $\lambda^{A'}_{(0)B'}$ is a spatial $SO(d)$ rotation, and $\lambda^{A'}_{(2)B'}$ a subleading spatial rotation. Combining the expansions of the transformation parameters~\eqref{eq:exp-trafo-params} with the transformations in~\eqref{eq:pre-carroll-trafos} leads to the following transformation properties for the fields that arise in the string Carroll expansion~\eqref{eq:string-carroll-exp}
\begin{equation}
\label{eq:vielbein-trafos}
    \begin{split}
        \delta \tau_\mu^A &= \pounds_\xi \tau_\mu^A + L_{(0)} \varepsilon^A{_B}\tau_\mu^B + \lambda^A_{(0)B'}e^{B'}_\mu\,,\\
        \delta m^A_\mu &= \pounds_\xi m^A_\mu + \pounds_\zeta \tau^A_\mu + L_{(0)} \varepsilon^A{_B}m_\mu^B + L_{(2)}\varepsilon^A{_B}\tau_\mu^B + \lambda^A_{(2)B'} e^{B'}_\mu + \lambda^A_{(0)B'}\varpi_\mu^{B'}\,,\\
        \delta e_\mu^{A'} &= \pounds_\xi e^{A'}_\mu + \lambda^{A'}_{(0)B'}e_\mu^{B'}\,,\\
        \delta \varpi_\mu^{A'} &= \pounds_\xi \varpi^{A'}_\mu + \pounds_\zeta e^{A'}_\mu + \lambda^{A'}_{(0)B'}\varpi^{B'}_\mu + \lambda^{A'}_{(2)B'}e^{B'}_\mu + \lambda^{A'}_{(0)B}\tau_\mu^B\,,\\
        \delta v^\mu_A &= \pounds_\xi v^\mu_A + L_{(0)}\varepsilon_A{^B}v^\mu_B\,,\\
        \delta M^\mu_A &= \pounds_\xi M^\mu_A + \pounds_\zeta M^\mu_A + \lambda_{(0)A}{^{B'}}e^\mu_{B'} + L_{(0)}\varepsilon_A{^B}M^\mu_B + L_{(2)}\varepsilon_A{^B}v^\mu_B\,,\\
        \delta N^\mu_A &= \pounds_\xi N^\mu_A + \pounds_\zeta M^\mu_A + \pounds_\vartheta v^\mu_A + L_{(0)} \varepsilon_A{^B}N^\mu_B + L_{(2)}\varepsilon_A{^B}M^\mu_B + L_{(4)}\varepsilon_A{^B}v^\mu_B\\
        &\quad + \lambda_{(0)A}{^{B'}}\pi^\mu_{B'} + \lambda_{(2)A}{^{B'}} e^\mu_{B'}\,,\\
        \delta e^\mu_{A'} &= \pounds_\xi e^\mu_{A'} + \lambda_{(0)A'}{^B}v^\mu_B + \lambda_{(0)A'}{^{B'}}e^\mu_{B'}\,,\\
        \delta \pi^\mu_{A'} &= \pounds_\xi \pi^\mu_{A'} + \pounds_\zeta e^\mu_{A'} + \lambda_{(0)A'}{^B}M^\mu_B + \lambda_{(2)A'}{^B}v^\mu_B + \lambda_{(0)A'}{^{B'}}\pi^\mu_{B'} + \lambda_{(2)A'}{^{B'}}e^\mu_{B'}\,.
    \end{split}
\end{equation}
In particular, these transformations tell us that $v^{\mu\nu} = \eta^{AB}v^\mu_Av^\nu_B$ and $h_{\mu\nu}$ are invariant under local tangent space transformations. 

\section{Carroll expansion of point particles}
\label{app:particle-exps}
In this appendix, we review the particle Carrollian expansion and derive the Lagrangians for the magnetic and electric Carroll particles from the perspective of $c^2$ expansions. What we call the ``electric Carroll particle'' appeared in~\cite{Bergshoeff:2014jla}, while the magnetic Carroll particle was discussed in~\cite{deBoer:2021jej}.

\subsection{Particle \texorpdfstring{$c^2$}{c} expansion and Carrollian geometry}
The particle $c^2$ expansion closely mirrors the string $c^2$ expansion that we developed in Section~\ref{sec:string-carroll-exp}, with the only difference being that only the timelike direction is singled out; in other words, the longitudinal space is one-dimensional. Following~\cite{Hansen:2021fxi}, we consider a $(d+1)$-dimensional manifold with metric $g_{\mu\nu}$. In this case, the pre-Carrollian decompositions of the metric and its inverse~\eqref{eq:metric-and-inverse-pre-Carroll} read
\begin{equation}
    g_{\mu\nu} = -c^2T_\mu T_\nu + \Pi_{\mu\nu}\,,\qquad g^{\mu\nu} = -\frac{1}{c^2}V^\mu V^\nu + \Pi^{\mu\nu}\,,
\end{equation}
The transverse $\Pi$'s may be decomposed in terms of transverse vielbeine
\begin{equation}
    \Pi_{\mu\nu} = \delta_{A'B'}\mathcal{E}^{A'}_\mu \mathcal{E}^{B'}_\nu\,,\qquad \Pi^{\mu\nu} = \delta^{A'B'}\mathcal{E}^\mu_{A'}\mathcal{E}^\nu_{B'}\,,
\end{equation}
where in this appendix the transverse indices range over $A',B'=1,\dots,d$.  Exactly paralleling the pre-Carrollian stringy variables of~\eqref{eq:string-carroll-exp}, the quantities above expand in powers of $c^2$ according to
\begin{equation}
\begin{split}
\label{eq:string-carroll-exp-app}
    V^\mu &= v^\mu + c^2 M^\mu  + \mathcal{O}(c^4)\,,\\
    T_\mu &= \tau_\mu + \mathcal{O}(c^2)\,,\\
    \mathcal{E}^\mu_{A'} &= e^\mu_{A'}  +\mathcal{O}(c^2)\,,\\
    \mathcal{E}_\mu^{A'} &= e_\mu^{A'} + c^2\varpi_\mu^{A'} +\mathcal{O}(c^4)\,
\end{split}
\end{equation}
which leads to the particle analogue of~\eqref{eq:metric-exp}, i.e.,
\begin{equation}
\label{eq:metric-exp-particle}
    \begin{split}
        g_{\mu\nu} &= h_{\mu\nu} - c^2  \tau_{\mu}\tau_\nu + c^2 \Phi_{\mu\nu} + \mathcal{O}(c^4)\,,\\
        g^{\mu\nu} &= -\frac{1}{c^2}v^{\mu}v^\nu + \bar h^{\mu\nu} + \mathcal{O}(c^2)\,,
    \end{split}
\end{equation}
where now
\begin{equation}
    \Phi_{\mu\nu} = 2\delta_{A'B'}e^{A'}_{(\mu}\varpi^{B'}_{\nu)}\,,\qquad \bar h^{\mu\nu} = h^{\mu\nu} - 2v^{(\mu} M^{\nu)}\,.
\end{equation}
One may read off the transformation of the particle Carroll variables from those of their stringy counterparts~\eqref{eq:vielbein-trafos} by considering the longitudinal index $A$ to only range over the time direction, and as such we refrain from reproducing these expressions here. 

The phase space Lagrangian for a point particle of mass $m$ moving in a Lorentzian background is (cf.~the relativistic phase space action for the string~\eqref{eq:rel-phase-space-action})
\begin{equation}
\label{eq:particle-PS-Lagrangian}
    L = \dot X^\mu P_\mu - \frac{1}{2} e (c^2 g^{\mu\nu}(X)P_\mu P_\nu + m^2 c^4)\,.
\end{equation}
The embedding fields and the momenta now only depend on the time $\sigma^0=:\tau$ along the worldline, while the embedding fields and momenta expand as in~\eqref{eq:emb-field-exp} and~\eqref{eq:field-expansions}, respectively. Just as in~\eqref{eq:metric-with-embedding-fields}, the the expansion of $g_{\mu\nu}(X)$ and its inverse acquire additional terms due to the expansion of the embedding fields. For a particle Carroll geometry, we have
\begin{equation}
\label{eq:metric-exp-app-part}
    \begin{split}
        g_{\mu\nu}(X) &= h_{\mu\nu}(x) - c^2\tau_\mu \tau_\nu(x) + c^2\Phi_{\mu\nu}(x,y) + \mathcal{O}(c^4)\,,\\
        g^{\mu\nu}(X) &= -c^{-2}v^\mu (x) v^\nu (x) + \bar h^{\mu\nu}(x,y) + \mathcal{O}(c^2)\,,
    \end{split}
\end{equation}
where
\begin{equation}
    \begin{split}
        \Phi_{\mu\nu}(x,y) &= \Phi_{\mu\nu}(x) + y^\rho \D_\rho h_{\mu\nu}(x)\,,\\
        \bar h^{\mu\nu}(x,y) &= h^{\mu\nu}(x) - 2v^{(\mu}(x)M^{\nu)}(x) - 2 y^\rho v^{(\mu}(x) \D_\rho v^{\nu)}(x)\,,
    \end{split}
\end{equation}
which have exactly the same form as their stringy counterparts in~\eqref{eq:stringy-objects}.

\subsection{Magnetic Carroll particle}
Just as for the magnetic Carroll string that we discussed in Sec.~\ref{sec:magnetic-string}, the starting point for the expansion that leads to the magnetic Carroll particle involves defining the following combinations 
\begin{equation}\label{eq:magneticparticle}
    \tilde e := c^2 e\,,\qquad \tilde p := mc\,,
\end{equation}
which are finite in the limit $c\to 0$ (cf.~the stringy analogues in~\eqref{eq:magnetic-string-combos}). This means that the phase space Lagrangian for the particle~\eqref{eq:particle-PS-Lagrangian} takes the form
\begin{equation}
    \tilde L = \dot X^\mu P_\mu - \frac{1}{2} \tilde e \left( g^{\mu\nu}(X)P_\mu P_\nu + p^2 \right)\,.
\end{equation}
The magnetic Lagrange multiplier $\tilde e$ expands in the same way as for the string~\eqref{eq:alternative-expansion}, while the expansions of $X^\mu$ and $P_\mu$ are as in~\eqref{eq:emb-field-exp} and~\eqref{eq:field-expansions}. Combining those expansions with those of the metric and its inverse~\eqref{eq:metric-exp-app}, we find that the Lagrangian $\tilde L$ expands in powers of $c^2$ as
\begin{equation}
    \tilde L = c^{-2}\tilde L_{\text{LO}} + \tilde L_{\text{NLO}} + \mathcal{O}(c^2)\,,
\end{equation}
where
\begin{equation}
\begin{split}
\tilde L_{\text{LO}} &= \frac{1}{2}\tilde e_{(0)} (v^\mu P_{(0)\mu})^2\,,\\ 
\tilde L_{\text{NLO}} &= \dot x^\mu P_{(0)\mu} - \frac{1}{2}\tilde e_{(0)}\left( \bar h^{\mu\nu}(x,y)P_{(0)\mu}P_{(0)\nu} + p^2 \right) + \tilde e_{(0)} v^\mu v^\nu P_{(0)\mu} P_{(2)\nu} + \frac{1}{2}\tilde e_{(2)}(v^\mu P_{(0)\mu})^2\,.
\end{split}
\end{equation}
Just as for the string, the LO theory is trivial. The NLO theory can equivalently be written as
\begin{equation}
    \tilde L_{\text{NLO}} = \dot x^\mu P_{(0)\mu} - \frac{1}{2}\tilde e_{(0)}\left( h^{\mu\nu}P_{(0)\mu}P_{(0)\nu} + p^2 \right)+\lambda v^\mu P_{(0)\mu}\,,
\end{equation}
where $\lambda$ is a Lagrange multiplier. This the action obtained in \cite{deBoer:2021jej}.
The constraint imposed by $\tilde e_{(0)}$ is now 
\begin{equation}
    h^{\mu\nu}P_{(0)\mu}P_{(0)\nu} = - p^2\,,
\end{equation}
but since $h^{\mu\nu}$ is positive semi-definite on spatial slices, this constraint is only nontrivial if $p^2 < 0$; in other words, the NLO theory is only nontrivial if the relativistic theory is tachyonic. We found exactly the same thing for the string, where the NLO magnetic phase space theory describes embeddings of a Lorentzian worldsheet into a target space with a positive semi-definite metric, which are necessarily trivial. To get something nontrivial we must go to NNLO\footnote{Alternatively, we could consider strings for which $\tilde T^2<0$ so that $\gamma_{(0)\alpha\beta}$ is Euclidean, but we will not do so here.}, although we refrain from giving the details. 

\subsection{Electric Carroll particle}
To get the electric Carroll particle, we define the following quantities (cf.~\eqref{eq:electric-string-combos})
\begin{equation}\label{eq:electricparticle}
    E_0 := mc^2\,,\qquad \hat e := e\,,
\end{equation}
which are kept finite as $c \to 0$, and where $\hat e$ expands in the same way as for the string~\eqref{eq:hat-e-exp}. With these, the particle Lagrangian~\eqref{eq:particle-PS-Lagrangian} takes the form
\begin{equation}
   \hat L =  \dot X^\mu P_\mu - \frac{1}{2} \hat e (c^2 g^{\mu\nu}(X)P_\mu P_\nu + E_0^2)\,,
\end{equation}
which expands in powers of $c^2$ as
\begin{equation}
    \hat L = \hat L_{\text{LO}} + \mathcal{O}(c^2)\,,
\end{equation}
where
\begin{equation}
    \hat L_{\text{LO}} = \dot x^\mu P_{(0)\mu} + \frac{1}{2}\hat e_{(0)}\left( (v^\mu P_{(0)\mu})^2 - E_0^2\right)\,.
\end{equation}
This action was found in~\cite{Bergshoeff:2014jla}. In flat Carroll spacetime we may choose Cartesian coordinates $x^\mu = (t,x^i)$ such that
\begin{equation}
    v^\mu = -\delta^\mu_t\,,\qquad h_{\mu\nu} = \delta^i_\mu \delta^i_\nu\,.
\end{equation}
If we furthermore decompose the momentum according to $P_{(0)\mu} = (-E,p_i)$, we may express the LO electric Lagrangian as
\begin{equation}
\begin{split}
\hat L_{\text{LO}} &= -E\dot t + p_i \dot x^i + \frac{1}{2}\hat e_{(0)}\left( E + E_0\right) \left( E - E_0\right)\\
&=-E\dot t + p_i \dot x^i + \lambda \left( E - E_0\right)\,,
\end{split}
\end{equation}
where we assumed that $E,E_0>0$ and defined $\lambda := \frac{1}{2}\hat e_{(0)}\left( E + E_0\right) $ to rewrite the Lagrangian to the form that appears in Eq.~(3.9) of~\cite{deBoer:2021jej}. The equations of motion of this Lagrangian describes a Carroll particle at rest with constant energy $E_0$. 

\section{Phase space formulation of the magnetic Carroll string}
\label{app:phase-space-formulation-magnetic-string}
In this appendix, we provide the details of the phase space formulation of the magnetic Carroll string of Section~\ref{sec:magnetic-string}. The starting point for the magnetic Carroll expansion of the string is the Lagrangian $\tilde L$ that appears in~\eqref{eq:magnetic-start-Lagrangian}, where the $c^2$-expansions of $\tilde e, u$ and $P_\mu$ feature in~\eqref{eq:alternative-expansion} and~\eqref{eq:field-expansions}. In addition to the $c^2$-expansion of the metric $g_{\mu\nu}(X)$ that appears in~\eqref{eq:metric-with-embedding-fields}, we also require the $c^2$-expansion of the inverse metric $g^{\mu\nu}(X)$, which has the form
\begin{equation}
g^{\mu\nu}(X) = \frac{1}{c^2}v^{\mu\nu}(x) + \bar h^{\mu\nu}(x,y) + c^2\Psi^{\mu\nu}(x,y,z) + \mathcal{O}(c^4)\,,
\end{equation}
where
\begin{equation}
\label{eq:stringy-objects-app}
\begin{split}
    %\Phi_{\mu\nu}(x,y) &= \Phi_{\mu\nu}(x) + y^\rho\D_\rho h_{\mu\nu}(x)\,,\\
    \bar h^{\mu\nu}(x,y) &= h^{\mu\nu}(x) + 2\eta^{AB} v^{(\mu}_{A}(x) M^{\nu)}_{B}(x) + y^\rho\D_\rho v^{\mu\nu}(x)\,,\\
    \Psi^{\mu\nu}(x,y,z) &= \eta^{AB}M^\mu_A(x) M_B^\nu(x) + 2\eta^{AB}v^{(\mu}_A(x) N^{\nu)}_B(x) + 2\delta^{A
    B'}e^{(\mu}_{A'}(x)\pi^{\nu)}_{B'}(x)\\
    &\quad+ y^\rho\D_\rho\bar h_{\mu\nu}(x) + z^\rho\D_\rho v^{\mu\nu}(x) + \frac{y^\rho y^\sigma}{2}\D_\rho\D_\sigma v^{\mu\nu}(x)\,.
\end{split}
\end{equation}
Combining all these ingredients allows us to explicitly write down the Lagrangians that appear at each order in $c^2$ in the expansion of $\tilde L$, cf.~\eqref{eq:magnetic-L-exp}, starting with $\tilde L_{\text{LO}}$:
\subsection{LO magnetic phase space theory}
The LO Lagrangian in the expansion~\eqref{eq:magnetic-L-exp} reads
\begin{equation}
\label{eq:LO-lag}
    \tilde L_{\text{LO}} = -\frac{1}{2}\oint d\sigma^1 \tilde e_{(0)}v^{\mu\nu} P_{(0)\mu} P_{(0)\nu}\,.
\end{equation}
Varying $P_{(0)\mu}$ gives
\begin{equation}
\label{eq:LO-P-eom}
    \tilde e_{(0)}v^{\mu\nu}P_{(0)\nu} = 0\,,
\end{equation}
which, assuming that $\tilde e_{(0)}\neq 0$, implies that $P_{(0)A} := v^\mu_A P_{(0)\mu} = 0$, and hence integrating out $P_{(0)A}$ gives zero for $\tilde L_{\text{LO}}$. The equation of motion for $\tilde e_{(0)}$ is $v^{\mu\nu} P_{(0)\mu} P_{(0)\nu} = 0$, and is identically satisfied when~\eqref{eq:LO-P-eom} holds. Hence, the LO theory in the expansion of phase space Lagrangian is trivial.

\subsection{NLO magnetic phase space theory}
The NLO Lagrangian in~\eqref{eq:magnetic-L-exp} is
\begin{equation}
\label{eq:NLO-mag-Lag}
\begin{split}
    \tilde L_{\text{NLO}} &= \oint d\sigma^1\bigg[ \dot x^\mu P_{(0)\mu} - \frac{1}{2}\tilde e_{(0)} \left( \tilde T^2 h_{\mu\nu} x'^\mu x'^\nu + \bar h^{\mu\nu}(x,y)P_{(0)\mu}P_{(0)\nu} + 2v^{\mu\nu}P_{(0)\mu}P_{(2)\nu} \right)\\
    &\quad - \frac{1}{2}\tilde e_{(2)} v^{\mu\nu} P_{(0)\mu} P_{(0)\nu} - u_{(0)} x'^\mu P_{(0)\mu} \bigg]\\
    &=  \oint d\sigma^1\bigg[ \dot x^\mu P_{(0)\mu} - \frac{1}{2}\tilde e_{(0)} \left( \tilde T^2 h_{\mu\nu} x'^\mu x'^\nu + \bar h^{\mu\nu}(x)P_{(0)\mu}P_{(0)\nu} \right)  - u_{(0)} x'^\mu P_{(0)\mu} \bigg]\\
    &\quad + \tilde e_{(2)}\frac{\delta \tilde L_{\text{LO}}}{\delta\tilde e_{(0)}} + P_{(2)\mu}\frac{\delta \tilde L_{\text{LO}}}{\delta P_{(0)\mu}} + y^\mu \frac{\delta \tilde L_{\text{LO}}}{\delta x^\mu}\,,
\end{split}
\end{equation}
where we made explicit the fact that $\bar h^{\mu\nu}(x,y)$ includes the subleading embedding field $y^\mu$. In the second equality, we highlighted that subleading fields source lower-order equations of motion (see~\cite{Hartong:2022dsx} for details). This action has \textit{trivial dynamics}: assuming as above that $\tilde e_{(0)}\neq 0$, the equation of motion for $P_{(2)A}$ says that $P_{(0)A} = 0$, which was the equation of motion for $P_{(0)\mu}$ at LO. Then the constraint imposed by $\tilde e_{(0)}$ implies that $h_{\mu\nu} x'^\nu =0$ and $h^{\mu\nu}P_{(0)\nu} = 0$, implying that there is no dynamics. This statement is the phase space analogue of the result obtained in~\cite{Bagchi:2023cfp} in the Polyakov formulation, where the triviality of the dynamics is a consequence of the fact that a Lorentzian worldsheet cannot be embedded in a space with a positive semi-definite metric. We will now demonstrate that the magnetic NLO phase space Lagrangian~\eqref{eq:NLO-mag-Lag} is equivalent to the LO Polyakov Lagrangian obtained in~\cite{Bagchi:2023cfp} and presented in Eq.~\eqref{eq:LO-Polyakov-Lagrangian} of the main text, which we obtain by integrating out the momenta $P_{(0)\mu}$ and $P_{(2)\mu}$.

Integrating out $P_{(2)\mu}$ in the NLO Lagrangian gives~\eqref{eq:LO-P-eom}, implying that $P_{(0)A} = 0$, which in turn implies that the LO equation of motion for $x^\mu$ vanishes identically, and so we may ignore the terms involving $y^\mu$. Imposing these in the Lagrangian~\eqref{eq:NLO-mag-Lag} leads to
\begin{equation}
\begin{split}
    \tilde L_{\text{NLO}} &= \oint d\sigma^1\bigg[ \dot x^\nu P_{(0)\mu}  h^\mu_\nu  - \frac{1}{2}\tilde e_{(0)} \left( \tilde T^2 h_{\mu\nu} x'^\mu x'^\nu + h^{\mu\nu}P_{(0)\mu}P_{(0)\nu}\right)\\
    &\quad - u_{(0)} x'^\nu P_{(0)\mu}  h^\mu_\nu\bigg] \,.
\end{split}
\end{equation}
Integrating out $P_{(0)\mu}$ gives an equation for $P_{(0)\mu}$
\begin{equation}
\label{eq:first-integrating-out}
    \begin{split}
      h^{\mu\nu}  P_{(0)\nu} &= \frac{1}{\tilde e_{(0)}}h^\mu_\nu\left( \dot x^\nu - u_{(0)}x'^\nu \right)\,,
    \end{split}
\end{equation}
which leads to the Polyakov action 
\begin{equation}
\label{eq:Carroll-Polyakov-1}
    \tilde S_{\text{NLO}} = \frac{1}{2}\int d\sigma^0\oint d\sigma^1 \frac{1}{\tilde e_{(0)}}\left[ h_{\mu\nu} \dot x^\mu\dot x^\nu + (u_{(0)}^2 - \tilde e^2_{(0)}\tilde T^2 )h_{\mu\nu} x'^\mu x'^\nu - 2u_{(0)}h_{\mu\nu}\dot x^\mu x'^\nu \right]\,.
\end{equation}
Note in particular that although the phase space action involves the subleading field $M^\mu_A$, this does not feature in the Polyakov action~\eqref{eq:Carroll-Polyakov-1}. To bring this to the same form as the LO Polyakov action~\eqref{eq:LO-Polyakov-Lagrangian}, we define
\begin{equation}
\label{eq:tilde-gamma}
    \tilde \gamma^{\alpha\beta}_{(0)} = \frac{1}{\tilde T  \tilde e_{(0)}}\begin{pmatrix} -1 ~&~ u_{(0)} \\ u_{(0)} ~&~ \tilde T^2 \tilde e_{(0)}^2 - u_{(0)}^2\end{pmatrix}\,,
\end{equation}
satisfying $\det (\tilde\gamma_{(0)}^{\alpha\beta}) = -1$, allowing us to write
\begin{equation}
\label{eq:tilde-gamma-det}
    \tilde\gamma^{\alpha\beta}_{(0)} = \sqrt{-\det (\gamma_{(0)\delta\gamma})} \gamma^{\alpha\beta}_{(0)}\,,
\end{equation}
where $\gamma_{(0)\alpha\beta}$ is a fiducial \textit{Lorentzian} worldsheet metric. With this, we recover the LO Polyakov Lagrangian~\eqref{eq:LO-Polyakov-Lagrangian}.

\subsection{NNLO magnetic phase space theory}
The NNLO phase space Lagrangian appearing in~\eqref{eq:magnetic-L-exp} has the form
\begin{equation}
\hspace{-1cm}
\label{eq:NNLO-mag-lag}
    \begin{split}
        &\tilde L_{\text{NNLO}} = \oint d\sigma^1 \bigg[ \dot y^\mu P_{(0)\mu} + \dot x^\mu P_{(2)\mu} - \frac{1}{2}\tilde e_{(4)} v^{\mu\nu}P_{(0)\mu}P_{(0)\nu} - \frac{1}{2}\tilde e_{(0)} \big\{ \Psi^{\mu\nu}(x,y,z)P_{(0)\mu}P_{(0)\nu}\\
        & + 2\bar h^{\mu\nu}(x,y)P_{(0)\mu}P_{(2)\nu} + v^{\mu\nu}P_{(2)\mu}P_{(2)\nu} + 2v^{\mu\nu}P_{(4)\mu}P_{(0)\nu} + \tilde T^2 (\tau_{\mu\nu} + \Phi_{\mu\nu}(x,y))x'^\mu x'^\nu\\
        &+ 2\tilde T^2 h_{\mu\nu}x'^\mu y'^\nu \big\} - \frac{1}{2}\tilde e_{(2)} \big( \tilde T^2 h_{\mu\nu} x'^\mu x'^\nu + \bar h^{\mu\nu}(x,y)P_{(0)\mu}P_{(0)\nu} + 2v^{\mu\nu}P_{(0)\mu}P_{(2)\nu} \big)\\
        &- u_{(2)}x'^\mu P_{(0)\mu} - u_{(0)}\left(y'^\mu P_{(0)\mu} + x'^\mu P_{(2)\mu}\right) \bigg]\,.
    \end{split}
\end{equation}
As we saw above, the role of the subleading fields is to remember equations of motion that appear at lower order. Therefore, we may equivalently express the NNLO Lagrangian as
\begin{equation}
\begin{split}
    \tilde L_{\text{NNLO}} &= -\frac{1}{2}\oint d\sigma^1 \tilde{e}_{(0)} \bigg[\Psi^{\mu\nu}(x)P_{(0)\mu}P_{(0)\nu} +\tilde T^2 (\tau_{\mu\nu} + \Phi_{\mu\nu}(x))x'^\mu x'^\nu  \bigg]\\
    &\quad + P_{(4)\mu}\frac{\delta\tilde L_{\text{LO}}}{\delta P_{(0)\mu}} + P_{(2)\mu}\frac{\delta\tilde L_{\text{NLO}}}{\delta P_{(0)\mu}} + \frac{1}{2}P_{(2)\mu}P_{(2)\nu}\frac{\delta\tilde L_{\text{LO}}}{\delta P_{(0)\mu}\delta P_{(0)\nu}}\\
    &\quad + \tilde e_{(4)} \frac{\delta \tilde L_{\text{LO}}}{\delta \tilde e_{(0)}} + \tilde e_{(2)} \frac{\delta \tilde L_{\text{NLO}}}{\delta \tilde e_{(0)}} + u_{(2)} \frac{\delta \tilde L_{\text{NLO}}}{\delta u_{(0)}} + z^\mu \frac{\delta \tilde L_{\text{LO}}}{\delta x^\mu} + y^\mu \frac{\delta \tilde L_{\text{NLO}}}{\delta x^\mu} + \frac{1}{2}y^\mu y^\nu\frac{\delta\tilde L_{\text{LO}}}{\delta x^\mu \delta x^\nu}\,.
\end{split}
\end{equation}
Integrating out $P_{(4)\mu}$, which now acts as a Lagrange multiplier for the LO equation of motion for $P_{(0)\mu}$, gives rise to~\eqref{eq:LO-P-eom} and thus implies that $v^{\mu\nu}P_{(0)\nu} = 0$. Imposing this in the NNLO Lagrangian~\eqref{eq:NNLO-mag-lag} causes all terms involving $\tilde e_{(4)}$ to drop out of the Lagrangian, which then reduces to
\begin{equation}
\hspace{-1cm}
    \begin{split}
        &\tilde L_{\text{NNLO}} = \oint d\sigma^1 \bigg[ \dot y^\mu P^\perp_{(0)\mu} + \dot x^\mu P_{(2)\mu}  - \frac{1}{2}\tilde e_{(0)} \big\{ \Psi^{\mu\nu}(x,y,z) P^\perp_{(0)\mu}P^\perp_{(0)\nu}\\
        & + 2\bar h^{\mu\nu}(x,y)P^\perp_{(0)\mu}P_{(2)\nu} + v^{\mu\nu}P_{(2)\mu}P_{(2)\nu} + \tilde T^2 (\tau_{\mu\nu} + \Phi_{\mu\nu}(x,y))x'^\mu x'^\nu + 2\tilde T^2 h_{\mu\nu}x'^\mu y'^\nu \big\}\\
        & - \frac{1}{2}\tilde e_{(2)} \big( \tilde T^2 h_{\mu\nu} x'^\mu x'^\nu +  h^{\mu\nu}P^\perp_{(0)\mu}P^\perp_{(0)\nu} \big) - u_{(2)}x'^\mu P^\perp_{(0)\mu} - u_{(0)}\left(y'^\mu P^\perp_{(0)\mu} + x'^\mu P_{(2)\mu}\right) \bigg]\,.
    \end{split}
\end{equation}
where we split the momentum into longitudinal ($\parallel$) and transverse ($\perp$) components according to
\begin{equation}
    P_{(0)\mu} = v^\nu_\mu P_{(0)\nu} + h^\nu_\mu P_{(0)\nu} =: P^\parallel_{(0)\mu} + P^\perp_{(0)\mu} = P^\perp_{(0)\mu}\,,
\end{equation}
where we used the equation of motion for $P_{(4)\mu}$ in the last equality. Next, we simultaneously integrate out $P_{(2)\mu}$ and $P_{(0)\mu}$: integrating out $P_{(2)\mu}$ gives equations for $h_\mu^\nu P_{(0)\nu}$ and $P_{(2)A} := v^\mu_A P_{(2)\mu}$. The spatial projection of the equation for $P_{(2)\mu}$ reproduces~\eqref{eq:first-integrating-out}, while the temporal part gives an equation for the longitudinal components $P^A_{(2)} = \eta^{AB}P_{(2)B}$
\begin{equation}
    P^A_{(2)} = \frac{1}{\tilde e_{(0)}}(D_0 x^A - u_{(0)}D_1 x^A) \,,
\end{equation}
where we used~\eqref{eq:first-integrating-out} and where we defined
\begin{equation}
    D_\alpha x^A = \tau_\mu^A \D_\alpha x^\mu -\eta^{AB} M^\nu_B(x,y) h_{\mu\nu}\D_\alpha x^\mu\,.
\end{equation}
A tedious calculation shows that plugging these back into the phase space Lagrangian leads to the result
\begin{equation}
\begin{split}
    \tilde L_{\text{NNLO}} &= \oint d\sigma^1\,\frac{1}{2\tilde e_{(0)}}\bigg[ \big(\dot x^\mu \dot x^\nu - 2u_{(0)}\dot x^\mu x'^\nu + (u_{(0)}^2 - \tilde e_{(0)}^2\tilde T^2) x'^\mu x'^\nu \big)(\tau_{\mu\nu} + \Phi_{\mu\nu}(x,y))\\
    &\quad + 2h_{\mu\nu}\big(\dot x^\mu \dot y^\nu - 2u_{(0)}\dot x^\mu y'^\nu + (u_{(0)}^2 - \tilde e_{(0)}^2\tilde T^2) x'^\mu y'^\nu\big)\\
    &\quad + h_{\mu\nu}\frac{\tilde e_{(2)}}{\tilde e_{(0)}} \big(-\dot x^\mu \dot x^\nu + 2u_{(0)}\dot x^\mu x'^nu + (-u_{(0)}^2 - \tilde e_{(0)}^2\tilde T^2) x'^\mu x'^\nu\big)\\
    &\quad + 2 h_{\mu\nu}u_{(2)}\big( -\dot x^\mu \dot x^\nu + u_{(0)} x'^\mu x'^\nu \big)\bigg]\,.
\end{split}
\end{equation}
Identifying $\gamma_{(0)}$ as in~\eqref{eq:tilde-gamma} and~\eqref{eq:tilde-gamma-det} above and defining
\begin{equation}
\label{eq:gamma-2}
\begin{split}
\gamma^{\alpha\beta}_{(2)} &= \begin{pmatrix}
       0 ~ &~ -u_{(2)}\\
       -u_{(2)}~&~2u_{(0)}u_{(2)} - 2\tilde T^2 \tilde e_{(0)}\tilde e_{(2)}
   \end{pmatrix}\,,\\
   \gamma_{(2)\alpha\beta} &= \begin{pmatrix}
       \frac{2u_{(0)} u_{(2)} }{\tilde T^2 \tilde e_{(0)}^2} ~ & ~ \frac{u_{(2)}}{\tilde T^2 \tilde e_{(0)}^2} - \frac{2u_{(0)} \tilde e_{(2)}}{\tilde T^2 \tilde e_{(0)}^3}\\
       \frac{u_{(2)}}{\tilde T^2 \tilde e_{(0)}^2} - \frac{2u_{(0)} \tilde e_{(2)}}{\tilde T^2 \tilde e_{(0)}^3} ~ & ~ - \frac{2 \tilde e_{(2)}}{\tilde T^2 \tilde e_{(0)}^3}
   \end{pmatrix}\,,     
\end{split}
\end{equation}
we obtain the NLO Polyakov Lagrangian derived in~\cite{Bagchi:2023cfp} that appears in~\eqref{eq:P-NLO-lagrangian} of the main text.

\section{Near-horizon geometry of a non-extremal 4D Kerr black hole}\label{Kerr}

In Boyer--Lindquist coordinates the 4D Kerr metric is given by (with $c=1$)
\begin{equation}
    ds^2=-\frac{\Delta}{\Sigma}\left( dt-a\sin^2\Theta d\phi\right)^2+\frac{\sin^2\Theta}{\Sigma}\left(\left(R^2+a^2\right)d\phi-adt\right)^2+\frac{\Sigma}{\Delta}dR^2+\Sigma d\Theta^2\,,
\end{equation}
where
\begin{equation}
\begin{split}
    \Sigma & =  R^2+a^2\cos^2\theta\,,\\
    \Delta & =  R^2+a^2-r_s R\,,
\end{split}
\end{equation}
with $r_s=2MG$ and $a=J/M$. The mass and angular momentum are $M$ and $J$. In here $\Theta$ and $\phi$ are oblate spherical coordinates. The outer and inner horizons are the real zeros of $\Delta$ which we can factorise as 
\begin{equation}
    \Delta=(R-R_-)(R-R_+)\,,\qquad R_\pm=\frac{1}{2}r_s\pm\frac{1}{2}\sqrt{r_s^2-4a^2}\,.
\end{equation}
Reality of the two zeros leads to the Kerr bound $r_s^2\ge 4a^2$. We will assume that $r_s$ is strictly larger than $2a$. 

We next expand this metric near the outer horizon by writing
\begin{equation}
    R=R_++\lambda\varepsilon\rho^2\,,
\end{equation}
and expanding the metric components to leading order in $\rho$. In here $\lambda$ is a constant which we will fix to be
\begin{equation}
    \lambda=\frac{R_+-R_-}{4\left(R_+^2+a^2\right)}\,.
\end{equation}
The result of this expansion is most conveniently expressed in terms a new angular coordinate $\tilde\phi$ that is defined by
\begin{equation}
    \tilde\phi=\phi-\frac{a}{R_+^2+a^2}t\,.
\end{equation}
This new angular variable inherits the periodicity of $\phi$. After some rewriting we can show that near $R_+$ we can express the metric as
\begin{eqnarray}
    ds^2 & = & \frac{R_+^2+a^2\cos^2\Theta}{R_+^2+a^2}\varepsilon\left[d\rho^2-\frac{\left(R_+-R_-\right)^2}{4\left(R_+^2+a^2\right)^2}\rho^2\left(dt-a\frac{R_+^2+a^2}{R_+^2+a^2\cos^2\Theta}\sin^2\Theta d\tilde\phi\right)^2\right]\nonumber\\
    &&+\left(R_+^2+a^2\cos^2\Theta\right)\left[d\Theta^2+\frac{\left(R_+^2+a^2\right)^2}{\left(R_+^2+a^2\cos^2\Theta\right)^2}\sin^2\Theta d\tilde\phi^2\right]+\cdots\,,
\end{eqnarray}
where the dots denote higher order terms in $\rho$. If we fix a point on the transverse space, i.e. we set $\Theta$ and $\tilde\phi$ equal to any constant then we obtain a 2D Rindler spacetime. We thus see that the metric is a 2D Rindler space fibred over the 2D base manifold whose coordinates are $\Theta$ and $\tilde\phi$. We can write this in terms of string Carroll geometry by defining
\begin{eqnarray}
    \tau_{\mu\nu}dx^\mu dx^\nu & = & \frac{1}{F}\left[d\rho^2-\frac{\left(R_+-R_-\right)^2}{4\left(R_+^2+a^2\right)^2}\rho^2\left(dt-aF\sin^2\Theta d\tilde\phi\right)^2\right]\,,\\
    h_{\mu\nu}dx^\mu dx^\nu & = & \frac{R_+^2+a^2}{F}\left[d\Theta^2+F^2\sin^2\Theta d\tilde\phi^2\right]\,,
\end{eqnarray}
where we defined
\begin{equation}
    F(\Theta)=\frac{R_+^2+a^2}{R_+^2+a^2\cos^2\Theta}\,,
\end{equation}
and where $\tau_{\mu\nu}$ has signature $(-1,1,0,0)$ and $h_{\mu\nu}$ has signature $(0,0,1,1)$.

\newpage

\addcontentsline{toc}{section}{\refname}

\providecommand{\href}[2]{#2}\begingroup\raggedright\endgroup


\begin{thebibliography}{10}

\bibitem{deVega:1987veo}
H.~J. de~Vega and N.~G. Sanchez, ``{A New Approach to String Quantization in
  Curved Space-Times},''
  \href{http://dx.doi.org/10.1016/0370-2693(87)90392-3}{{\em Phys. Lett. B}
  {\bfseries 197} (1987) 320--326}.

\bibitem{Dijkgraaf:1991ba}
R.~Dijkgraaf, H.~L. Verlinde, and E.~P. Verlinde, ``{String propagation in a
  black hole geometry},''
  \href{http://dx.doi.org/10.1016/0550-3213(92)90237-6}{{\em Nucl. Phys. B}
  {\bfseries 371} (1992) 269--314}.

\bibitem{Susskind:1993if}
L.~Susskind, L.~Thorlacius, and J.~Uglum, ``{The Stretched horizon and black
  hole complementarity},''
  \href{http://dx.doi.org/10.1103/PhysRevD.48.3743}{{\em Phys. Rev. D}
  {\bfseries 48} (1993) 3743--3761},
  \href{http://arxiv.org/abs/hep-th/9306069}{{\ttfamily arXiv:hep-th/9306069}}.

\bibitem{Susskind:1994sm}
L.~Susskind and J.~Uglum, ``{Black hole entropy in canonical quantum gravity
  and superstring theory},''
  \href{http://dx.doi.org/10.1103/PhysRevD.50.2700}{{\em Phys. Rev. D}
  {\bfseries 50} (1994) 2700--2711},
  \href{http://arxiv.org/abs/hep-th/9401070}{{\ttfamily arXiv:hep-th/9401070}}.

\bibitem{Susskind:1994uu}
L.~Susskind and J.~Uglum, ``{Black holes, interactions, and strings},'' in {\em
  {Particles, Strings, and Cosmology (PASCOS 94)}}, pp.~0254--270.
\newblock 10, 1994.
\newblock \href{http://arxiv.org/abs/hep-th/9410074}{{\ttfamily
  arXiv:hep-th/9410074}}.

\bibitem{Kunduri:2007vf}
H.~K. Kunduri, J.~Lucietti, and H.~S. Reall, ``{Near-horizon symmetries of
  extremal black holes},''
  \href{http://dx.doi.org/10.1088/0264-9381/24/16/012}{{\em Class. Quant.
  Grav.} {\bfseries 24} (2007) 4169--4190},
  \href{http://arxiv.org/abs/0705.4214}{{\ttfamily arXiv:0705.4214 [hep-th]}}.

\bibitem{hartong:2015xda}
J.~Hartong, ``{Gauging the Carroll Algebra and Ultra-Relativistic Gravity},''
  \href{http://dx.doi.org/10.1007/JHEP08(2015)069}{{\em JHEP} {\bfseries 08}
  (2015) 069},
\href{http://arxiv.org/abs/1505.05011}{{\ttfamily arXiv:1505.05011 [hep-th]}}.
%%CITATION = ARXIV:1505.05011;%%.

\bibitem{Donnay:2019jiz}
L.~Donnay and C.~Marteau, ``{Carrollian Physics at the Black Hole Horizon},''
  \href{http://dx.doi.org/10.1088/1361-6382/ab2fd5}{{\em Class. Quant. Grav.}
  {\bfseries 36} no.~16, (2019) 165002},
  \href{http://arxiv.org/abs/1903.09654}{{\ttfamily arXiv:1903.09654
  [hep-th]}}.

\bibitem{Bagchi:2023cfp}
A.~Bagchi, A.~Banerjee, J.~Hartong, E.~Have, K.~S. Kolekar, and M.~Mandlik,
  ``{Strings near black holes are Carrollian},''
  \href{http://arxiv.org/abs/2312.14240}{{\ttfamily arXiv:2312.14240
  [hep-th]}}.

\bibitem{Bergshoeff:2022eog}
E.~Bergshoeff, J.~Figueroa-O'Farrill, and J.~Gomis, ``{A non-lorentzian
  primer},'' \href{http://dx.doi.org/10.21468/SciPostPhysLectNotes.69}{{\em
  SciPost Phys. Lect. Notes} {\bfseries 69} (2023) 1},
  \href{http://arxiv.org/abs/2206.12177}{{\ttfamily arXiv:2206.12177
  [hep-th]}}.

\bibitem{Oling:2022fft}
G.~Oling and Z.~Yan, ``{Aspects of Nonrelativistic Strings},''
  \href{http://dx.doi.org/10.3389/fphy.2022.832271}{{\em Front. in Phys.}
  {\bfseries 10} (2022) 832271},
  \href{http://arxiv.org/abs/2202.12698}{{\ttfamily arXiv:2202.12698
  [hep-th]}}.

\bibitem{Hartong:2022lsy}
J.~Hartong, N.~A. Obers, and G.~Oling, ``{Review on Non-Relativistic
  Gravity},'' \href{http://dx.doi.org/10.3389/fphy.2023.1116888}{{\em Front. in
  Phys.} {\bfseries 11} (2023) 1116888},
  \href{http://arxiv.org/abs/2212.11309}{{\ttfamily arXiv:2212.11309 [gr-qc]}}.

\bibitem{Baiguera:2023fus}
S.~Baiguera, ``{Aspects of non-relativistic quantum field theories},''
  \href{http://dx.doi.org/10.1140/epjc/s10052-024-12630-y}{{\em Eur. Phys. J.
  C} {\bfseries 84} no.~3, (2024) 268},
  \href{http://arxiv.org/abs/2311.00027}{{\ttfamily arXiv:2311.00027
  [hep-th]}}.

\bibitem{Bagchi:2010zz}
A.~Bagchi, ``{Correspondence between Asymptotically Flat Spacetimes and
  Nonrelativistic Conformal Field Theories},''
\href{http://dx.doi.org/10.1103/PhysRevLett.105.171601}{{\em Phys.Rev.Lett.}
  {\bfseries 105} (2010) 171601}.
%%CITATION = PRLTA,105,171601;%%.

\bibitem{Barnich:2012aw}
G.~Barnich, A.~Gomberoff, and H.~A. Gonzalez, ``{The Flat limit of three
  dimensional asymptotically anti-de Sitter spacetimes},''
  \href{http://dx.doi.org/10.1103/PhysRevD.86.024020}{{\em Phys. Rev. D}
  {\bfseries 86} (2012) 024020},
  \href{http://arxiv.org/abs/1204.3288}{{\ttfamily arXiv:1204.3288 [gr-qc]}}.

\bibitem{Bagchi:2012xr}
A.~Bagchi, S.~Detournay, R.~Fareghbal, and J.~Simon, ``{Holography of 3D Flat
  Cosmological Horizons},''
  \href{http://dx.doi.org/10.1103/PhysRevLett.110.141302}{{\em Phys.Rev.Lett.}
  {\bfseries 110} no.~14, (2013) 141302},
\href{http://arxiv.org/abs/1208.4372}{{\ttfamily arXiv:1208.4372 [hep-th]}}.
%%CITATION = ARXIV:1208.4372;%%.

\bibitem{Barnich:2012xq}
G.~Barnich, ``{Entropy of three-dimensional asymptotically flat cosmological
  solutions},'' \href{http://dx.doi.org/10.1007/JHEP10(2012)095}{{\em JHEP}
  {\bfseries 10} (2012) 095}, \href{http://arxiv.org/abs/1208.4371}{{\ttfamily
  arXiv:1208.4371 [hep-th]}}.

\bibitem{Bagchi:2012cy}
A.~Bagchi and R.~Fareghbal, ``{BMS/GCA Redux: Towards Flatspace Holography from
  Non-Relativistic Symmetries},''
  \href{http://dx.doi.org/10.1007/JHEP10(2012)092}{{\em JHEP} {\bfseries 1210}
  (2012) 092},
\href{http://arxiv.org/abs/1203.5795}{{\ttfamily arXiv:1203.5795 [hep-th]}}.
%%CITATION = ARXIV:1203.5795;%%.

\bibitem{Bagchi:2014iea}
A.~Bagchi, R.~Basu, D.~Grumiller, and M.~Riegler, ``{Entanglement entropy in
  Galilean conformal field theories and flat holography},''
  \href{http://dx.doi.org/10.1103/PhysRevLett.114.111602}{{\em Phys. Rev.
  Lett.} {\bfseries 114} no.~11, (2015) 111602},
  \href{http://arxiv.org/abs/1410.4089}{{\ttfamily arXiv:1410.4089 [hep-th]}}.

\bibitem{Hartong:2015usd}
J.~Hartong, ``{Holographic Reconstruction of 3D Flat Space-Time},''
  \href{http://dx.doi.org/10.1007/JHEP10(2016)104}{{\em JHEP} {\bfseries 10}
  (2016) 104}, \href{http://arxiv.org/abs/1511.01387}{{\ttfamily
  arXiv:1511.01387 [hep-th]}}.

\bibitem{Bagchi:2016bcd}
A.~Bagchi, R.~Basu, A.~Kakkar, and A.~Mehra, ``{Flat Holography: Aspects of the
  dual field theory},'' \href{http://dx.doi.org/10.1007/JHEP12(2016)147}{{\em
  JHEP} {\bfseries 12} (2016) 147},
  \href{http://arxiv.org/abs/1609.06203}{{\ttfamily arXiv:1609.06203
  [hep-th]}}.

\bibitem{Donnay:2022aba}
L.~Donnay, A.~Fiorucci, Y.~Herfray, and R.~Ruzziconi, ``{Carrollian Perspective
  on Celestial Holography},''
  \href{http://dx.doi.org/10.1103/PhysRevLett.129.071602}{{\em Phys. Rev.
  Lett.} {\bfseries 129} no.~7, (2022) 071602},
  \href{http://arxiv.org/abs/2202.04702}{{\ttfamily arXiv:2202.04702
  [hep-th]}}.

\bibitem{Bagchi:2022emh}
A.~Bagchi, S.~Banerjee, R.~Basu, and S.~Dutta, ``{Scattering Amplitudes:
  Celestial and Carrollian},''
  \href{http://dx.doi.org/10.1103/PhysRevLett.128.241601}{{\em Phys. Rev.
  Lett.} {\bfseries 128} no.~24, (2022) 241601},
  \href{http://arxiv.org/abs/2202.08438}{{\ttfamily arXiv:2202.08438
  [hep-th]}}.

\bibitem{Figueroa-Ofarrill:2021sxz}
J.~Figueroa-O'Farrill, E.~Have, S.~Prohazka, and J.~Salzer, ``{Carrollian and
  celestial spaces at infinity},''
  \href{http://dx.doi.org/10.1007/JHEP09(2022)007}{{\em JHEP} {\bfseries 09}
  (2022) 007}, \href{http://arxiv.org/abs/2112.03319}{{\ttfamily
  arXiv:2112.03319 [hep-th]}}.

\bibitem{Donnay:2022wvx}
L.~Donnay, A.~Fiorucci, Y.~Herfray, and R.~Ruzziconi, ``{Bridging Carrollian
  and celestial holography},''
  \href{http://dx.doi.org/10.1103/PhysRevD.107.126027}{{\em Phys. Rev. D}
  {\bfseries 107} no.~12, (2023) 126027},
  \href{http://arxiv.org/abs/2212.12553}{{\ttfamily arXiv:2212.12553
  [hep-th]}}.

\bibitem{Bagchi:2023fbj}
A.~Bagchi, P.~Dhivakar, and S.~Dutta, ``{AdS Witten diagrams to Carrollian
  correlators},'' \href{http://dx.doi.org/10.1007/JHEP04(2023)135}{{\em JHEP}
  {\bfseries 04} (2023) 135}, \href{http://arxiv.org/abs/2303.07388}{{\ttfamily
  arXiv:2303.07388 [hep-th]}}.

\bibitem{Saha:2023hsl}
A.~Saha, ``{Carrollian approach to 1 + 3D flat holography},''
  \href{http://dx.doi.org/10.1007/JHEP06(2023)051}{{\em JHEP} {\bfseries 06}
  (2023) 051}, \href{http://arxiv.org/abs/2304.02696}{{\ttfamily
  arXiv:2304.02696 [hep-th]}}.

\bibitem{Bagchi:2023cen}
A.~Bagchi, P.~Dhivakar, and S.~Dutta, ``{Holography in Flat Spacetimes: the
  case for Carroll},'' \href{http://arxiv.org/abs/2311.11246}{{\ttfamily
  arXiv:2311.11246 [hep-th]}}.

\bibitem{Mason:2023mti}
L.~Mason, R.~Ruzziconi, and A.~Yelleshpur~Srikant, ``{Carrollian amplitudes and
  celestial symmetries},''
  \href{http://dx.doi.org/10.1007/JHEP05(2024)012}{{\em JHEP} {\bfseries 05}
  (2024) 012}, \href{http://arxiv.org/abs/2312.10138}{{\ttfamily
  arXiv:2312.10138 [hep-th]}}.

\bibitem{Have:2024dff}
E.~Have, K.~Nguyen, S.~Prohazka, and J.~Salzer, ``{Massive carrollian fields at
  timelike infinity},'' \href{http://arxiv.org/abs/2402.05190}{{\ttfamily
  arXiv:2402.05190 [hep-th]}}.

\bibitem{Bondi:1962px}
H.~Bondi, M.~van~der Burg, and A.~Metzner, ``{Gravitational waves in general
  relativity. 7. Waves from axisymmetric isolated systems},''
\href{http://dx.doi.org/10.1098/rspa.1962.0161}{{\em Proc.Roy.Soc.Lond.}
  {\bfseries A269} (1962) 21--52}.
%%CITATION = PRSLA,A269,21;%%.

\bibitem{Sachs:1962wk}
R.~Sachs, ``{Gravitational waves in general relativity. 8. Waves in
  asymptotically flat space-times},''
\href{http://dx.doi.org/10.1098/rspa.1962.0206}{{\em Proc.Roy.Soc.Lond.}
  {\bfseries A270} (1962) 103--126}.
%%CITATION = PRSLA,A270,103;%%.

\bibitem{duval:2014uoa}
C.~Duval, G.~Gibbons, P.~Horvathy, and P.~Zhang, ``{Carroll versus Newton and
  Galilei: two dual non-Einsteinian concepts of time},''
  \href{http://dx.doi.org/10.1088/0264-9381/31/8/085016}{{\em
  Class.Quant.Grav.} {\bfseries 31} (2014) 085016},
\href{http://arxiv.org/abs/1402.0657}{{\ttfamily arXiv:1402.0657 [gr-qc]}}.
%%CITATION = ARXIV:1402.0657;%%.

\bibitem{Duval:2014lpa}
C.~Duval, G.~Gibbons, and P.~Horvathy, ``{Conformal Carroll groups},''
  \href{http://dx.doi.org/10.1088/1751-8113/47/33/335204}{{\em J.Phys.}
  {\bfseries A47} (2014) 335204},
\href{http://arxiv.org/abs/1403.4213}{{\ttfamily arXiv:1403.4213 [hep-th]}}.
%%CITATION = ARXIV:1403.4213;%%.

\bibitem{Bidussi:2021nmp}
L.~Bidussi, J.~Hartong, E.~Have, J.~Musaeus, and S.~Prohazka, ``{Fractons,
  dipole symmetries and curved spacetime},''
  \href{http://dx.doi.org/10.21468/SciPostPhys.12.6.205}{{\em SciPost Phys.}
  {\bfseries 12} no.~6, (2022) 205},
  \href{http://arxiv.org/abs/2111.03668}{{\ttfamily arXiv:2111.03668
  [hep-th]}}.

\bibitem{Marsot:2022imf}
L.~Marsot, P.~M. Zhang, M.~Chernodub, and P.~A. Horvathy, ``{Hall effects in
  Carroll dynamics},''
  \href{http://dx.doi.org/10.1016/j.physrep.2023.07.007}{{\em Phys. Rept.}
  {\bfseries 1028} (2023) 1--60},
  \href{http://arxiv.org/abs/2212.02360}{{\ttfamily arXiv:2212.02360
  [hep-th]}}.

\bibitem{Figueroa-OFarrill:2023vbj}
J.~Figueroa-O'Farrill, A.~P\'erez, and S.~Prohazka, ``{Carroll/fracton
  particles and their correspondence},''
  \href{http://dx.doi.org/10.1007/JHEP06(2023)207}{{\em JHEP} {\bfseries 06}
  (2023) 207}, \href{http://arxiv.org/abs/2305.06730}{{\ttfamily
  arXiv:2305.06730 [hep-th]}}.

\bibitem{Figueroa-OFarrill:2023qty}
J.~Figueroa-O'Farrill, A.~P\'erez, and S.~Prohazka, ``{Quantum Carroll/fracton
  particles},'' \href{http://dx.doi.org/10.1007/JHEP10(2023)041}{{\em JHEP}
  {\bfseries 10} (2023) 041}, \href{http://arxiv.org/abs/2307.05674}{{\ttfamily
  arXiv:2307.05674 [hep-th]}}.

\bibitem{Bagchi:2022eui}
A.~Bagchi, A.~Banerjee, R.~Basu, M.~Islam, and S.~Mondal, ``{Magic fermions:
  Carroll and flat bands},''
  \href{http://dx.doi.org/10.1007/JHEP03(2023)227}{{\em JHEP} {\bfseries 03}
  (2023) 227}, \href{http://arxiv.org/abs/2211.11640}{{\ttfamily
  arXiv:2211.11640 [hep-th]}}.

\bibitem{deBoer:2021jej}
J.~de~Boer, J.~Hartong, N.~A. Obers, W.~Sybesma, and S.~Vandoren, ``{Carroll
  Symmetry, Dark Energy and Inflation},''
  \href{http://dx.doi.org/10.3389/fphy.2022.810405}{{\em Front. in Phys.}
  {\bfseries 10} (2022) 810405},
  \href{http://arxiv.org/abs/2110.02319}{{\ttfamily arXiv:2110.02319
  [hep-th]}}.

\bibitem{Bagchi:2023ysc}
A.~Bagchi, K.~S. Kolekar, and A.~Shukla, ``{Carrollian Origins of Bjorken
  Flow},'' \href{http://dx.doi.org/10.1103/PhysRevLett.130.241601}{{\em Phys.
  Rev. Lett.} {\bfseries 130} no.~24, (2023) 241601},
  \href{http://arxiv.org/abs/2302.03053}{{\ttfamily arXiv:2302.03053
  [hep-th]}}.

\bibitem{Bagchi:2023rwd}
A.~Bagchi, K.~S. Kolekar, T.~Mandal, and A.~Shukla, ``{Heavy-ion collisions,
  Gubser flow, and Carroll hydrodynamics},''
  \href{http://dx.doi.org/10.1103/PhysRevD.109.056004}{{\em Phys. Rev. D}
  {\bfseries 109} no.~5, (2024) 056004},
  \href{http://arxiv.org/abs/2310.03167}{{\ttfamily arXiv:2310.03167
  [hep-th]}}.

\bibitem{Ciambelli:2018wre}
L.~Ciambelli, C.~Marteau, A.~C. Petkou, P.~M. Petropoulos, and K.~Siampos,
  ``{Flat holography and Carrollian fluids},''
  \href{http://dx.doi.org/10.1007/JHEP07(2018)165}{{\em JHEP} {\bfseries 07}
  (2018) 165}, \href{http://arxiv.org/abs/1802.06809}{{\ttfamily
  arXiv:1802.06809 [hep-th]}}.

\bibitem{Redondo-Yuste:2022czg}
J.~Redondo-Yuste and L.~Lehner, ``{Non-linear black hole dynamics and
  Carrollian fluids},'' \href{http://dx.doi.org/10.1007/JHEP02(2023)240}{{\em
  JHEP} {\bfseries 02} (2023) 240},
  \href{http://arxiv.org/abs/2212.06175}{{\ttfamily arXiv:2212.06175 [gr-qc]}}.

\bibitem{Ciambelli:2018xat}
L.~Ciambelli, C.~Marteau, A.~C. Petkou, P.~M. Petropoulos, and K.~Siampos,
  ``{Covariant Galilean versus Carrollian hydrodynamics from relativistic
  fluids},'' \href{http://dx.doi.org/10.1088/1361-6382/aacf1a}{{\em Class.
  Quant. Grav.} {\bfseries 35} no.~16, (2018) 165001},
  \href{http://arxiv.org/abs/1802.05286}{{\ttfamily arXiv:1802.05286
  [hep-th]}}.

\bibitem{Campoleoni:2018ltl}
A.~Campoleoni, L.~Ciambelli, C.~Marteau, P.~M. Petropoulos, and K.~Siampos,
  ``{Two-dimensional fluids and their holographic duals},''
  \href{http://dx.doi.org/10.1016/j.nuclphysb.2019.114692}{{\em Nucl. Phys. B}
  {\bfseries 946} (2019) 114692},
  \href{http://arxiv.org/abs/1812.04019}{{\ttfamily arXiv:1812.04019
  [hep-th]}}.

\bibitem{Petkou:2022bmz}
A.~C. Petkou, P.~M. Petropoulos, D.~R. Betancour, and K.~Siampos,
  ``{Relativistic fluids, hydrodynamic frames and their Galilean versus
  Carrollian avatars},'' \href{http://dx.doi.org/10.1007/JHEP09(2022)162}{{\em
  JHEP} {\bfseries 09} (2022) 162},
  \href{http://arxiv.org/abs/2205.09142}{{\ttfamily arXiv:2205.09142
  [hep-th]}}.

\bibitem{Freidel:2022bai}
L.~Freidel and P.~Jai-akson, ``{Carrollian hydrodynamics from symmetries},''
  \href{http://dx.doi.org/10.1088/1361-6382/acb194}{{\em Class. Quant. Grav.}
  {\bfseries 40} no.~5, (2023) 055009},
  \href{http://arxiv.org/abs/2209.03328}{{\ttfamily arXiv:2209.03328
  [hep-th]}}.

\bibitem{Freidel:2022vjq}
L.~Freidel and P.~Jai-akson, ``{Carrollian hydrodynamics and symplectic
  structure on stretched horizons},''
  \href{http://dx.doi.org/10.1007/JHEP05(2024)135}{{\em JHEP} {\bfseries 05}
  (2024) 135}, \href{http://arxiv.org/abs/2211.06415}{{\ttfamily
  arXiv:2211.06415 [gr-qc]}}.

\bibitem{deBoer:2023fnj}
J.~de~Boer, J.~Hartong, N.~A. Obers, W.~Sybesma, and S.~Vandoren, ``{Carroll
  stories},'' \href{http://dx.doi.org/10.1007/JHEP09(2023)148}{{\em JHEP}
  {\bfseries 09} (2023) 148}, \href{http://arxiv.org/abs/2307.06827}{{\ttfamily
  arXiv:2307.06827 [hep-th]}}.

\bibitem{Armas:2023dcz}
J.~Armas and E.~Have, ``{Carrollian Fluids and Spontaneous Breaking of Boost
  Symmetry},'' \href{http://dx.doi.org/10.1103/PhysRevLett.132.161606}{{\em
  Phys. Rev. Lett.} {\bfseries 132} no.~16, (2024) 161606},
  \href{http://arxiv.org/abs/2308.10594}{{\ttfamily arXiv:2308.10594
  [hep-th]}}.

\bibitem{Bergshoeff:2017btm}
E.~Bergshoeff, J.~Gomis, B.~Rollier, J.~Rosseel, and T.~ter Veldhuis,
  ``{Carroll versus Galilei Gravity},''
  \href{http://dx.doi.org/10.1007/JHEP03(2017)165}{{\em JHEP} {\bfseries 03}
  (2017) 165},
\href{http://arxiv.org/abs/1701.06156}{{\ttfamily arXiv:1701.06156 [hep-th]}}.
%%CITATION = ARXIV:1701.06156;%%.

\bibitem{Grumiller:2020elf}
D.~Grumiller, J.~Hartong, S.~Prohazka, and J.~Salzer, ``{Limits of JT
  gravity},'' \href{http://dx.doi.org/10.1007/JHEP02(2021)134}{{\em JHEP}
  {\bfseries 02} (2021) 134}, \href{http://arxiv.org/abs/2011.13870}{{\ttfamily
  arXiv:2011.13870 [hep-th]}}.

\bibitem{Hansen:2021fxi}
D.~Hansen, N.~A. Obers, G.~Oling, and B.~T. S\o{}gaard, ``{Carroll Expansion of
  General Relativity},''
  \href{http://dx.doi.org/10.21468/SciPostPhys.13.3.055}{{\em SciPost Phys.}
  {\bfseries 13} no.~3, (2022) 055},
  \href{http://arxiv.org/abs/2112.12684}{{\ttfamily arXiv:2112.12684
  [hep-th]}}.

\bibitem{Campoleoni:2022ebj}
A.~Campoleoni, M.~Henneaux, S.~Pekar, A.~P\'erez, and P.~Salgado-Rebolledo,
  ``{Magnetic Carrollian gravity from the Carroll algebra},''
  \href{http://dx.doi.org/10.1007/JHEP09(2022)127}{{\em JHEP} {\bfseries 09}
  (2022) 127}, \href{http://arxiv.org/abs/2207.14167}{{\ttfamily
  arXiv:2207.14167 [hep-th]}}.

\bibitem{Perez:2021abf}
A.~P\'erez, ``{Asymptotic symmetries in Carrollian theories of gravity},''
  \href{http://dx.doi.org/10.1007/JHEP12(2021)173}{{\em JHEP} {\bfseries 12}
  (2021) 173}, \href{http://arxiv.org/abs/2110.15834}{{\ttfamily
  arXiv:2110.15834 [hep-th]}}.

\bibitem{Ecker:2023uwm}
F.~Ecker, D.~Grumiller, J.~Hartong, A.~P\'erez, S.~Prohazka, and R.~Troncoso,
  ``{Carroll black holes},'' \href{http://arxiv.org/abs/2308.10947}{{\ttfamily
  arXiv:2308.10947 [hep-th]}}.

\bibitem{Aggarwal:2024gfb}
A.~Aggarwal, F.~Ecker, D.~Grumiller, and D.~Vassilevich, ``{Carroll Hawking
  effect},'' \href{http://arxiv.org/abs/2403.00073}{{\ttfamily arXiv:2403.00073
  [hep-th]}}.

\bibitem{VandenBleeken:2017rij}
D.~Van~den Bleeken, ``Torsional Newton-Cartan gravity from the large c
  expansion of general relativity,''
  \href{http://dx.doi.org/10.1088/1361-6382/aa83d4}{{\em Class. Quant. Grav.}
  {\bfseries 34} no.~18, (2017) 185004},
\href{http://arxiv.org/abs/1703.03459}{{\ttfamily arXiv:1703.03459 [gr-qc]}}.
%%CITATION = ARXIV:1703.03459;%%.

\bibitem{Hansen:2018ofj}
D.~Hansen, J.~Hartong, and N.~A. Obers, ``{Action Principle for Newtonian
  Gravity},'' \href{http://dx.doi.org/10.1103/PhysRevLett.122.061106}{{\em
  Phys. Rev. Lett.} {\bfseries 122} no.~6, (2019) 061106},
\href{http://arxiv.org/abs/1807.04765}{{\ttfamily arXiv:1807.04765 [hep-th]}}.
%%CITATION = ARXIV:1807.04765;%%.

\bibitem{Hansen:2019vqf}
D.~Hansen, J.~Hartong, and N.~A. Obers, ``{Gravity between Newton and
  Einstein},'' \href{http://dx.doi.org/10.1142/S0218271819440103}{{\em Int. J.
  Mod. Phys. D} {\bfseries 28} no.~14, (2019) 1944010},
  \href{http://arxiv.org/abs/1904.05706}{{\ttfamily arXiv:1904.05706 [gr-qc]}}.

\bibitem{Bergshoeff:2019ctr}
E.~Bergshoeff, J.~M. Izquierdo, T.~Ort\'\i{}n, and L.~Romano, ``{Lie Algebra
  Expansions and Actions for Non-Relativistic Gravity},''
  \href{http://dx.doi.org/10.1007/JHEP08(2019)048}{{\em JHEP} {\bfseries 08}
  (2019) 048}, \href{http://arxiv.org/abs/1904.08304}{{\ttfamily
  arXiv:1904.08304 [hep-th]}}.

\bibitem{Hansen:2020pqs}
D.~Hansen, J.~Hartong, and N.~A. Obers, ``{Non-Relativistic Gravity and its
  Coupling to Matter},'' \href{http://dx.doi.org/10.1007/JHEP06(2020)145}{{\em
  JHEP} {\bfseries 06} (2020) 145},
  \href{http://arxiv.org/abs/2001.10277}{{\ttfamily arXiv:2001.10277 [gr-qc]}}.

\bibitem{Hartong:2023yxo}
J.~Hartong, E.~Have, N.~A. Obers, and I.~Pikovski, ``{A coupling prescription
  for post-Newtonian corrections in Quantum Mechanics},''
  \href{http://dx.doi.org/10.21468/SciPostPhys.16.3.088}{{\em SciPost Phys.}
  {\bfseries 16} (2024) 088}, \href{http://arxiv.org/abs/2308.07373}{{\ttfamily
  arXiv:2308.07373 [gr-qc]}}.

\bibitem{Ergen:2020yop}
M.~Ergen, E.~Hamamci, and D.~Van~den Bleeken, ``{Oddity in nonrelativistic,
  strong gravity},''
  \href{http://dx.doi.org/10.1140/epjc/s10052-020-8112-6}{{\em Eur. Phys. J. C}
  {\bfseries 80} no.~6, (2020) 563},
  \href{http://arxiv.org/abs/2002.02688}{{\ttfamily arXiv:2002.02688 [gr-qc]}}.
  [Erratum: Eur.Phys.J.C 80, 657 (2020)].

\bibitem{Hartong:2023ckn}
J.~Hartong and J.~Musaeus, ``{Toward a covariant framework for post-Newtonian
  expansions for radiative sources},''
  \href{http://dx.doi.org/10.1103/PhysRevD.109.124058}{{\em Phys. Rev. D}
  {\bfseries 109} no.~12, (2024) 124058},
  \href{http://arxiv.org/abs/2311.07546}{{\ttfamily arXiv:2311.07546 [gr-qc]}}.

\bibitem{Hartong:2021ekg}
J.~Hartong and E.~Have, ``{Nonrelativistic Expansion of Closed Bosonic
  Strings},'' \href{http://dx.doi.org/10.1103/PhysRevLett.128.021602}{{\em
  Phys. Rev. Lett.} {\bfseries 128} no.~2, (2022) 021602},
  \href{http://arxiv.org/abs/2107.00023}{{\ttfamily arXiv:2107.00023
  [hep-th]}}.

\bibitem{Hartong:2022dsx}
J.~Hartong and E.~Have, ``{Nonrelativistic approximations of closed bosonic
  string theory},'' \href{http://dx.doi.org/10.1007/JHEP02(2023)153}{{\em JHEP}
  {\bfseries 02} (2023) 153}, \href{http://arxiv.org/abs/2211.01795}{{\ttfamily
  arXiv:2211.01795 [hep-th]}}.

\bibitem{Hartong:2024ydv}
J.~Hartong and E.~Have, ``{Non-relativistic expansion of open strings and
  D-branes},'' \href{http://arxiv.org/abs/2407.05985}{{\ttfamily
  arXiv:2407.05985 [hep-th]}}.

\bibitem{Cardona:2016ytk}
B.~Cardona, J.~Gomis, and J.~M. Pons, ``{Dynamics of Carroll Strings},''
  \href{http://dx.doi.org/10.1007/JHEP07(2016)050}{{\em JHEP} {\bfseries 07}
  (2016) 050}, \href{http://arxiv.org/abs/1605.05483}{{\ttfamily
  arXiv:1605.05483 [hep-th]}}.

\bibitem{Blair:2023noj}
C.~D.~A. Blair, J.~Lahnsteiner, N.~A.~J. Obers, and Z.~Yan, ``{Unification of
  Decoupling Limits in String and M Theory},''
  \href{http://dx.doi.org/10.1103/PhysRevLett.132.161603}{{\em Phys. Rev.
  Lett.} {\bfseries 132} no.~16, (2024) 161603},
  \href{http://arxiv.org/abs/2311.10564}{{\ttfamily arXiv:2311.10564
  [hep-th]}}.

\bibitem{Gomis:2023eav}
J.~Gomis and Z.~Yan, ``{Worldsheet Formalism for Decoupling Limits in String
  Theory},'' \href{http://arxiv.org/abs/2311.10565}{{\ttfamily arXiv:2311.10565
  [hep-th]}}.

\bibitem{Harksen:2024bnh}
M.~Harksen, D.~Hidalgo, W.~Sybesma, and L.~Thorlacius, ``{Carroll strings with
  an extended symmetry algebra},''
  \href{http://dx.doi.org/10.1007/JHEP05(2024)206}{{\em JHEP} {\bfseries 05}
  (2024) 206}, \href{http://arxiv.org/abs/2403.01984}{{\ttfamily
  arXiv:2403.01984 [hep-th]}}.

\bibitem{Casalbuoni:2024jmj}
R.~Casalbuoni, D.~Dominici, and J.~Gomis, ``{Non equivalence of Carroll limits
  in relativistic string theory},''
  \href{http://arxiv.org/abs/2403.02152}{{\ttfamily arXiv:2403.02152
  [hep-th]}}.

\bibitem{Bagchi:2015nca}
A.~Bagchi, S.~Chakrabortty, and P.~Parekh, ``{Tensionless Strings from
  Worldsheet Symmetries},''
  \href{http://dx.doi.org/10.1007/JHEP01(2016)158}{{\em JHEP} {\bfseries 01}
  (2016) 158}, \href{http://arxiv.org/abs/1507.04361}{{\ttfamily
  arXiv:1507.04361 [hep-th]}}.

\bibitem{Isberg:1993av}
J.~Isberg, U.~Lindstr\"om, B.~Sundborg, and G.~Theodoridis, ``{Classical and
  quantized tensionless strings},''
  \href{http://dx.doi.org/10.1016/0550-3213(94)90056-6}{{\em Nucl. Phys.}
  {\bfseries B411} (1994) 122--156},
\href{http://arxiv.org/abs/hep-th/9307108}{{\ttfamily arXiv:hep-th/9307108
  [hep-th]}}.
%%CITATION = HEP-TH/9307108;%%.

\bibitem{Bagchi:2013bga}
A.~Bagchi, ``{Tensionless Strings and Galilean Conformal Algebra},''
  \href{http://dx.doi.org/10.1007/JHEP05(2013)141}{{\em JHEP} {\bfseries 05}
  (2013) 141},
\href{http://arxiv.org/abs/1303.0291}{{\ttfamily arXiv:1303.0291 [hep-th]}}.
%%CITATION = ARXIV:1303.0291;%%.

\bibitem{Fontanella:2022gyt}
A.~Fontanella, ``{Non-extremal near-horizon geometries},''
  \href{http://dx.doi.org/10.1088/1361-6382/acd980}{{\em Class. Quant. Grav.}
  {\bfseries 40} no.~13, (2023) 135006},
  \href{http://arxiv.org/abs/2211.03861}{{\ttfamily arXiv:2211.03861 [gr-qc]}}.

\bibitem{Bagchi:2021ban}
A.~Bagchi, A.~Banerjee, S.~Chakrabortty, and R.~Chatterjee, ``{A Rindler road
  to Carrollian worldsheets},''
  \href{http://dx.doi.org/10.1007/JHEP04(2022)082}{{\em JHEP} {\bfseries 04}
  (2022) 082}, \href{http://arxiv.org/abs/2111.01172}{{\ttfamily
  arXiv:2111.01172 [hep-th]}}.

\bibitem{deVega:1987um}
H.~J. de~Vega and N.~G. Sanchez, ``{String Quantization in Accelerated Frames
  and Black Holes},''
  \href{http://dx.doi.org/10.1016/0550-3213(88)90374-4}{{\em Nucl. Phys. B}
  {\bfseries 299} (1988) 818}.

\bibitem{Ganor:1994rm}
O.~Ganor, J.~Sonnenschein, and S.~Yankielowicz, ``{Folds in 2-D string
  theories},'' \href{http://dx.doi.org/10.1016/0550-3213(94)90275-5}{{\em Nucl.
  Phys. B} {\bfseries 427} (1994) 203--244},
  \href{http://arxiv.org/abs/hep-th/9404149}{{\ttfamily arXiv:hep-th/9404149}}.

\bibitem{Bars:1994sv}
I.~Bars and J.~Schulze, ``{Folded strings falling into a black hole},''
  \href{http://dx.doi.org/10.1103/PhysRevD.51.1854}{{\em Phys. Rev. D}
  {\bfseries 51} (1995) 1854--1868},
  \href{http://arxiv.org/abs/hep-th/9405156}{{\ttfamily arXiv:hep-th/9405156}}.

\bibitem{Bardeen:1975gx}
W.~A. Bardeen, I.~Bars, A.~J. Hanson, and R.~D. Peccei, ``{A Study of the
  Longitudinal Kink Modes of the String},''
  \href{http://dx.doi.org/10.1103/PhysRevD.13.2364}{{\em Phys. Rev. D}
  {\bfseries 13} (1976) 2364--2382}.

\bibitem{Bagchi:2020fpr}
A.~Bagchi, A.~Banerjee, S.~Chakrabortty, S.~Dutta, and P.~Parekh, ``{A tale of
  three \textemdash{} tensionless strings and vacuum structure},''
  \href{http://dx.doi.org/10.1007/JHEP04(2020)061}{{\em JHEP} {\bfseries 04}
  (2020) 061}, \href{http://arxiv.org/abs/2001.00354}{{\ttfamily
  arXiv:2001.00354 [hep-th]}}.

\bibitem{Susskind:1993ki}
L.~Susskind, ``{String theory and the principles of black hole
  complementarity},'' \href{http://dx.doi.org/10.1103/PhysRevLett.71.2367}{{\em
  Phys. Rev. Lett.} {\bfseries 71} (1993) 2367--2368},
  \href{http://arxiv.org/abs/hep-th/9307168}{{\ttfamily arXiv:hep-th/9307168}}.

\bibitem{figueroa-ofarrill:2022mcy}
J.~Figueroa-O'Farrill, E.~Have, S.~Prohazka, and J.~Salzer, ``{The gauging
  procedure and carrollian gravity},''
  \href{http://dx.doi.org/10.1007/JHEP09(2022)243}{{\em JHEP} {\bfseries 09}
  (2022) 243}, \href{http://arxiv.org/abs/2206.14178}{{\ttfamily
  arXiv:2206.14178 [hep-th]}}.

\bibitem{Bergshoeff:2023rkk}
E.~Bergshoeff, J.~Figueroa-O'Farrill, K.~van Helden, J.~Rosseel, I.~Rotko, and
  T.~ter Veldhuis, ``{-brane Galilean and Carrollian geometries and
  gravities},'' \href{http://dx.doi.org/10.1088/1751-8121/ad4c62}{{\em J. Phys.
  A} {\bfseries 57} no.~24, (2024) 245205},
  \href{http://arxiv.org/abs/2308.12852}{{\ttfamily arXiv:2308.12852
  [hep-th]}}.

\bibitem{Barducci:2018wuj}
A.~Barducci, R.~Casalbuoni, and J.~Gomis, ``{Confined dynamical systems with
  Carroll and Galilei symmetries},''
  \href{http://dx.doi.org/10.1103/PhysRevD.98.085018}{{\em Phys. Rev. D}
  {\bfseries 98} no.~8, (2018) 085018},
  \href{http://arxiv.org/abs/1804.10495}{{\ttfamily arXiv:1804.10495
  [hep-th]}}.

\bibitem{Bergshoeff:2020xhv}
E.~Bergshoeff, J.~M. Izquierdo, and L.~Romano, ``{Carroll versus Galilei from a
  Brane Perspective},'' \href{http://dx.doi.org/10.1007/JHEP10(2020)066}{{\em
  JHEP} {\bfseries 10} (2020) 066},
  \href{http://arxiv.org/abs/2003.03062}{{\ttfamily arXiv:2003.03062
  [hep-th]}}.

\bibitem{Bergshoeff:2014jla}
E.~Bergshoeff, J.~Gomis, and G.~Longhi, ``{Dynamics of Carroll Particles},''
  \href{http://dx.doi.org/10.1088/0264-9381/31/20/205009}{{\em
  Class.Quant.Grav.} {\bfseries 31} no.~20, (2014) 205009},
\href{http://arxiv.org/abs/1405.2264}{{\ttfamily arXiv:1405.2264 [hep-th]}}.
%%CITATION = ARXIV:1405.2264;%%.

\end{thebibliography}
\end{document}